\pdfoutput=1
\documentclass[10pt,twocolumn,preprintnumbers,amsmath,amssymb,nofootinbib
,superscriptaddress]{revtex4-1}
\usepackage{graphicx,longtable,mathrsfs,color,array}
\usepackage[hidelinks]{hyperref}
\usepackage[usenames,dvipsnames]{xcolor} 
\usepackage{amssymb,amsmath,mathtools,mathrsfs,slashed} 
\usepackage{epsfig,subfigure,placeins,float} 
\usepackage{booktabs,longtable,ctable,multirow} 
\usepackage{exscale,relsize} 
\usepackage[normalem]{ulem} 
\usepackage{enumerate}
\usepackage{times, mathptmx} 
\usepackage[utf8]{inputenc}
\usepackage{color}
\usepackage{graphicx}
\usepackage{color}
\usepackage{graphicx,graphics}
\usepackage{bm}
\usepackage{tabularx}
\usepackage{rotating}
\usepackage{units}
\allowdisplaybreaks[1]
\usepackage{hhline}
\usepackage{longtable}
\usepackage{placeins}
\usepackage{natbib}
\begin{document}
\title[CCSN Dust Masses]{Dust masses for a large sample of core-collapse supernovae from optical emission line asymmetries: dust formation on 30-year timescales}



\author{Maria Niculescu-Duvaz}
\affiliation{Dept. of Physics \& Astronomy, University College London, Gower St, London WC1E 6BT,UK}
\author{Michael J. Barlow}
\affiliation{Dept. of Physics \& Astronomy, University College London, Gower St, London WC1E 6BT,UK}
\author{Antonia Bevan}
\affiliation{Dept. of Physics \& Astronomy, University College London, Gower St, London WC1E 6BT,UK}
\author{Roger Wesson}
\affiliation{Dept. of Physics \& Astronomy, University College London, Gower St, London WC1E 6BT,UK}
\author{Danny Milisavljevic}{
\affiliation{Department of Physics and Astronomy, Purdue University, 525 Northwestern Ave., West Lafayette, IN 47907, USA}
\author{Ilse De Looze}
\affiliation{Sterrenkundig Observatorium, Ghent University, Krijgslaan 281 - S9, 9000 Gent, Belgium}
\author{Geoff C. Clayton}
\affiliation{Department of Physics \& Astronomy, Louisiana State University, Baton Rouge, LA 70803, USA}
\author{Kelsie Krafton}
\affiliation{Department of Physics \& Astronomy, Louisiana State University, Baton Rouge, LA 70803, USA}
\author{Mikako Matsuura}
\affiliation{School of Physics \& Astronomy, Cardiff University, Cardiff, Wales, UK}
\author{Ryan Brady}
\affiliation{Dept. of Physics \& Astronomy, University College London, Gower St, London WC1E 6BT,UK}

\begin{abstract}

Modelling the red-blue asymmetries seen in the broad emission lines of core-collapse supernovae (CCSNe) is a powerful technique to quantify total dust mass formed in the ejecta at late times ($>5$ years after outburst) when ejecta dust temperatures become too low to be detected by mid-IR instruments. Following our success in using the Monte Carlo radiative transfer code {\sc damocles} to measure the dust mass evolution in SN~1987A and other CCSNe, we present the most comprehensive sample of dust mass measurements yet made with {\sc damocles}, for CCSNe aged between four and sixty years after outburst. Our sample comprises of multi-epoch late-time optical spectra taken with the Gemini GMOS and VLT X-Shooter spectrographs, supplemented by archival spectra. For the fourteen CCSNe that we have modelled, we confirm a dust mass growth with time that can be fit by a sigmoid curve which is found to saturate beyond an age of $\sim30$~years, at a mass of 0.23$^{+0.17}_{-0.12}$~M$_\odot$. An expanded sample including dust masses found in the literature for a further eleven CCSNe and six CCSN remnants, the dust mass at saturation is found to be 0.42$^{+0.09}_{-0.05}$~M$_\odot$. Uncertainty limits for our dust masses were determined from a Bayesian analysis using the affine invariant Markov Chain Monte Carlo ensemble sampler {\em emcee} with {\sc damocles}. The best-fitting line profile models for our sample all required grain radii between 0.1 and 0.5~$\mu$m. Our results are consistent with CCSNe forming enough dust in their ejecta to significantly contribute to the dust budget of the Universe.

\end{abstract}
\maketitle



\section{Introduction}

Since the discovery of large amounts of dust in highly redshifted young galaxies \citep[e.g.][]{Bertoldi2003, Watson2015a, Laporte2017b}, the origin of the bulk of the cosmic dust in the Universe has been debated.
It has been proposed that a significant fraction of cosmic dust, particularly at high redshifts, is formed in the ejecta of core-collapse supernovae (CCSNe), with \citet{Morgan2003a} and \citet{Dwek} estimating that each CCSN would need to produce $\geq$0.1 M$_\odot$ of dust for this to be the case. Theoretical models predict that CCSNe can form $>0.1$~M$_\odot$ of dust \citep[][]{Nozawa2003, Sarangi2015}. However, in theoretical simulations the amount of ejecta-formed dust found to be destroyed by the reverse shock has varied between 0-100~per~cent depending on the details of the simulation \citep{Nath, Silvia2010, Bocchio2014, Micelotta2016, Kirchschlager2019, Slavin2020, Priestley2021}. A more comprehensive knowledge of dust masses, grain sizes and radial locations is needed for a range of different types of CCSNe of different ages in order to inform dust destruction simulations.

Over the last few decades, the capability of CCSNe to form large amounts of dust has been well established.
Model fits to the SEDs of warm dust emitting at mid-IR wavelengths as measured with the {\em Spitzer Space Telescope} found dust masses of only around $10^{-4}-10^{-2}$~M$_\odot$ to be present in CCSNe and SNRs, less than the theoretical predictions \citep{Sugerman2006, Rho2009, Fabbri2011a}.
This changed with the work of \citet{Matsuura}, who used {\em Herschel Space Observatory} observations of SN~1987A taken 23 years after outburst to probe previously undetectable cold dust emitting at far-IR wavelengths and found a cold dust mass of $\sim$0.5~M$_\odot$. Follow-up ALMA observations of SN~1987A \citep{Indebetouw2014}, resolved this dust component to be in the centre of the remnant. Since then, far-IR data has been utilised to detect cold dust masses of 0.04 - 1.0~M$_\odot$ in several supernovae remnants, including Cas A, with ages ranging from several hundred to several thousand years old \citep{Gomez2012, DeLooze2017, Temim2017, Chawner2019, Priestley2019, DeLooze2019, Niculescu-Duvaz2021}.

There is a current lack of far-IR or submillimetre telescopes with sensitivities sufficient to detect supernovae beyond the Local Group, and existing observations have only been able to detect cold dust in nearby Galactic and Magellanic Cloud CC-SNRs. However, \citet{Lucy1989} showed for SN~1987A that dust located within the emitting gas can produce a red-blue line asymmetry in the broad line profiles. This effect is caused by light from the receding red-shifted side of a CCSNe being attenuated by more dust than light from the approaching blue-shifted side. Since then, many authors have noted the presence of red-blue asymmetries in the line profiles of CCSNe (e.g. \citet{Smith2008} for SN 2006jc, \citet{Mauerhan2012} for SN 1998S, \citet{Gall2014} for SN 2010jl), while \citet{Milisavljevic2012} presented a sample of 10 late-time CCSN spectra that all displayed varying degrees of blue-shifted emission peaks in their optical line profiles.

We note that \citet{Anderson2014} discussed the observed blue-shifting of H$\alpha$ emission peaks during the very early post-explosion phases of Type~II CCSNe and attributed this to the H$\alpha$ emission line formation radius during such early phases being at or within the effective photospheric radius of the expanding ejecta, with much of the red-shifted H$\alpha$ emission consequently occulted by the photospheric disk. However, they showed from theoretical modelling that by epochs greater than $\sim$150 days post-explosion the effective photospheric radius should have moved within the formation zone for H$\alpha$ emission line photons, with no blue-shifting of H$\alpha$ emission peaks by this mechanism predicted beyond this epoch, in agreement with observations. Any subsequent blue-shifting of emission line peaks can therefore be attributed to the effects of newly formed dust within the ejecta.

To increase the number of CCSNe with known dust mass determinations, \citet{bevan2016} presented the Monte Carlo radiative transfer code {\sc damocles} which quantifies the amount of newly formed dust causing absorption and scattering of SN ejecta line emission. It is able to treat arbitrary dust/gas geometries, a range of velocity and density distributions and dust and gas clumping configurations, as well as handling a wide range of grain species and radii. 
\citet{bevan2016} used {\sc damocles} to model the line profiles in the optical spectra of SN~1987A between 714 and 3604 days, allowing them to determine the ejecta-condensed dust mass. Their results were in agreement with the work of \citet{Wesson2015} for epochs in common, showing a steady increase in dust mass with time and that by day 3604 the dust mass in SN 1987A was $\sim$0.1 M$_{\odot}$. Since the dust mass on days 8500-9200 had reached 0.6-0.8~M$_\odot$ \citep{Matsuura, Wesson2015}, most of the dust in this CCSN must have formed after day 3604. \citet{Bevan2017} also applied {\sc damocles} to model the line profiles of three other objects: SN~1980K, SN~1993J and Cas~A, finding ejecta dust masses ranging from 0.1~M$_\odot$ for SN~1993J to 1.1~M$_\odot$ for Cas~A.

\citet{Wesson2015} tracked the dust mass evolution in SN~1987A over its first 25 years by modelling its optical to far-IR SEDs, where they found the dust mass grew from $3\times10^{-3}$~M$_{\odot}$ at 615 days to 0.8~M$_{\odot}$ at 9200 days.  \citet{Wesson2015} noted that large grains $>2~\mu$m in radius were required to fit the SED of SN~1987A at late epochs. \citet{Wesson2015} were able to fit a sigmoid curve to SN~1987A's slow dust mass evolution with time.
The dust mass evolution over a period of ten years in SN~2005ip, quantified from red-blue line asymmetry fits by \citet{Bevan2019}, confirmed the dust-growth trend seen for SN~1987A by \citet{Wesson2015}. A similar trend for SN~2010jl at early epochs is also seen \citep{Gall2014, bevan2020}. Theoretical predictions of dust mass growth in CCSNe do not agree with these trends, with e.g. \citet{Sarangi2015} and \citet{Sluder2018} predicting that most of the dust should form quite rapidly, within the first three years post-explosion.

\citet{Dwek2019} have argued that this dichotomy can be resolved if ejecta dust masses actually grow to their final values within two to three years, with most of the dust initially hidden in optically thick clumps - with the continued expansion of the ejecta these clumps would eventually become optically thin, revealing the  full dust mass. However, by combining {\sc damocles} optical line profile modelling with {\sc mocassin} optical-IR SED modelling, \citet{wesson_bevan2021} have shown that SN~1987A cannot have formed a dust mass larger than 0.01~M$_\odot$ by day 1000, even with optically thick clumps. Dust masses larger than this value that can fit the observed SED cannot match the observed optical line profiles and dust masses larger than 0.01~M$_\odot$ that can fit the observed line profiles cannot fit the observed SED.

Independent evidence for the slow
growth of supernova dust with time has come from the work of \citet{Liu2018}, whose study of pre-solar SiC dust grains from supernovae used the radio-active $^{49}$V-$^{49}$Ti chronometer to show that the dust grains formed at least two years, and most likely ten years, after the parent star had exploded. This timescale is consistent with the work of \citet{ott2019}, who measured barium isotope ratios in supernova-condensed dust grains found in primitive meteorites, and inferred that the dust grains had condensed about 20 years after explosion.

The time at which CCSN observational dust mass growth plateaus or saturates is still unclear, as the dust-mass growth curve is still sparsely sampled in the evolutionary stage between a supernova and a supernova remnant, at ages between 20-100 years. It is around this stage where the effect of the reverse shock on the ejecta dust also starts to become significant, so it is particularly important to determine the dust masses in CCSNe of a similar age to and older than SN 1987A. It can also be argued that even for earlier times there are not yet enough CCSN dust mass estimates available to be able to discern correlations between dust mass, grain radius and CCSN properties such as progenitor mass and SN sub-type. In this work, we aim to increase the sample of CCSNe that have derived dust masses with robustly quantified uncertainties across a range of ages. We also try, where possible, to constrain the dust grain radius and composition.

The spectra that are modelled in this paper are drawn from a Gemini GMOS and VLT X-Shooter late-time survey of CCSN spectra (Wesson et al. 2022, in preparation). For their sample they selected CCSNe that had been discovered before 2013 and which, with a few exceptions, had occurred in host galaxies at
distances $<$ 30 Mpc. From the 306 initially selected CCSNe, they retained the CCSNe covered by {\em Hubble Space Telescope ({\em HST})} optical images taken after the explosion date. In addition, in order to avoid confusion by dense star fields, they only selected CCSNe which were located in more isolated galaxy parts. 
The final X-Shooter plus GMOS sample consisted of fifty-five CCSNe that had exploded between 1957 and 2012. Broad emission lines were detected from fourteen of these supernovae. The modelling of the line profiles of twelve of these objects using {\sc damocles} is reported in the current paper. {\sc damocles} models for a thirteenth, SN~2005ip, have already been published by \citet{Bevan2019},
while {\sc damocles} models for the fourteenth, SN~1995N, are reported by Wesson et al. (2022, MNRAS submitted).
For eight of the CCSNe that had X-Shooter or GMOS spectral detections, we were able to supplement those data with archival spectra, which also provided us with late-time spectra of two further CCSNe,
SN~1993J and SN~1998S, that have broad-line detections. 

As discussed by Wesson et al. (2022), a key characteristic that appears to be required for very late time emission from CCSNe to be detectable is the presence of strong interactions between the supernova ejecta and circumstellar material \citep{Fesen1999, Milisavljevic2012}. Such interactions may make broad-line emission from the ejecta detectable through one or all of these mechanisms: (a)  
irradiation of ejecta material by X-ray and UV photons emitted from shocked interaction regions (Wesson et al. 2022); (b) emission from a reverse shock propagating back into the ejecta \citep{Fesen2020}; or (c) ejecta emission from heating by a pulsar wind nebula \citep[e.g. the Crab Nebula or SN~1957D,][]{owen2015, Long2012}. The dust responsible for blue-shifting the emission line peaks must be located either in, or interior to, the line-emitting material.

\section{Observations}
Most of the optical spectra we model in this work are from the Gemini GMOS and VLT X-Shooter late-time spectroscopic survey of CCSNe presented by Wesson et al. (2022, in preparation), where fuller details of the observations can be found. A summary of the GMOS and X-Shooter spectra modelled in this work can be found in Tables~\ref{table:gmos-dat} and \ref{table:xshoot-dat}.
For some objects we also model archival data or unpublished spectra, which are summarised in Table~\ref{table:archiv-obs-sum}. 

The Gemini GMOS-S and GMOS-N spectra generally cover the range 4400-7500~\AA. All spectra were obtained in long-slit mode using the B600 grating, with a slit width of 0.75~arcsec. The spectra were taken at two or three central wavelength settings and co-added to prevent important spectral features from falling in detector gaps. The spectra have a resolution of 3.5~\AA\ at a wavelength of 6000~\AA. The 2D spectra were bias-corrected, flat-fielded and wavelength calibrated using the {\sc iraf} {\em gemini} package, and corrected for cosmic rays using the {\sc lacos} package of \citet{vanDokkum2001}.
The sky subtraction regions were determined by visual inspection and the spectra were extracted using 15 rows centered on the supernova's position.

\begin{table*}
	\caption{Gemini GMOS Observations - details of targets modelled in this work.}
\begin{tabular}{ccccccccc}
Name & Date & Epoch (d) & Program  & Exp.Time (s) & Central $\lambda$ (\AA ) & Galaxy & D(Mpc) & SN Type \\
	\hline
SN 1957D & 24 Apr 2015 & 20,949 & GS-2015A-Q-53 & 3$\times$900 & 5970 & NGC 5236 & 4.6 & II\\
 " & 7 Mar 2018 & 22,010 & GS-2018A-Q-311 & 6$\times$1200 & " & "  \\
SN 1970G & 22 May 2016 & 16,733 & GN-2016A-Q-85 & 4$\times$900 & 5225 & NGC 5457 & 6.7 & II-L  \\
SN 1979C & 16 Apr 2015 & 13,150 & GN-2015A-Q-56 & 3$\times$600 & 5970 & NGC 4321 & 15.0 & II-L \\
 " & 12 May 2017 & 13,907 &  GN-2017A-Q-72 & 12$\times$600 & "    \\
SN 1980K & 23;24 Apr 2016 & 12,988 & GN-2016A-Q-85 & 4$\times$900 & 5225 & NGC 6946   & 7.7 & II-L \\
 " & 8;12 May 2018 &  13,734 & GN-2018A-Q-313 & 6$\times$1200 & 5970 & \\
SN 1986E & 9 Apr 2015 & 10,588  &  GN-2015A-Q-56 & 3$\times$900 & 5970 & NGC 4302 & 17.0 & II-L \\
 " & 17 Apr 2018 & 11,692 & GN-2018A-Q-313 & 6$\times$900 & " & "  \\
SN 2004et & 28 Apr 2015 & 3849 & GN-2015A-Q-56 & 3$\times$900 & 5225 & NGC 6946 & 7.7 & II-P  \\
 "  & 25 Apr 2017 & 4577 & GN-2017A-Q-72 & 12$\times$600 & 5970 & " & \\
SN 2007it & 19 Apr 2015 & 2775 &  GS-2015A-Q-53 & 3$\times$900 & 5970 & NGC 5530 & 12.0 & II \\
 " & 1 Mar 2018 & 3822 & GS-2018A-Q-311 & 6$\times$1200 & 5970 & " & \\ 
 SN 2010jl & 22 Jan 2016 & 1906 & GN-2016A-Q-85 & 4$\times$900 & 5225 & UGC 5189A & 49 & IIn \\
SN 2011ja & 26 Feb 2014 & 807 & GS-2014A-Q-70 & 6$\times$900.48 & 5970 & NGC 4945 & 3.6 & II  \\
 " & 9 Jun 2015 & 1275 & GS-2015A-Q-53 & 3$\times$900 & 5970 & " & \\
 " & 17 Feb 2018 & 2259 & GS-2018A-Q-311 & 6$\times$1200 & 5970 & " & \\
 
    \hline
  \end{tabular}
    \label{table:gmos-dat}
\end{table*}

\begin{table*}
	\caption{ESO VLT X-Shooter Observations - details of targets modelled in this work.}
	
  \begin{tabular}{cccccccccc}

Name & Date & Epoch (d) & OB ID  & UVB (s) & VIS (s) & NIR (s) & Galaxy & D(Mpc) & SN Type\\ 
\hline
Programme 097.D-0525(A)\\
\hline
SN 1957D & 08 Jun 2016 & 21,360 &  1342702 & 1145.76 & 1051.76 & 6$\times$200 & NGC 5236 & 4.6 &  II\\
 " & 26 Jan 2017 &  21,592 & 1342707 & " & " & " & " & \\
 " & 27 Jan 2017 & 21,593 & 1342710 & " & " & " & " & \\
 " & 28 Jan 2017 & 21,598 & 1342713 & " & " & 
" & " & \\
SN 1996cr & 10 May 2016 & 7360 & 1343004 & 1145.76 & 1051.76 & 6$\times$200 &  ESO 097-13 & 4.2 & IIn: \\
 " & 12 May 2016 & 7362 & 1343009 & " & " & " & " & \\
 " & 08 Jun 2016 & 7389 & 1343012 & " & " & " & " & \\
SN 2011ja & 10 May 2016 & 1611 & 1342667 & 1145.76 & 1051.76 & 6$\times$200 & NGC 4945 & 3.6 & II  \\
 " & 10 May 2016 & 1611 & 1342671 & " & " & " & " & \\
 " & 10 May 2016 & 1611 & 1342674 & " & " & " & " & \\
\hline
Programme 0103.D-0281(A)\\
\hline
SN 1996cr & 01 May 2019 & 8446 & 2292225 & 1522.00 & 1500.00 & 5$\times$300 & ESO 097-13 & 4.2 & IIn:  \\
 " & 29 May 2019 & 8474 & 2291103 & " & " & " & " & \\
    \hline
  \end{tabular}
  \label{table:xshoot-dat}
\end{table*}

\begin{table*}
\caption{Summary of Archival/supplementary supernova spectra modelled in this work.}
\begin{tabular}{cccccc}
SN     & Epoch (d) & Type & Host Galaxy & D(Mpc) & Reference \\
\hline
1957D  & 11371         & II   & NGC 5236    & 4.0& \citet{Long1989}           \\
       & 19459   &      &      &             & \citet{Long2012}          \\
1970G  & 16693         & II-L  & NGC 5457    & 6.7& This paper  \\
1979C  & 5146         & II-L  & NGC 4321    & 15.0& \citet{Fesen1999} \\
& 10575 & & & & \citet{Milisavljevic2009}\\
1980K  &    12977         & II-L  & NGC 6946    & 7.7& This paper          \\
1986E  & 3712          & II-L  & NGC 4302    &    17.0&        \citet{Cappellaro1995}\\
1993J  & 8417          & IIn  & M81         & 3.6 & This paper          \\
1996cr & 3603 & IIn  & ESO 097-13  &4.2 & \citet{Bauer2008}         \\
1998S  & 258          & IIn  & NGC 3877    &17.0 & \citet{Fransson2005}          \\
  & 440           &   &             & & \citet{Fransson2005}          \\
  & 1170          &   &             & & \citet{Pozzo2004}          \\
  & 2148          &   &             & & \citet{Fransson2005}          \\
  & 6574          &   &             & & \citet{Smith2016}         \\
 2004et & 646 & II-P & NGC 6946 & 7.7 & \citet{Fabbri2011a}\\ 
2012au & 2277          & Ib/c & NGC 4790    & 24.0 & \citet{Milisavljevic2018} \\
\hline
\label{table:archiv-obs-sum}
\end{tabular}
\end{table*}

The X-Shooter spectra were taken in IFU mode, with a field of view of 4$\times$1.8~arcsec, using an effective slit width of 0.6~arcsec. The spectral resolving powers were 8600 for the 3000-5600~\AA\ UVB region and 13500 for the 5500-10200~\AA\ VIS region.
Objects were acquired using blind offsets from nearby stars, and extracted from a circular aperture with a radius of 1~arcsec, and sky-subtracted using an annulus with inner-outer radii of 1-2~arcsec. The data were reduced using the software package Reflex \citep{Freudling2013}, which is implemented in the Kepler workflow engine and automated the data reduction process.

For each spectrum the continuum was normalised to unity throughout, using the interactive Starlink package {\sc dipso} \citep{Howarth2014}. 

\section{Modelling red-blue line profile asymmetries using {\sc damocles}}
\subsection{Methodology}
This work is motivated by the overwhelming presence of blue-shifted asymmetries in the emission line profiles of CCSNe. Out of a sample of 10 supernovae, \citet{Milisavljevic2012} noted that all the SN emission lines showed some degree of blue-shifting, which they suggested was due to internal dust obscuration. Similarly, upon a visual inspection of our sample of 14 CCSNe, all SNe either exhibited a red scattering wing and/or a blue-shifted peak, which can both be caused by dust absorption and scattering. If dust formation in supernovae was less prevalent, and if emission line asymmetries were instead due to physical asymmetries in the ejecta, then for CCSNe older than approximately 300 days we would expect to observe roughly equal numbers having red-shifted or blue-shifted emission line peaks. This is not the case. Given the statistical significance of this finding, we assume that if the emission line peaks of a supernova are blue-shifted, or an extended red scattering wing is present, this is caused by dust internal to the supernova, which we then quantify with the Monte Carlo radiative transfer code {\sc damocles}. 

{\sc damocles} is written in {\sc Fortran} 95, and parallelised with {\sc openmp}. It models line photons subjected to scattering and absorption by dust in expanding ejecta. Total energy conservation is not a requirement, as it is assumed that any packet absorbed by dust would be re-emitted outside the wavelength range of interest and thus no longer contributes to the resulting line profile. In addition to this, the absorption and scattering of radiation by dust is assumed to be independent of the dust temperature. It has been benchmarked \citep{bevan2016} against analytic models of theoretical line profiles based on work by \citet{Gerasimovic1933a}, and also against numerical models of SN 1987A produced by \citet{Lucy1989}. {\sc damocles} is able to treat any arbitrary dust/gas geometry, a range of velocity and density distributions and dust and gas clumping configurations, as well as handling a wide range of grain species and radii.
Our approach to modelling the line profiles follows the same principles as described for SN~1987A by \citet{bevan2016}, where a more detailed description of {\sc damocles} can also be found. In all cases, the parameter space was examined manually to find the best fitting model. The free parameters of the models were the dust mass M$_d$, grain radius $a$, outer expansion velocity V$_{max}$, emitting shell radius ratio R$_{in}$/R$_{out}$, and the density profile index $\beta$, such that the density profile  $\rho$ $\propto$ r$^{- \beta}$.

The manual fitting process can be briefly described as follows: V$_{\rm max}$ is determined from the point at which the observed profile vanishes on the blue side, while V$_{\rm min}$ is constrained by the width of the ``flat-top'' region of the emission line profile, i.e. the region between the velocity of the blue-shifted peak found at -V$_{\rm min}$ and an inflection point in the profile found at +V$_{\rm min}$. Thus the R$_{\rm in}$/R$_{\rm out}$ value is set so that the model line profile matches various inflection points of the observed profile, and the density profile $\beta$ is identified from the gradient of the observed line profile wings. Having fixed these values, we then iterate over the grain radius and dust mass to fit the observed profile (in the case of the Bayesian fitting process described below, the above procedures are done automatically).

We assume that at these late epochs the line-emitting gas is optically thin, with the emissivity distribution proportional to the square of the local gas density. The gas was kept smoothly distributed throughout. We assume that the supernova ejecta is in free expansion, such that V$_{max}t$ = R$_{out}$, where $t$ is the age of the supernova. However, this assumption does not hold for the Intermediate Width Components (IWCs) or for some hydrogen-emitting shells in some CCSNe models, and the R$_{out}$ of the components is a free parameter in such cases. Unless otherwise specified, we keep the dust and the gas coupled in our simulations (so the smoothly distributed dust and the gas have the same V$_{\rm max}$, $R_{\rm in}/R_{\rm out}$ and $\beta$ values). We adopted intrinsic line flux ratios for the [O~{\sc iii}] 4959,5007$\AA$, [O~{\sc ii}] 7319,7330$\AA$ and [O~{\sc i}] 6300,6363$\AA$ doublets fitted by the {\sc damocles} models of 2.98, 1.23 and 3.13, respectively \citep{Storey2000a, ZEIPPEN1987, Baluja1988}.

We modelled the line profiles using smooth or clumped dust distributions. Clumps were adopted to have radii equal to R$_{\rm out}$/40.
The clumped dust models used the same V$_{\rm max}$ and $R_{\rm in}/R_{\rm out}$ parameters as the coupled gas distribution, although distributed by a $\beta_{\rm clump}$ parameter, the clump distribution power law index, instead of the $\beta$ value used for the gas as well as for smoothly distributed dust models. From a 10D Bayesian model of SN 1987A 714 days past explosion \citep[][]{Bevan2018}, values of 3 were found to be preferred for the clump distribution power law index as well as filling factors of <0.25. We fixed the clump distribution power law index to 3 and set the filling factor to 0.10 in order to reduce the number of free parameters required in our models. Varying the clump number density distribution power law index over values of 1-5 was found to change the dust mass needed to provide a good fit by a factor of $\sim$3. 

We modelled a few SNe where instead of smoothly distributed gas, we used clumped gas coupled to the clumped dust, in order to determine the degree to which gas clumping would affect the derived dust mass. The amount by which the dust mass required to fit the line profiles changed was dependent on the difference between the best-fitting $\beta$ index, derived from the smoothly distributed dust gas model, and the fixed dust clump number density distribution index of 3. SN 1996cr, for which models required the highest value  of $\beta$ out of the SNe in our sample, was found to require a dust mass 3.5$\times$ smaller in order to fit the line profiles when the gas and dust were co-located in clumps. SN~1970G, on the other hand, which was best fitted with a smoothly distributed gas $\beta$ value of -0.1, required a factor of two larger dust mass to fit the H$\alpha$ line profile when the gas and dust were co-located in clumps.

A clumped dust model will present less of a scattering wing and attenuate the red wing of the line profile less than for a smooth dust model, given no change in other model parameters between the two cases. This is because when the dust is
located in clumps, radiation is subject to less scattering as well as to less absorption. Therefore, for a clumped dust model to match the line profile generated by a smooth dust model, both a small modification to the grain radius to produce a larger albedo and a dust mass larger than for the smooth case by a factor of 1.5-3 are required. Several authors found that dust needed to be present in optically thick clumps in order to reproduce early observations of SN~1987A \citep{Lucy1989,Lucy1991,Bouchet1996,Kozma1998b}. So, for all objects in this work, the final dust masses we report are those found from clumped models, on the grounds that dust is more likely to exist in clumps for the CCSNe in our sample, rather than in a smoothly distributed medium. 

Most line profiles can be fit using either 100~per~cent amorphous carbon (AmC) or 100~per~cent silicate dust with grain sizes with matching albedos, apart from cases where a large red scattering wing requires the dust species to have an albedo greater than 0.7, which could only be achieved by using silicate dust with grain radii in the range 0.1-1.0~$\mu$m.
For CCSNe where we cannot confirm the grain species but can roughly constrain the grain radius for each species, we present our final dust mass for the object using a 50:50 AmC to astronomical silicate dust species mixture, as some proportion of both carbon grains and silicates are often found to be needed from SED modelling of SN dust, e.g. for Cas~A see
\citet{Rho2008, Arendt_2014,DeLooze2017}, while for 1E0102-7912 see \citet{Rho2009, sandstrom2009}.

All models were convolved to the spectral resolution of the optical data. For all figures showing line profile fits to optical data, "dust-free" and dusty model line profiles are shown. These model line profiles have the same gas distribution, but the dusty model couples this gas distribution to a dust distribution and the dust-free model does not. This is to aid the reader to visualise how adding dust to the simulation affects the shape of the model line profile. 
The parameters for smooth and clumped AmC and silicate dust species are included in the tables listing model parameters. As was demonstrated in \citet{Niculescu-Duvaz2021}, the choice of silicate grain species barely affects the dust optical depth in the optical regime, hence we use astronomical silicate species with the optical constants of \citet{draine1984ApJ...285...89D} in our models. Our AmC models use the BE amorphous carbon optical constants of \citet{zubko1996}. For some CCSNe in our sample, the grain radius of the species could not be constrained, often due to insufficient signal-to-noise in the observed line profiles, which also meant an upper limit could not be established on the dust mass.
In these cases, we present results using conservative grain sizes for both the smoothly distributed and clumped 100~per~cent AmC and 100~per~cent silicate dust models to provide a rough estimate of possible dust mass values. For these CCSNe, we provide lower limits to the dust masses, given by the dust mass required to attenuate the line profiles using the grain radius that required the lowest dust masses, which was 0.1~$\mu$m for a 50:50 AmC to silicate ratio.

In order to quantify errors on the model parameters, we also conducted a Bayesian analysis on all supernova spectra, with the exception of those that exhibited multiple emission components. The Bayesian approach to modelling line profiles with {\sc damocles} is described by \citet{Bevan2018}. It combines an affine invariant Markov Chain Monte Carlo (MCMC) ensemble sampler, {\sc emcee} \citep{Goodman2010, Foreman-Mackey2013} with {\sc damocles} to sample the posterior probability distribution of the input parameters, which is defined by Bayes' Theorem: 
\begin{equation} \label{eq:bayeseq}
P(\theta|D) \propto P(\theta)P(D|\theta)
\end{equation}
In this equation, D is the data, $\theta$ is the set of parameters of the model, P($\theta$) is our prior understanding of the probability of the parameters, and P(D|$\theta$) is the likelihood, which is the probability of obtaining the data for a given set of parameters.
The likelihood function is proportional to $\exp{-\chi^2_{red}}$, where $\chi^2_{red}$=$\chi^2$/$\upsilon$, and $\upsilon$ is the number of degrees of freedom and $\chi^2$ is expressed in equation (\ref{eq:likelihoodeq}):
\begin{equation} \label{eq:likelihoodeq}
\chi^2 = \sum^{n}_{i=1}
\frac{(f_{mod,i} - f_{obs,i})^2}{\sigma_i^2} 
\end{equation}
In this equation, f$_{mod,i}$ is the modelled flux in bin i, f$_{obs,i}$ is the
observed flux in frequency bin i, and $\sigma_i$ is the combined Monte Carlo and observational uncertainty in bin i. The modelled line profile from which f$_{mod,i}$ is sampled has its peak normalised to the peak flux of the observed line profile. The priors for each model are given in uniform space, apart from the dust mass and grain radius which span several magnitudes, so they are given in log-uniform space.

\begin{figure*}
\centering

\includegraphics[width=0.9\linewidth]{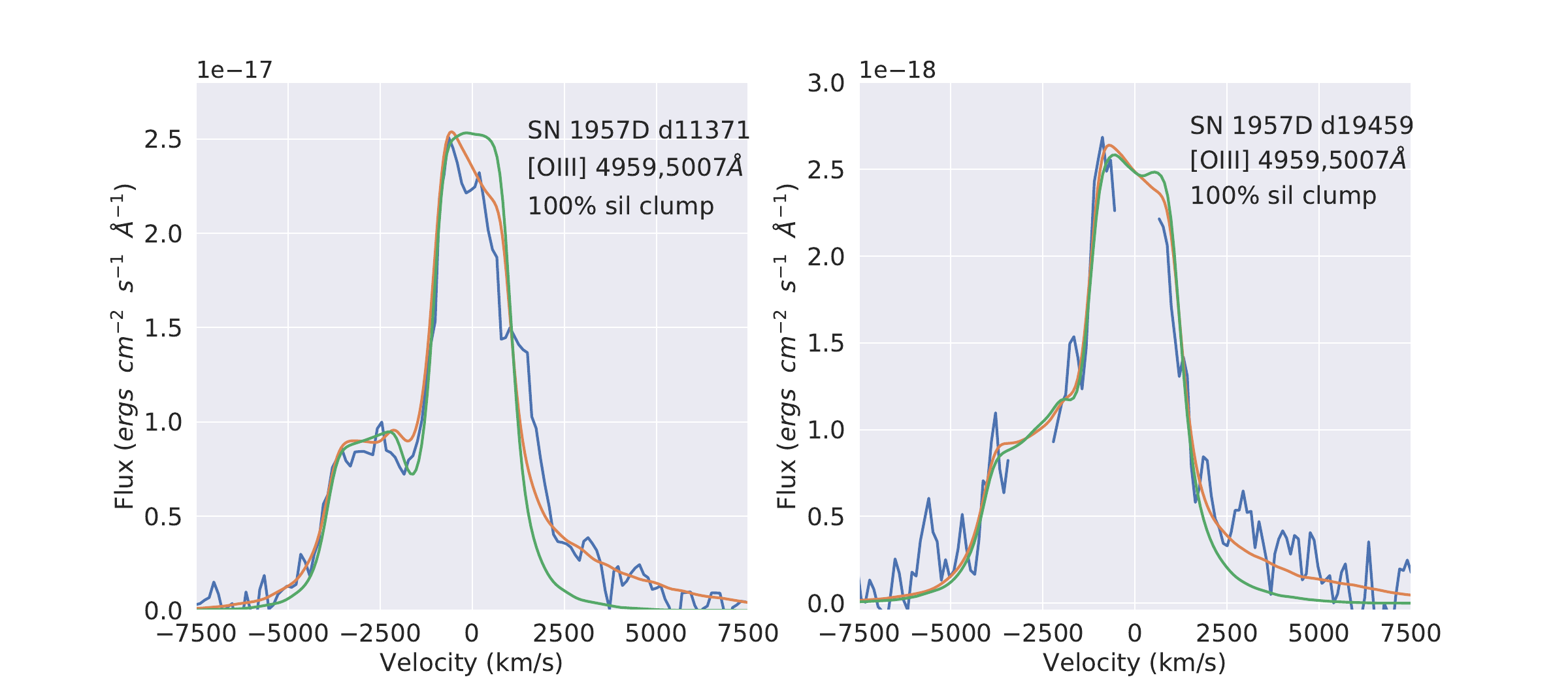}

\caption[{\sc damocles} models of the {[}O~{\sc iii}{]} 4959,5007~\AA\ profiles of SN~1957D 11371 and 19459 days post-explosion.]{{\sc damocles} models of the {[}O~{\sc iii}{]} 4959,5007~\AA\ profiles of SN~1957D 11371 and 19459 days post-explosion. The green line is the dust-free model, the orange line is the dust-affected model and the blue line is the observed spectral line. Clumped dust models with 100~per~cent astronomical silicate are shown.}
\label{fig:57d-89-11-fits}
\end{figure*}

\begin{figure*}
\centering

\includegraphics[scale=0.31]{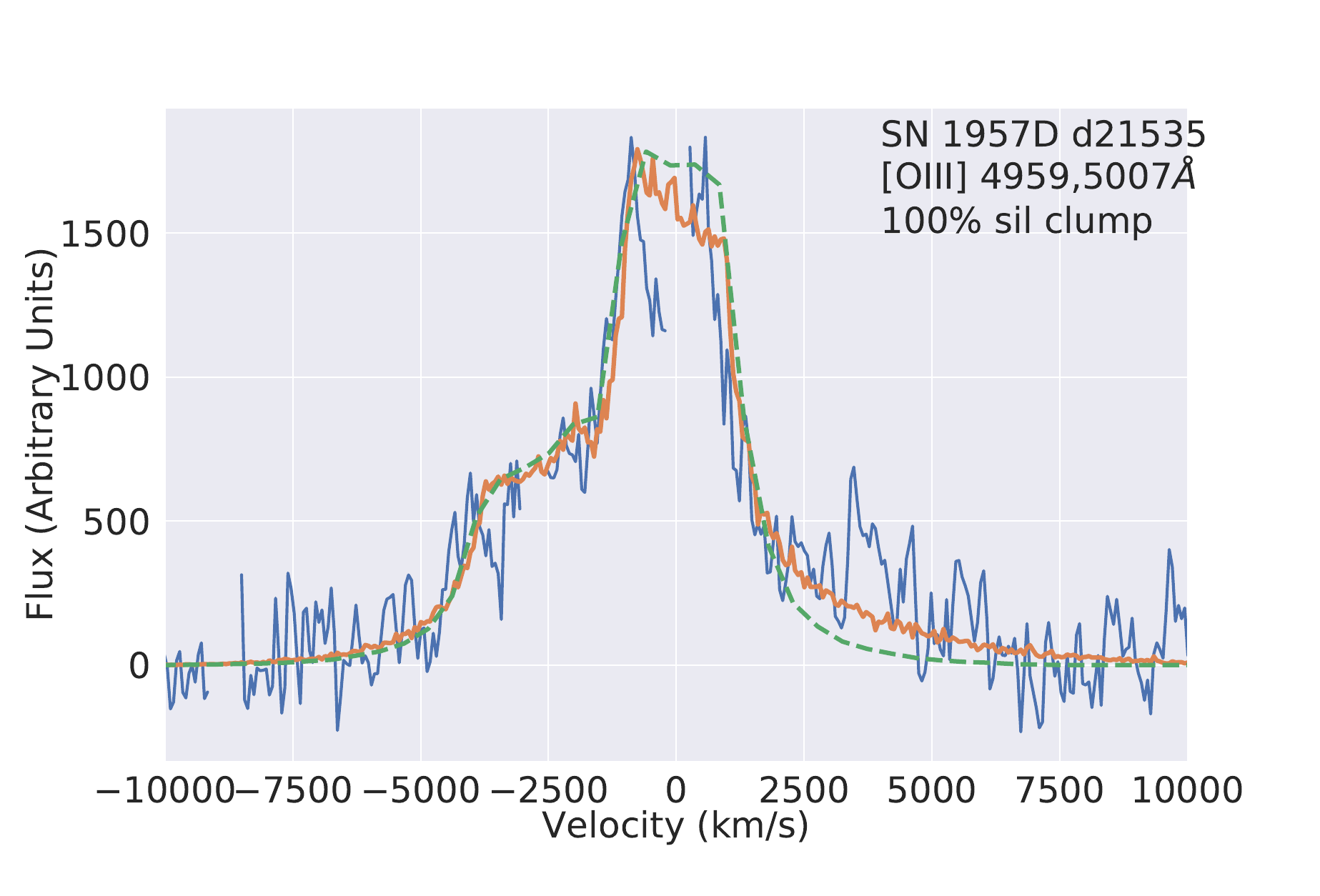}

\caption[{\sc damocles} model of the {[}O~{\sc iii}{]} profiles of SN~1957D 21535 days post-explosion.]{{\sc damocles} model of the {[}O~{\sc iii}{]} profiles of SN~1957D 21535 days post-explosion. The green line is the dust-free model, the orange line is the dust-affected model and the blue line is the observed spectral line. Clumped dust models with 100~per~cent astronomical silicate are shown.}
\label{fig:57d-18-fits}
\end{figure*}

The final posterior probability distributions are presented as a series of 2D contour plots, where each pair of parameters are
marginalised over the other parameters. A 1-D
marginalised posterior probability distribution for each parameter is also presented. The "best fitting" parameter value from the Bayesian analysis is given as the median of the marginalised 1-D probability distribution, as many deviated from a Gaussian distribution. The lower and upper limits represent the 16th and 84th quartiles for the same 1-D probability distribution. The Bayesian model fits presented in this work usually use either a 100~per~cent AmC or 100~per~cent silicate dust composition, and in cases where we present a final dust mass using a 50:50 AmC to silicate dust ratio, which is determined manually, we find the percentage errors on the dust mass from the limits found by the Bayesian analysis and scale them to that value.
As we had to run many models, we restricted our parameter space to 5 dimensions, explored by 250 walkers. For each parameter, the number of iteration steps for the autocorrelation function to initially decay down towards zero can be estimated to be one autocorrelation time. This value was checked for every parameter in every Bayesian model. We checked that each model was run for 5 or more autocorrelation times to ensure convergence. We also checked the acceptance fraction for each simulation, which averaged at around 0.3.

For emission line profiles where intermediate width components took up a large part of the profile, we could not use a Bayesian analysis to evaluate the errors, as the Bayesian version of {\sc damocles} can only model single component emission lines. In these cases we used $\chi^2$ estimates to evaluate the goodness of fit, and compared the $\chi^2$ values of dusty and dust-free models. We present a $\chi^2$ value for every dusty model. The uncertainty limits on our final dust mass are then given by a 35~per~cent variation in $\chi^2$ when varying only the dust mass and fixing all other parameters, as this value leads to uncertainties on the dust mass that are of a similar magnitude to the uncertainties found from Bayesian inference.

\subsection{Application to CCSNe}
\subsubsection{SN 1957D}

SN 1957D is the oldest supernova modelled in this work. It was first discovered by H. Gates on 28/12/1957 in M~83, when it was well past maximum light, so its exact explosion date is unknown. We adopt an M~83 redshift of 0.00115 for SN 1957D (\citet{Meyer2004}).  

\begin{figure*}
\centering

\includegraphics[width=\linewidth]{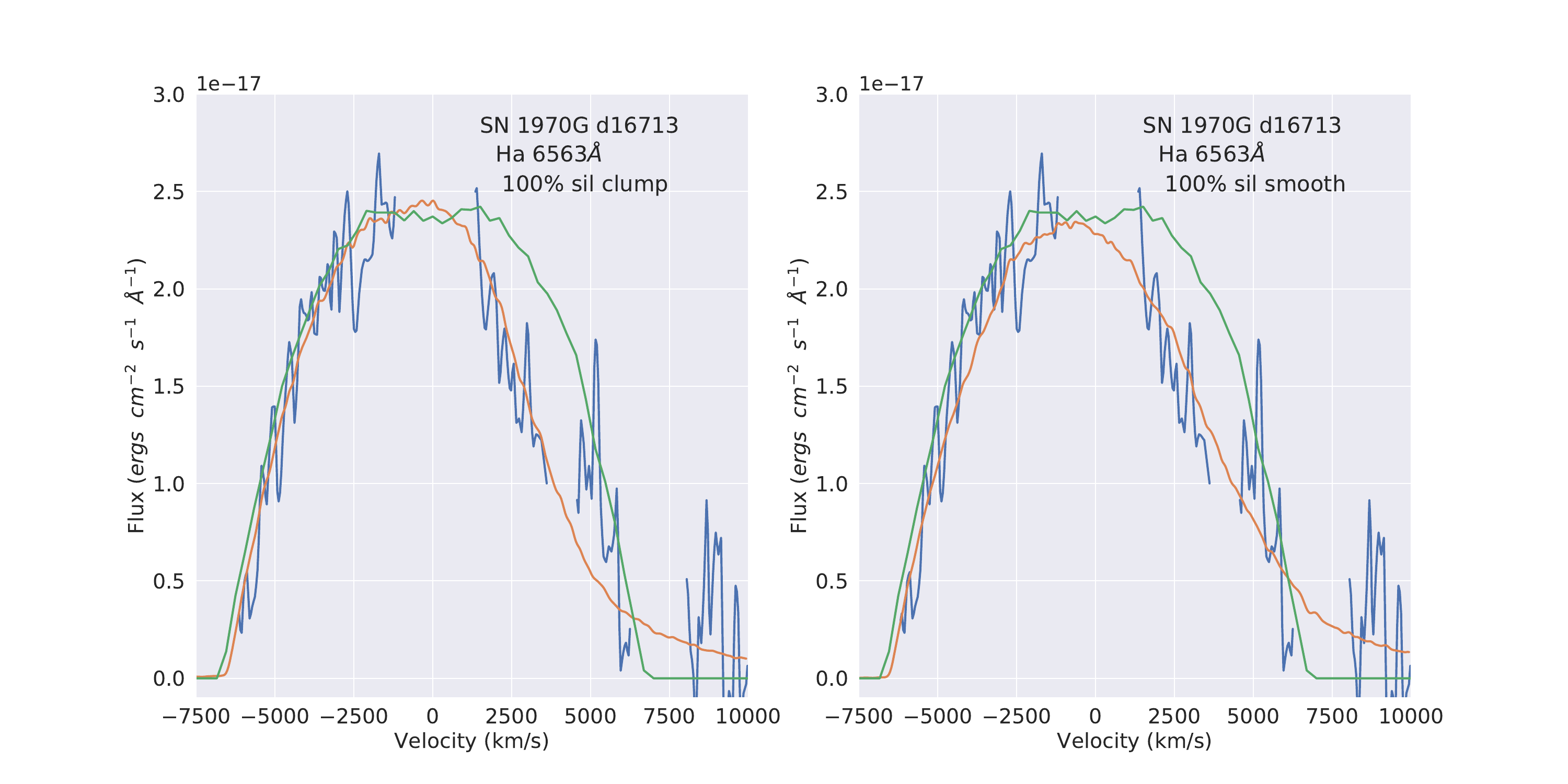}

\caption{{\sc damocles} models of the H$\alpha$ profile of SN 1970G 16713 days post-explosion. The green line is the dust-free model, the orange line is the dust-affected model and the blue line is the observed spectral line. Clumped and smooth dust models of 100~per~cent astronomical silicates are shown.}
\label{fig:1970g-fits}
\end{figure*}

We modelled the [O~{\sc iii}] 4959,5007-$\AA$ emission from SN~1957D at 11371, 19459 and 21535 days past explosion. Observational data for the first two epochs are summarised in Table~\ref{table:archiv-obs-sum}. We combined all the GMOS and X-Shooter spectra of SN~1957D (Tables~\ref{table:gmos-dat} and
\ref{table:xshoot-dat}) to make the nominal day 21535 spectrum. As we could not constrain the dust species, smooth and clumped AmC and silicate dust models for all epochs can be found in Table~\ref{table:4sn-params}. Our clumped silicate dust-affected [O~{\sc iii}]~4959,5007-$\AA$ line profile models can be found in Figures \ref{fig:57d-89-11-fits} and \ref{fig:57d-18-fits} for all epochs.
From a manual investigation of the parameter space, we found the best fitting models to the line profile shapes required similar V$_{max}$, $\beta$ and R$_{in}$/R$_{out}$ for the three epochs, where there appeared to be a possible small increase of dust mass over time. Due to the presence of a persistent red scattering wing, we found that 
our clumped dusty models required a grain radius
of 0.4~$\mu$m for 100~per~cent AmC grains, 100~per~cent silicate dust grains, or a 50:50 carbon to silicate mixture in order to provide the best model fits to the [O~{\sc iii}] doublet at all epochs. The derived dust masses at all epochs in SN 1957D were found to be low compared to the dust masses that we find for SN~1979C and SN~1980K at similar epochs. 
Table~\ref{table:4sn-params}
lists the derived dust parameters for 100~per~cent AmC or 100~per~cent silicate compositions.
For a 50:50 carbon to silicate mixture the best-fitting dust masses were 0.012$^{+0.048}_{-0.012}$, 0.035$^{+0.075}_{-0.030}$ and 0.05$^{+0.50}_{-0.048}$ M$_{\odot}$ for days 11371, 19459 and 21535.

In order to constrain the uncertainties on the model parameters, we ran a Bayesian MCMC analysis for days 11371 and 21535 of SN~1957D using 100~per~cent carbon dust. The resulting corner plot for the day 11371 model can be found in Figure A1. We were able to constrain the AmC grain radius at 11371 days to be $>0.1\mu$m, where the median value of the grain radius 1-D posterior probability distribution was 0.3~$\mu$m, very similar to our previously estimated grain radius of 0.4~$\mu$m. As this was the epoch with the best signal to noise, we fixed the grain radius derived at this epoch to be the value adopted for our manual models at the other epochs. 
The dust mass that we had derived for an AmC carbon grain species, from a manual examination of parameter space, was a factor of 2.5 smaller than the median value derived from the 1-D posterior probability distribution for the dust mass. We applied the percentage errors taken from the Bayesian analysis to our best fit dust mass derived from a manual fit for 50:50 AmC to silicate dust using a grain radius of 0.4~$\mu$m, to derive absolute upper and lower limits. The parameters were less well constrained at day 21535 due to a lower signal-to-noise [O~{\sc iii}] profile, and the median dust mass was 5$\times$ less than our best fitting model for clumped carbon grains, but our manually estimated value was well within the error uncertainties on the dust mass from the Bayesian analysis. 

We considered the signal-to-noise of the day 19459 spectrum to be too low to conduct a Bayesian analysis, and instead derived a dust mass of 0.035$^{+0.065}_{-0.030}$ M$_{\odot}$
for a 50:50 AmC to silicate mixture with a grain radius of 0.3~$\mu$m, where a dust-free model gave a $\chi^2$ of 1.4, while the dusty models returned an average $\chi^2$ of 0.8.

\begin{table*}
\centering
\caption{
Parameters used in the {\sc damocles} models of the broad emission lines of SN 1957D, SN 1970G, SN 1980K and SN 1993J for spherically symmetric smooth and clumped dust models. "a" is the dust grain radius. The value of $a$ derived at day 21535 for SN 1957D has been fixed for all epochs. The "\% Sil" column stands for the percentage of the dust species that is astronomical silicate, where the remainder is amorphous carbon dust. The optical depth is calculated from R$_{in}$ to R$_{out}$ at the central line wavelengths ([O~{\sc iii}]=5007~$\AA$,[O~{\sc i}]=6300~$\AA$, H$\alpha$=6563~$\AA$). } 
\centering
\begin{tabular}{cccccccccccccc}
\hline
SN & Line & Epoch & Clumped? & \% Sil & a & V$_{max}$ & V$_{min}$ & ~~~~$\beta_{gas}$ & R$_{out}$~~~~~~ & R$_{in}$ & M$_{dust}$ & $\tau$ & $\chi^2$   \\
 & & days & & & $\mu$m & km~s$^{-1}$ & km~s$^{-1}$ & & 10$^{15}$~cm & 10$^{15}$~cm & $10^{-2}$~M$_\odot$ & & \\
\hline
1957D & [O~{\sc iii}] & 11371 & no  & 0 & 0.4 & 5800 & 870  & 2.5 & 1079 & 140.3 & 0.5 & 0.17 &                   \\
1957D & [O~{\sc iii}] & 11371 & yes & 0 & 0.4 & 5800 & 870  &  2.7 & 1079 & 140.3 & 1.0   & 0.31 &                   \\
1957D & [O~{\sc iii}] &  11371 & no  & 100 & 0.2 & 5800 & 870  & 2.5 & 1079 & 140.3 & 0.4 & 0.32 &                   \\
1957D & [O~{\sc iii}] &  11371 & yes & 100 & 0.4 & 5800 & 870  & 2.5 & 1079 & 140.3 & 1.4 & 0.18 &                   \\
1957D & [O~{\sc iii}] &  19459 & no  & 0 & 0.4 & 6800 & 1020 & 2.4 & 1265 & 164.5 & 1.5 & 0.16 & 0.89 \\
1957D & [O~{\sc iii}] &  19459 & yes & 0 & 0.4 & 6800 & 1020 & 2.4 & 1265 & 164.5 & 2.5 & 0.41 & 0.76 \\
1957D & [O~{\sc iii}] &  19459 & no  & 100 & 0.2 & 6800 & 1020 & 2.4 & 1265 & 164.5 & 0.8 & 0.11 & 0.84 \\
1957D & [O~{\sc iii}] &  19459 & yes & 100 & 0.4 & 6800 & 1020 & 2.4 & 1265 & 164.5 & 3.5 & 0.14 & 0.83 \\
1957D & [O~{\sc iii}] &  21535 & no  & 0 & 0.4 & 7500 & 962  & 2.5 & 1377 & 179.0 & 3.0   & 0.15 &                   \\
1957D & [O~{\sc iii}] &  21535 & yes & 0 & 0.4 & 7500 & 962  & 2.5 & 1377 & 179.0 & 5.0   & 0.27 &                   \\
1957D & [O~{\sc iii}] &  21535 & no  & 100 & 0.2 & 7500 & 962  & 2.5 & 1377 & 179.0 & 2.0   & 0.22 &                   \\
1957D & [O~{\sc iii}] &  21535 & yes & 100 & 0.4 & 7500 & 962  & 2.5 & 1377 & 179.0 & 7.0   & 0.19 & \\
 & & & & & & & & & & & & & \\
1970G & H$\alpha$ & 16713 & no  & AmC & 0.20 & 6700 & 2546 & -0.3 & 968.6 & 290.6 & 1.5  & 0.34 & \\
1970G & H$\alpha$ & 16713 & yes & AmC & 0.25 & 6700 & 2546 & -0.3 & 968.6 & 290.6 & 1.5  & 0.34 & \\
1970G & H$\alpha$ & 16713 & no  & sil & 0.20 & 6600 & 3960 & -0.3 & 954.2 & 572.5 & 7.0  & 0.61 & \\
1970G & H$\alpha$ & 16713 & yes & sil & 0.20 & 6600 & 3960 & -0.1 & 954.2 & 572.5 & 10.0 & 1.12 & \\
 & & & & & & & & & & & & & \\
1980K & H$\alpha$       & 13169 & no  & 0   & 3.50 & 5300 & 4505 & 1.5 & 593.1 & 504.0 & 40.0 & 0.53 & \\
1980K & H$\alpha$       & 13169 & yes & 0   & 3.50 & 5300 & 4505 & 1.5 & 593.1 & 504.0 & 60.0 & 0.81  \\
1980K & H$\alpha$       & 13169 & no  & 100 & 0.10 & 5300 & 4505 & 1.5 & 593.1 & 504.0 & 20.0 & 0.92 & \\
1980K & H$\alpha$       & 13169 & yes & 100 & 0.10 & 5300 & 4505 & 1.5 & 593.1 & 504.0 & 60.0 & 2.91 & \\
1980K & [O~{\sc i}] & 13169 & no  & 0   & 3.50 & 5300 & 3445 & 1.5 & 593.1 & 385.5 & 40.0 & 0.67 & \\
1980K & [O~{\sc i}] & 13169 & yes & 0   & 3.50 & 5300 & 3445 & 1.5 & 593.1 & 385.5 & 60.0 & 0.97 & \\
1980K & [O~{\sc i}] & 13169 & no  & 100 & 0.10 & 5300 & 3445 & 1.5 & 593.1 & 385.5 & 20.0 & 1.43 & \\
1980K & [O~{\sc i}] & 13169 & yes & 100 & 0.10 & 5300 & 3445 & 1.5 & 593.1 & 385.5 & 60.0 & 4.13 & \\
 & & & & & & & & & & & & & \\
1993J  & [O~{\sc iii}] & 8417  & no  & 0   & 0.20  & 5800  & 4408 & 5    & 421.8  & 320.6 & 1.3   & 0.86 & \\
1993J  & [O~{\sc iii}] & 8417  & yes & 0   & 0.20  & 5800  & 4408 & 5    & 421.8  & 320.6 & 2.5   & 1.66 & \\
1993J  & [O~{\sc iii}] & 8417  & no  & 100 & 0.04 & 5800  & 4408 & 5    & 421.8  & 320.6 & 25.0  & 0.82 & \\
1993J  & [O~{\sc iii}] & 8417  & yes & 100 & 0.04 & 5800  & 4408 & 5    & 421.8  & 320.6 & 50.0  & 1.64 & \\
\hline
\label{table:4sn-params}
\end{tabular}
\end{table*}

\subsubsection{SN 1970G}
SN 1970G was discovered on July 30th 1970 \citep[][]{detre1970}. It is located in the galaxy M~101, \citep[for which we adopted a redshift of z=0.000811 from][]{Sabater2012}. 
It was classified as a type II-L supernova from its light curve \citep[][]{Young1989} although \citet{Barbon1979} considered that it could be a transitional object between II-P and II-L, 

As the red wing of the H$\alpha$ profile was cut off in our GMOS-N spectrum of SN~1970G taken at 16733 days post-explosion, we combined it with an MMT spectrum taken 16693 days post-explosion, averaging to an epoch of 16713 days, Both spectra had detections of broad H$\alpha$ 6563~$\AA$ and [O~{\sc i}] 6300,6363~$\AA$, with a weaker detection of [O~{\sc iii}] 4959,5007~$\AA$. The GMOS spectrum was convolved to the resolution of the MMT spectrum of 7$\AA$. Unfortunately, part of the red wing in the MMT spectrum was tainted by what appears to be an absorption feature and by nebular [S~{\sc ii}] 6717, 6731-$\AA$ emission. These were removed from the H$\alpha$ line before modelling with {\sc damocles}. We omit models of the [O~{\sc i}] 6300,6363~$\AA$ line due to low signal to noise and the blending of its red wing with the H$\alpha$ line.

Our {\sc damocles} models for SN~1970G were constructed for an age of 16713 days, with the best fitting clumped silicate dust models for the H$\alpha$ profile shown in Figure~\ref{fig:1970g-fits}. The parameters for the smooth and for the clumped AmC and silicate dust models are listed in  Table~\ref{table:4sn-params}. The grain radius or grain species cannot be well constrained from the H$\alpha$ line, due to gaps in the line profile in the red wing. We thus present conservative grain radius estimates for both smoothly distributed and clumped 100~per~cent AmC and 100~per~cent silicate dust models. The clumped dust mass derived from the best fits to the H$\alpha$ profile ranges from 0.02~M$_{\odot}$ for a 100~per~cent AmC dust distribution to 0.1~M$_{\odot}$ for a 100~per~cent silicate distribution. All the dusty models had a lower $\chi^2$ value than for the dust free case ($\chi^2$=1.58). We found a marginal improvement of the $\chi^2$ value for a 100~per~cent silicate dust model over a 100~per~cent amorphous carbon model. Therefore, our preferred dust mass for SN~1970G is $0.10^{+1.22}_{-0.097}$~M$_{\odot}$, where the limits are taken from a Bayesian inference.

\begin{figure*}
\centering

\includegraphics[width=\linewidth]{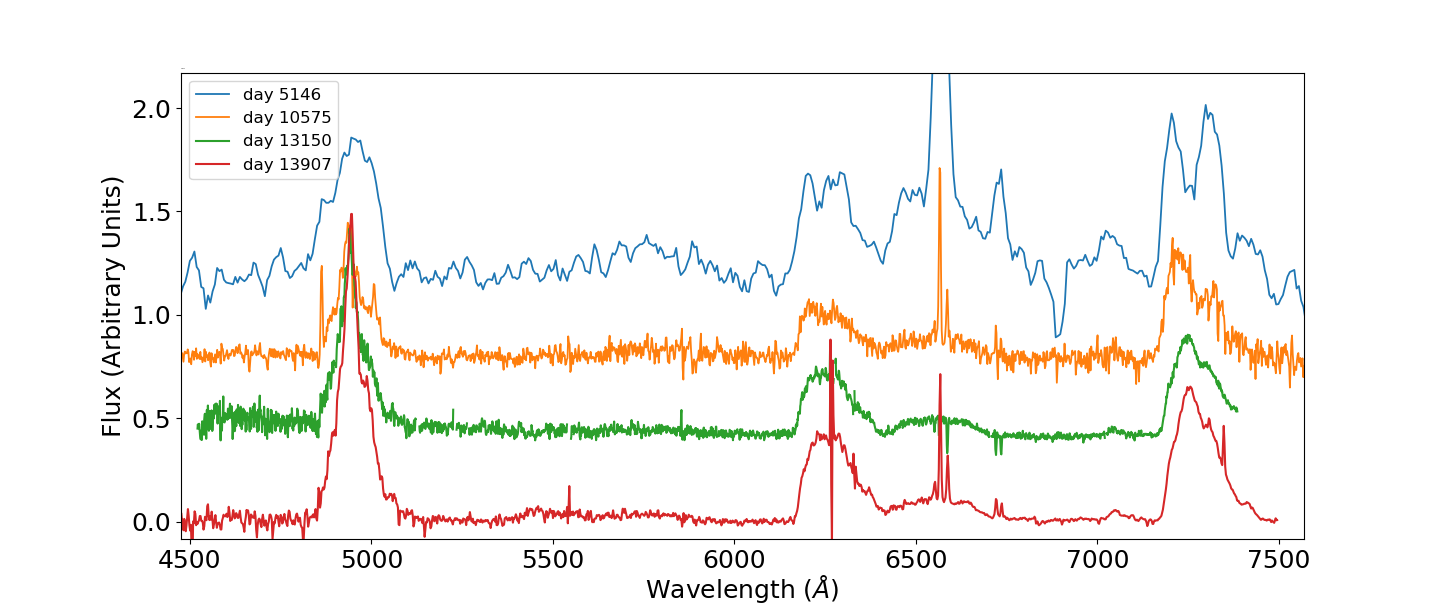}

\caption{The evolution of the optical spectrum of SN 1979C. Information on the spectra is summarised in Tables~\ref{table:gmos-dat} and \ref{table:archiv-obs-sum}. }
\label{fig:1979C-evoln}
\end{figure*} 
A similar $\chi^2$ value can be obtained by decoupling the H$\alpha$ emitting gas from the dust, which is distributed using the parameters found for the freely expanding [O~{\sc i}]-emitting ejecta model. No additional dust component coupled to the H$\alpha$ emitting region is required. The $\chi^2$ value is minimized when there is a small overlap between the radii of the  H$\alpha$ shell and the ejecta dust shell, so as to induce some dust absorption on the red wing of the H$\alpha$ line.

The H$\alpha$ and [O~{\sc i}] emission line models require different V$_{max}$, R$_{in}$/R$_{out}$ and $\beta$ values, indicating that the two emitting species are not co-located. The assumption of an H$\alpha$ distribution that is expanding freely and coupled with the dust can create a model which fits the profile well, but leads to derived dust masses that are of a factor of $\sim$5 less than the dust masses derived from the [O~{\sc i}] modelling. 

\subsubsection{SN~1979C}
\label{subsubsec:1979c}
SN 1979C was discovered on April 19th 1979 in M~100 \citep[][]{mattei1979}. It was thought to be a Type II-L supernova from the fast decline of its light curve and absence of a plateau \citep{Panagia1980}. The explosion was considered unusually bright when compared to other Type II-Ls, with a peak M$_{B}$ of
–20 mag \citep{youngbranch}.



Figure~\ref{fig:1979C-evoln} shows the evolution of the spectra through 5146, 10575, 13150 and 13907 days. The spectra have been corrected for the redshift of the host galaxy of  z = 0.00525 \citep{vandriel2016}. At the earliest epoch, the oxygen doublets are double peaked, most noticeably the [O~{\sc ii}] 7319,7330-$\AA$ profile, with a weaker peak centred at 0~km~s$^{-1}$.
The central peak declines in brightness over 10,000 days until it is barely visible at 13907 days. The H$\alpha$ profiles at day~10575 and beyond notably lack this central peak, and are far more symmetric than the forbidden oxygen emission lines. We interpreted the central peak to be an intermediate width component (IWC), possibly formed by an ejecta-CSM interaction, while we term the broader underlying profile to be the broad component (BC), which represents the fast-expanding ejecta. We assumed a simple model setup for the oxygen lines at days 5146 and 13150 and for the [O~{\sc iii}] doublet at day 13907, in which the dust has formed in the ejecta, and is therefore coupled with the BC, and the separate component emitting the IWC is located where R$_{in,IWC}$ $\approx$ R$_{out,BC}$, and whose
red-shifted emission is affected by absorption by the ejecta dust. We found that no separate dust component co-located with the IWC was necessary, but note that the shape of the IWC on day 5146 is unclear due to the lower resolution, while at later epochs the IWC is faint, so we cannot rule out this possibility. 

\begin{figure*}
\centering

\includegraphics[width=\linewidth]{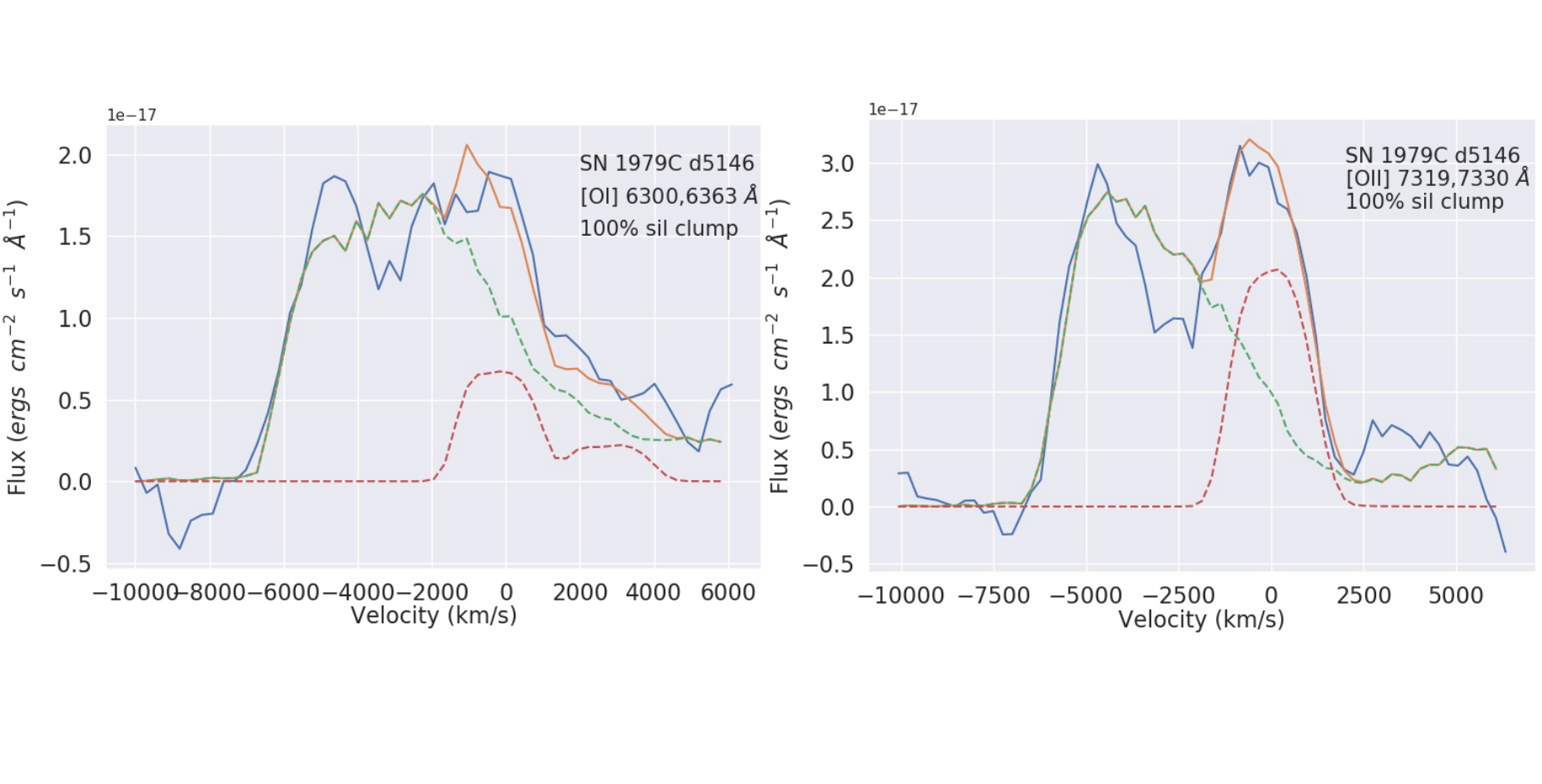}

\caption{{\sc damocles} models of the [O~{\sc i}] and [O~{\sc ii}] profiles of SN~1979C. The blue lines are from the observed day 5146 spectrum presented by \citet{Fesen1999}. The red and green dashed lines are the {\sc damocles} models of the Intermediate Width Component (IWC) and Broad Component (BC), and the orange line is the sum of these two components. Clumped dust models consisting of 100~per~cent astronomical silicates are shown.}
\label{fig:1979C-93fits}
\end{figure*}

\begin{figure*}
\centering

\includegraphics[width=\linewidth]{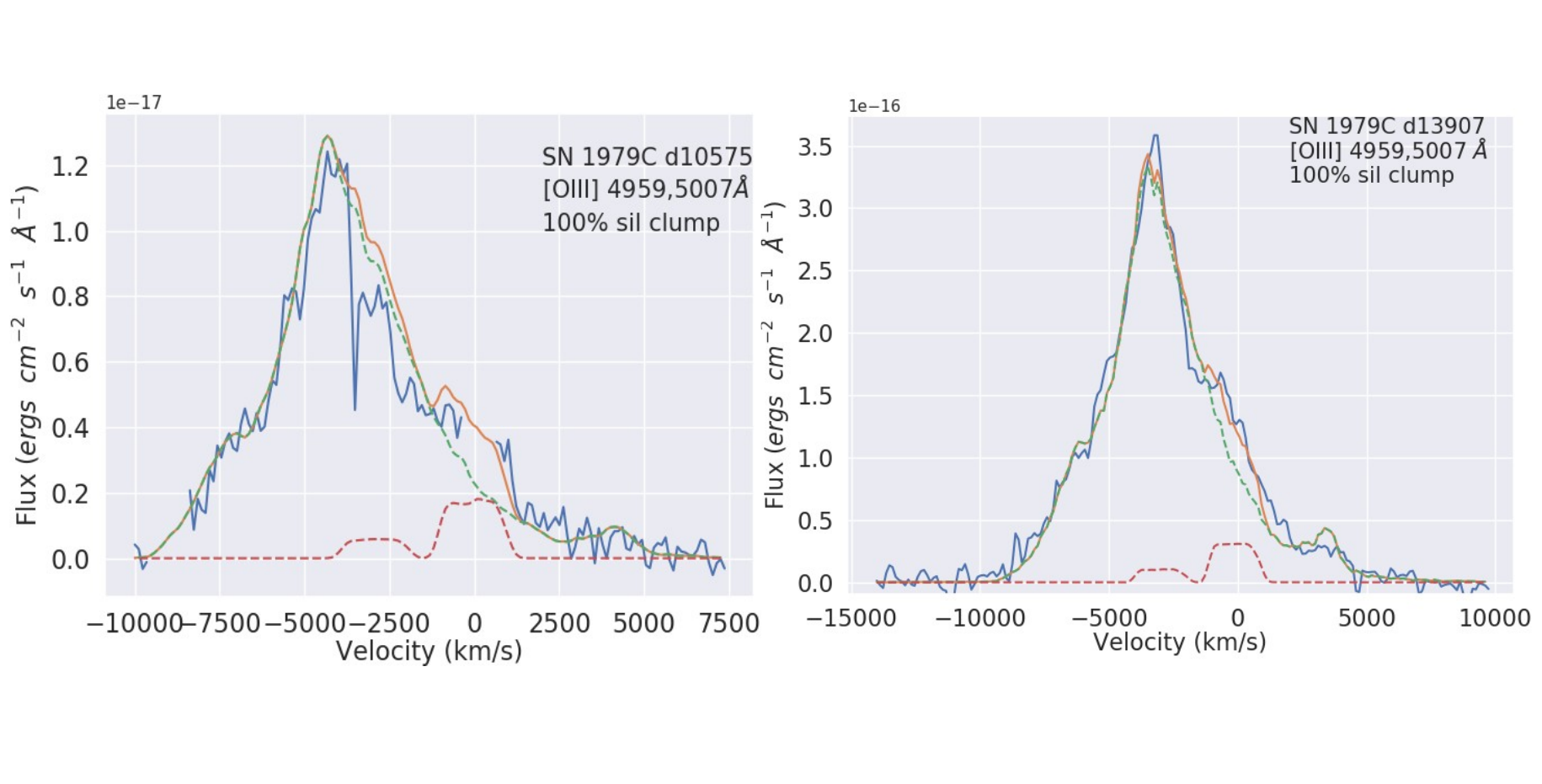}

\caption{{\sc damocles} models of the [O~{\sc iii}] profile of SN~1979C at 10575 and 13907 days post-explosion. The red and green dashed lines are the {\sc damocles} models of the IWC and BC, and the orange line is the sum of these two components. Clumped dust models of 100~per~cent astronomical silicates are shown.}
\label{fig:1979C-oiii-2008-2017-fits}
\end{figure*}

Models of the [O~{\sc i}] 6300,6363-$\AA$ and [O~{\sc ii}] 7319,7330-$\AA$ doublets at an epoch of 5146 days are shown in Figure~\ref{fig:1979C-93fits}. We did not model the [O~{\sc iii}] 4959,5007-$\AA$ doublet at 5146 days as there were several poorly subtracted narrower nebular lines present which, given the low resolution, obscured the line profile shape. The dust was optically thick at this epoch. We found that for a dust clump of 100\% astronomical silicate $\tau=4.4$ at a line wavelength of 7319~\AA. 

We modelled the [O~{\sc i}] 6300,6363-$\AA$, [O~{\sc ii}] 7319,7330-$\AA$ and [O~{\sc iii}] 4959,5007-$\AA$ doublets at days 10575 and 13907 post-explosion. The best-fitting clumped silicate dust models are shown in Figures \ref{fig:1979C-oiii-2008-2017-fits} and \ref{fig:1979C-oii-oi-2008-2017-fits}. 

We modelled the H$\alpha$ line profile at day 13150, as it was slightly less blended with the [O~{\sc i}] profile at this epoch than at day 13907, which can be seen in Figure \ref{fig:1979C-hafits}. 
We used a dust clumping mass fraction of 0.95 for all our clumped models at all epochs. A diagram showing the adopted distributions of the H$\alpha$ and oxygen emitting regions for SN~1979C can be seen in Figure~A7.


\begin{table*}
\caption{Parameters used in the {\sc damocles} models of SN 1979C at various epochs for smooth and clumped  for 100~per~cent AmC dust (AmC) or 100~per~cent astronomical silicate dust (sil) of radius $a$, for the broad and intermediate width components. $a$ is derived from fitting the oxygen lines at day 13907 and is fixed at the other epochs. The optical depth is calculated from R$_{in}$ to R$_{out}$ at the central line wavelength ([O~{\sc i}]=6300 $\AA$, [O~{\sc ii}]=7319~$\AA$), [O~{\sc iii}]=5007~$\AA$, Ha=6563~$\AA$)} 
\centering

\begin{tabular}{cp{0.5cm}ccp{0.4cm}cp{1.4cm}p{1.1cm}p{0.4cm}p{1.1cm}p{1.1cm}p{0.7cm}p{0.7cm}p{0.5 cm}}
\hline
Epoch & Line &  Comp. & Clumped? & Species & a & V$_{max}$ & V$_{min}$ & $\beta_{gas}$ & R$_{out}$ & R$_{in}$ & $\tau$ & M$_{dust}$ & $\chi^2$   \\
days & & & & & $\mu$m & km~s$^{-1}$ & km~s$^{-1}$ & & 10$^{15}$~cm & 10$^{15}$~cm & & M$_{\odot}$ & \\
\hline
5146  & [O~{\sc ii}]  & BC  & Yes & AmC & 0.20 & 6700 & 5360 & 4.5 & 302.0  & 241.0 & 22.0 & 0.15  & 3.43 \\
5146  & [O~{\sc ii}]  & BC  & No  & AmC & 0.13 & 6700 & 5360 & 4.5 & 297.0  & 237.0 & 6.2  & 0.03  & 4.32 \\
5146  & [O~{\sc ii}]  & BC  & Yes & sil & 0.06 & 6800 & 5440 & 4.5 & 297.0  & 237.0 & 10.6 & 2.40  & 3.38 \\
5146  & [O~{\sc ii}]  & BC  & No  & sil & 0.05 & 6800 & 5440 & 4.5 & 297.0  & 237.0 & 2.8  & 1.00  & 4.00 \\
5146  & [O~{\sc ii}]  & IWC &     &     &      & 1800 & 1260 & 0.1 & 430.0  & 301.0 &      &       &      \\
5146  & [O~{\sc i}]   & BC  & Yes & AmC & 0.20 & 7000 & 5320 & 4.5 & 311.2  & 236.5 & 12.9 & 0.10  & 1.72 \\
5146  & [O~{\sc i}]   & BC  & No  & AmC & 0.13 & 7000 & 5320 & 4.5 & 311.2  & 236.5 & 4.1  & 0.02  & 2.33 \\
5146  & [O~{\sc i}]   & BC  & Yes & sil & 0.06 & 7000 & 5320 & 4.5 & 311.2  & 236.5 & 17.3 & 2.40  & 1.70 \\
5146  & [O~{\sc i}]   & BC  & No  & sil & 0.05 & 7000 & 5320 & 4.5 & 311.2  & 236.5 & 3.8  & 0.70  & 2.15 \\
5146  & [O~{\sc i}]   & IWC &     &     &      & 1800 & 1260 & 0.1 & 430.0  & 301.0 &      &       &      \\
 &&&&&&&&&&&&& \\
10575 & [O~{\sc iii}] & BC  & Yes & AmC & 0.15 & 6900 & 4485 & 6.0 & 630.4  & 409.8 & 16.2 & 0.35  & 1.96 \\
10575 & [O~{\sc iii}] & BC  & No  & AmC & 0.13 & 6900 & 4485 & 6.0 & 630.4  & 409.8 & 4.2  & 0.08  & 1.96 \\
10575 & [O~{\sc iii}] & BC  & Yes & sil & 0.04 & 6900 & 4485 & 6.0 & 630.4  & 409.8 & 16.5 & 12.0 & 1.94 \\
10575 & [O~{\sc iii}] & BC  & No  & sil & 0.04 & 6900 & 4485 & 6.0 & 630.4  & 409.8 & 3.4  & 2.50  & 1.92 \\
10575 & [O~{\sc iii}] & IWC &     &     &      & 1300 & 910  & 0.1 & 1000 & 700.0 &      &       &      \\
10575 & [O~{\sc ii}]  & BC  & Yes & AmC & 0.20 & 7000 & 4410 & 4.5 & 639.6  & 402.9 & 16.0 & 0.45  &      \\
10575 & [O~{\sc ii}]  & BC  & No  & AmC & 0.13 & 6800 & 4080 & 4.5 & 621.3  & 372.8 & 5.4  & 0.09  &      \\
10575 & [O~{\sc ii}]  & BC  & Yes & sil & 0.06 & 6600 & 4950 & 5.5 & 603.0  & 452.3 & 11.7 & 10.0 &      \\
10575 & [O~{\sc ii}]  & BC  & No  & sil & 0.05 & 6600 & 4950 & 5.5 & 603.0  & 452.3 & 7.0  & 3.00  &      \\
10575 & [O~{\sc ii}]  & IWC &     &     &      & 1400 & 980  & 0.1 & 1000 & 700.0 &      &       &      \\
10575 & [O~{\sc i}]   & BC  & Yes & AmC & 0.20 & 6800 & 4760 & 4.0 & 621.3  & 434.9 & 6.9  & 0.45  & 0.46 \\
10575 & [O~{\sc i}]   & BC  & No  & AmC & 0.13 & 6800 & 4760 & 4.0 & 621.3  & 434.9 & 2.9  & 0.06  & 0.47 \\
10575 & [O~{\sc i}]   & BC  & Yes & sil & 0.06 & 6800 & 4964 & 5.5 & 621.3  & 453.6 & 19.2 & 10.0 & 0.47 \\
10575 & [O~{\sc i}]   & BC  & No  & sil & 0.05 & 6800 & 4964 & 5.5 & 621.3  & 453.6 & 2.7  & 2.00  & 0.48 \\
10575 & [O~{\sc i}]   & IWC &     &     &      & 1400 & 980  & 0.1 & 1000 & 700.0 &      &       &      \\
 &&&&&&&&&&&&& \\
13150 & H$\alpha$         & BC  & Yes & AmC & 0.20 & 8900 & 4005 & 1.5 & 1710 & 855.0 & 3.6  & 0.16  & 0.90 \\
13150 & H$\alpha$         & BC  & No  & AmC & 0.13 & 8900 & 4005 & 1.5 & 1710 & 855.0 & 3.8  & 0.08  & 0.87 \\
13150 & H$\alpha$         & BC  & Yes & sil & 0.06 & 8900 & 4005 & 1.5 & 1610 & 805.0 & 4.9  & 3.40  & 0.95 \\
13150 & H$\alpha$         & BC  & No  & sil & 0.05 & 8900 & 4005 & 1.5 & 1610 & 805.0 & 1.8  & 2.45  & 0.88 \\
 &&&&&&&&&&&&& \\
13907 & [O~{\sc iii}] & BC  & Yes & AmC & 0.15 & 6700 & 3618 & 4.6 & 805.0  & 434.7 & 10.2 & 0.30  & 2.50 \\
13907 & [O~{\sc iii}] & BC  & No  & AmC & 0.13 & 6700 & 3618 & 4.6 & 805.0  & 434.7 & 3.1  & 0.08  & 2.90 \\
13907 & [O~{\sc iii}] & BC  & Yes & sil & 0.04 & 6700 & 3618 & 4.9 & 805.0  & 434.7 & 8.1  & 8.50  & 2.90 \\
13907 & [O~{\sc iii}] & BC  & No  & sil & 0.04 & 6700 & 3618 & 4.9 & 805.0  & 434.7 & 2.6  & 2.70  & 3.08 \\
13907 & [O~{\sc iii}] & IWC &     &     &      & 1400 & 980  & 0.1 & 1190 & 833.0 &      &       &      \\
13907 & [O~{\sc ii}]  & BC  & Yes & AmC & 0.20 & 6600 & 3696 & 2.2 & 793.0  & 444.1 & 3.0  & 0.16  &      \\
13907 & [O~{\sc ii}]  & BC  & No  & AmC & 0.13 & 6600 & 3696 & 2.2 & 793.0  & 444.1 & 1.9  & 0.08  &      \\
13907 & [O~{\sc ii}]  & BC  & Yes & sil & 0.06 & 6600 & 3696 & 2.2 & 793.0  & 444.1 & 3.6  & 4.40  &      \\
13907 & [O~{\sc ii}]  & BC  & No  & sil & 0.05 & 6600 & 3696 & 2.2 & 793.0  & 444.1 & 1.6  & 2.45  &      \\
13907 & [O~{\sc i}]   & BC  & Yes & AmC & 0.20 & 6800 & 3808 & 2.2 & 817.1  & 457.6 & 3.7  & 0.16  & 0.37 \\
13907 & [O~{\sc i}]   & BC  & No  & AmC & 0.13 & 6800 & 3808 & 2.2 & 817.1  & 457.6 & 1.9  & 0.06  & 0.44 \\
13907 & [O~{\sc i}]   & BC  & Yes & sil & 0.06 & 6800 & 3808 & 2.2 & 817.1  & 457.6 & 5.7  & 4.40  & 0.43 \\
13907 & [O~{\sc i}]   & BC  & No  & sil & 0.05 & 6800 & 3808 & 2.2 & 817.1  & 457.6 & 2.0  & 2.00  & 0.43 \\
\hline

\end{tabular}
\label{table:1979C-params}
\end{table*}

\begin{figure*}
\centering

\includegraphics[width=0.9\linewidth]{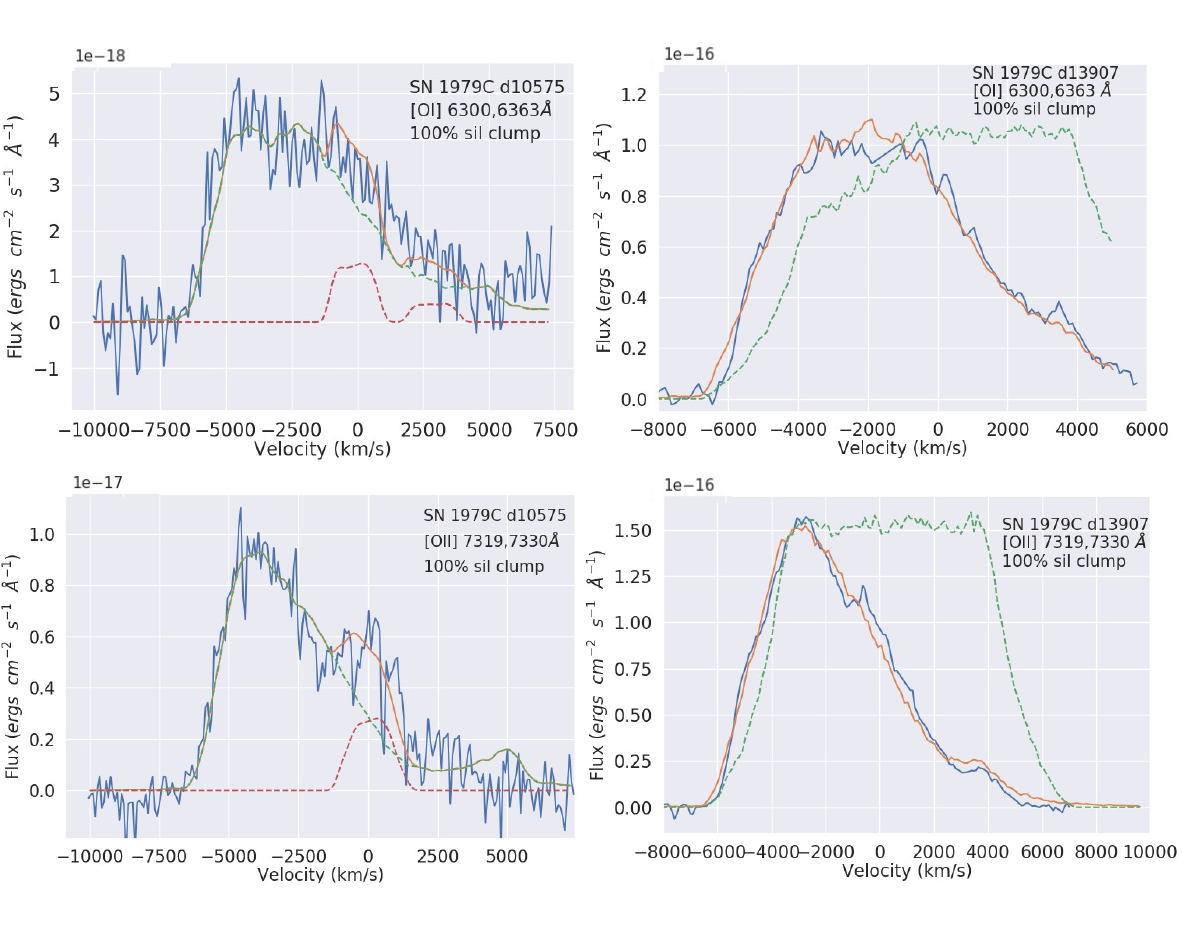}
\caption{{\sc damocles} models of the [O~{\sc i}] and [O~{\sc ii}] doublets in SN 1979C at 10575 and 13907 days post-explosion. For day 10575, the red and green dashed lines are the {\sc damocles} models of the Intermediate Width and Broad Component, and the orange line is the sum of these two components. For day 13907, as no IWC component is required the orange line is the BC. The green dashed line is the dust free model. Clumped dust models of 100~per~cent astronomical silicates are shown.}
\label{fig:1979C-oii-oi-2008-2017-fits}
\end{figure*}

\begin{figure*}
\centering

\includegraphics[scale=0.5]{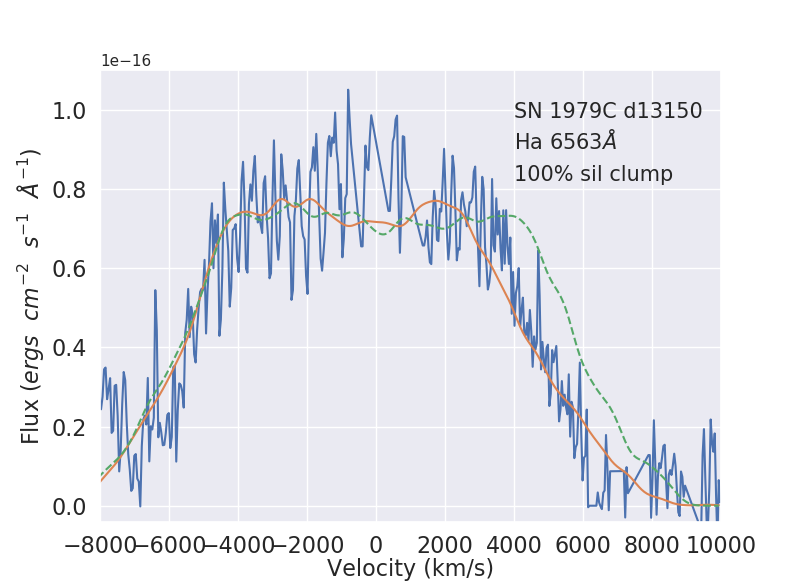}

\caption{{\sc damocles} model of the  H$\alpha$ profile of SN 1979C at 13150 days post-explosion. The green dashed line is the dust free model. The orange line represents the best-fitting dusty model, where the dust is assumed to consist of 100~per~cent silicate dust clumps, located with the freely expanding [O~{\sc ii}] ejecta at day 13907, where the parameters for this dust distribution are given in Table \ref{table:1979C-params}. }
\label{fig:1979C-hafits}
\end{figure*}

The grain radius was originally constrained for a smooth AmC dust model from the scattering wing of the [O~{\sc ii}] 7319,7330-$\AA$ line profile at day 13907, which had the best signal to noise of all the line profiles for any epoch. The albedo for this dust grain radius was 0.48. For a clumped amorphous carbon dust composition, as well as for a smooth and clumped silicate dust composition, we found dust grain radii that best matched this albedo. These grain sizes, along with the rest of the model parameters, are listed for all the modelled lines in Table~\ref{table:1979C-params}. The grain radius derived at this epoch was fixed for the other epochs.The dust masses derived using 100~per~cent silicate compositions are much higher than for the 100~per~cent AmC models, leading us to favour dust models for SN~1979C that have a silicate proportion of 50~per~cent or less.
For consistency with the other CCSNe modelled in our sample, we use a 50:50 AmC to silicate dust composition with a single grain radius, for which in this case the best fitting grain radius was 0.1~$\mu$m, to report our final preferred dust masses.

From a manual parameter space examination,
the [O~{\sc i}], [O~{\sc ii}] and [O~{\sc iii}] lines at day 10575 all required similarly large dust masses, where the [O~{\sc i}] and [O~{\sc ii}] doublets could be well fit with models of matching parameters. We note that the [O~{\sc iii}] distribution differs from that of [O~{\sc i}] and [O~{\sc ii}] in requiring a higher $\beta$ value, which implies the [O~{\sc iii}] gas is more densely distributed than the other oxygen ion species. A higher $\beta$ value requires more dust to provide the same amount of absorption as a model with a lower $\beta$, which has led to slightly higher dust masses derived from modelling the [O~{\sc iii}] doublet in comparison to the other oxygen ion lines. These trends also apply to the spectral lines at day 13907 past explosion.

Using a Bayesian analysis, we evaluated the errors on the ejecta dust masses at day 10575 and 13907 that were derived from the [O~{\sc ii}] doublet, where the resulting corner plots can be found in Figures~A2 and A3. We removed the IWC component to the [O~{\sc ii}] line at both epochs prior to modelling. The oxygen line profiles appear significantly red-shifted between days 10575 and 13907, so quantifying the errors on the derived dust masses is particularly important.
Simulations for both epochs were run with 100~per~cent clumped AmC dust, where the median dust mass at both epochs agreed very well with our initial estimates for the 100~per~cent AmC dust masses shown in Table \ref{table:1979C-params}. We therefore extrapolated the dust mass uncertainties to our 50:50 AmC to silicate dust masses for a 0.1~$\mu$m single-size grain radius, leading to dust masses of 0.65$^{+0.85}_{-0.43}$ and 0.30$^{+0.13}_{-0.15}$~M$_\odot$ at days 10575 and 13907. Whilst the median values would indicate dust destruction between the two epochs, the large error bars on the dust mass at day 10575 mean we cannot definitely conclude that dust destruction is happening. However, this possibility makes SN~1979C a particularly interesting target for continuing observations.
We also note that both Bayesian runs favoured grain radii of 0.03-0.63~$\mu$m for 100~per~cent AmC grains, where the 1-D probability distribution at day 13907 peaked at the same grain radii found with the manual fitting process, namely $\sim$0.2~$\mu$m.

To model the H$\alpha$ line at day 13150, we decoupled the dust and gas distributions, where the dust distribution shared that of the [O~{\sc ii}] and [O~{\sc i}] models for day 13907, and the gas was located such that R$_{in,Ha}$ = R$_{out,O}$. We found a good fit to the H$\alpha$ line with this simple model, seen in Figure~\ref{fig:1979C-hafits}. The $\chi^2$ for the dusty models averaged 0.9, while it was 1.5 for the dust-free model.

\subsubsection{SN 1980K}
SN 1980K was discovered by P. Wild on 1980 October 28 in NGC~6946, and reached a peak brightness of V = 11.4 mag a few days later \citep{Buta94}. \citet{Montes11998} estimated an explosion date for SN 1980K of 2nd October 1980, which we use in this work.
The detection of a broad H$\alpha$ line in early spectra and a linearly decaying light curve after peak brightness resulted in its classification 
as a Type II-L SN \citep{Barbon1979}. 
The emergence of a near-IR flux excess in 1981 led \citet{dwek1983} to surmise that there could be dust in 1980K, but they could not resolve whether the dust was newly formed in the ejecta or pre-existing grains in a CSM. \citet{Milisavljevic2012} presented a spectrum of SN~1980K obtained 30 years after explosion (on day 10964) and postulated that there could be dust present in the ejecta based on the observed blue-shifting of their H$\alpha$ and [O~{\sc i}] line profiles. The profiles of the day 10964 lines were modelled using {\sc damocles} by \citet{Bevan2017} (B17), who derived a dust mass of $\sim$0.2~M$_{\odot}$ for a dust composition that was deduced to be dominated by silicate grains, from the presence of extended red scattering wings.

\begin{figure*}
\centering

\includegraphics[width=\linewidth]{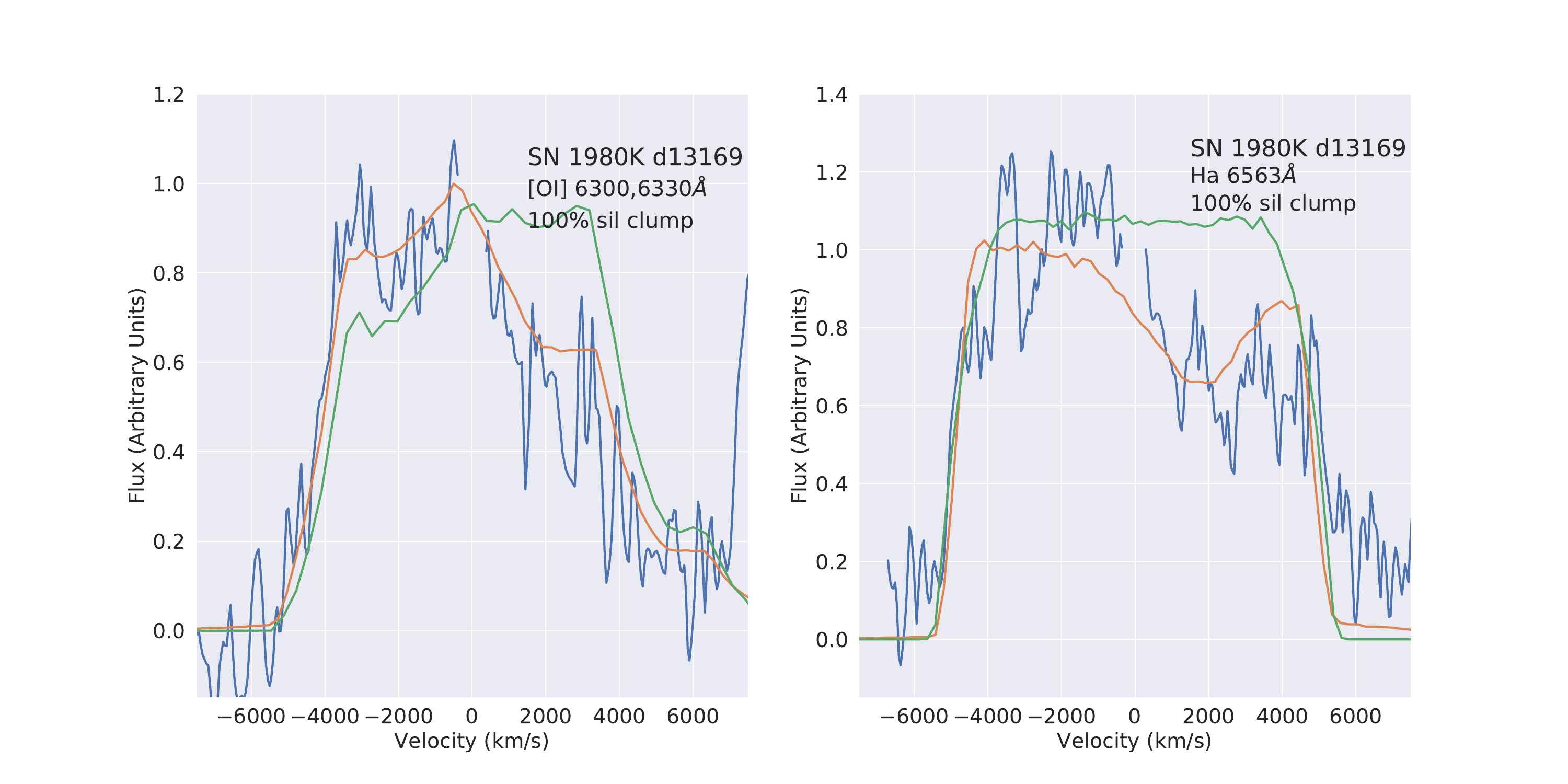}

\caption{{\sc damocles} models of the [O~{\sc i}] (left) and H$\alpha$ (right) line profiles of SN~1980K at 13169 days post-explosion. The green line is the dust-free {\sc damocles} model, and the orange line is the dust-affected model. Clumped dust models consisting of 100~per~cent astronomical silicate grains of 0.1~$\mu$m radius are shown.}
\label{fig:1980k-fits}
\end{figure*}

\begin{figure*}
\centering

\includegraphics[width=\linewidth]{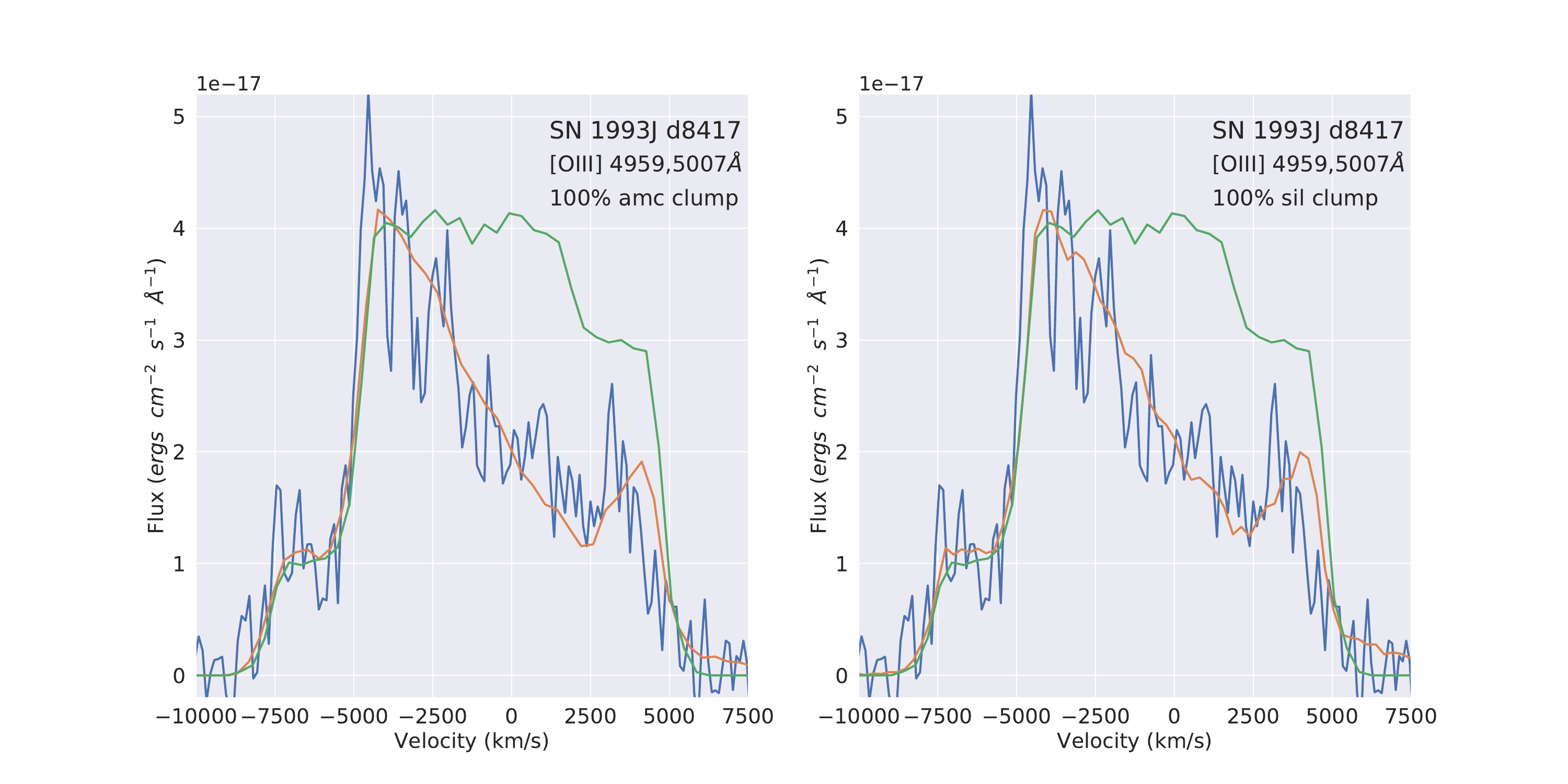}

\caption{{\sc damocles} models of the [O~{\sc iii}] 4959,5007-\AA\ doublet profile of SN~1993J at 8417 days post-explosion. The green line is the dust-free {\sc damocles} model, and the orange line is the dust-affected model. Clumped dust models for 100~per~cent AmC (left) and 100~per~cent astronomical  silicates (right) are shown.}
\label{fig:1993j-fits}
\end{figure*}

We have summed the day~12977 MMT spectrum of SN~1980K (Table~\ref{table:archiv-obs-sum}) with our GMOS spectra taken at a mean epoch of 13361 days after explosion (Table~\ref{table:gmos-dat}), correcting for the small recessional velocity of the host galaxy (40~km~s$^{-1}$). The resulting `day~13169' spectrum of SN~1980K has a slightly higher S/N than the day~10964 spectrum taken six years earlier by \citet{Milisavljevic2012} and modelled by B17, but not a high enough S/N to conduct a Bayesian analysis. We have modelled the [O~{\sc i}] 6300,6363-$\AA$ and H$\alpha$ lines using the same species and grain radius values as B17. Our best fit clumped silicate dust models are shown in Figure~\ref{fig:1980k-fits}, with the parameters listed in Table~\ref{table:4sn-params}. The model parameters for the [O~{\sc i}] line we found were similar to those of B17, although the density exponent of $\beta = 1.5$ required for our best-fitting models was lower than their day 10964 best-fitting value of $\beta=4.0$. The best-fitting dust mass required to attenuate the [O~{\sc i}] line profile, for a 100~per~cent clumped silicate model with a grain radius of 0.1~$\mu$m, was 0.60$^{+3.29}_{-0.57}$~M$_\odot$. It was possible to fit the H$\alpha$ line profile with similar parameters as those used to model the [O~{\sc i}] line. Therefore, we ran a Bayesian analysis of the H$\alpha$ and [O~{\sc i}] line simultaneously using a silicate dust grain species, which provided the dust mass uncertainty limits on the value referenced above.
The day 10964 dust mass of 0.3~M$_\odot$ derived by B17 using the same grain parameters is within our calculated uncertainties. 

\begin{figure*}
\centering

\includegraphics[width=\linewidth]{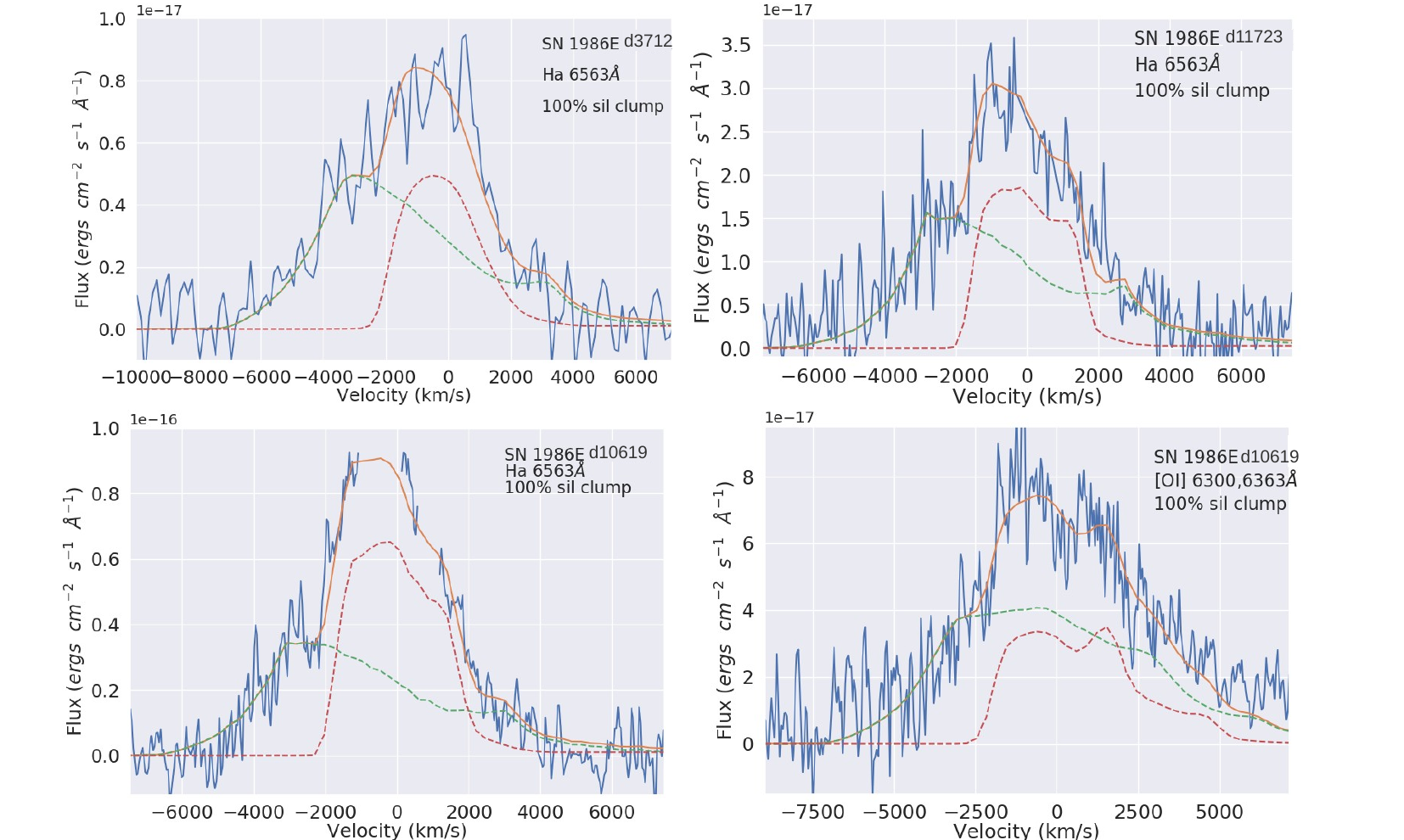}

\caption{{\sc damocles} models of the H$\alpha$ line profiles of SN~1986E at 3712, 10619 and 11723 days post-explosion, and of the [O~{\sc i}] 6300,6363-$\AA$ doublet 10619 days past explosion. The green line is the dust-affected BC, the red line is the dust-affected IWC, and the orange line is the sum of the green and red lines. Clumped dust models consisting of 100~per~cent astronomical silicates are shown.}
\label{fig:86e-allfits}
\end{figure*}
 
Setting the R$_{in}$ value for the H$\alpha$-emitting gas to be equal to the R$_{out}$ value of the [O~{\sc i}] distribution, as in the case of SN~1979C, could not replicate the observed H$\alpha$ line profile, as then the absorption induced by the dust that is coupled with the [O~{\sc i}]-emitting ejecta produced an over-absorption of the red wing of the H$\alpha$ line. The adopted distributions of the H$\alpha$ and oxygen emitting regions in SN 1980K can be seen in Figure A7.

\subsubsection{SN 1993J}
SN 1993J is located in the nearby M~81 galaxy and was discovered on 28/03/1993 \citep[][]{ripero1993}. \citet{baron1993} established an explosion date of 27/03/1993. 
The asymmetries in the broad oxygen lines in the late-time spectra led \citet{Fransson2005} and \citet{Milisavljevic2012} to speculate on the presence of dust in the SN, and modelling by B17 of the red-blue asymmetries in the latter's day 6101 spectrum using {\sc damocles} yielded an ejecta silicate dust mass of $\sim$0.1~M$_\odot$. \citet{zsiros2022} modelled archival 3.6-24-$\mu$m {\em Spitzer} SEDs of SN~1993J, obtained between 3875 and 5359 days post-explosion, and found there could be either $\sim5\times10^{-3}$ M$_{\odot}$ of silicate dust or $\sim1\times10^{-3}$~M$_\odot$ of amorphous carbon dust, emitting at T$\sim$200~K, significantly less than the $\sim$0.1~M$_\odot$ found by B17, which we attribute to the insensitivity of the {\em Spitzer} measurements to colder dust emitting longwards of 24~$\mu$m.

For SN~1993J we modelled the [O~{\sc iii}] 4959,5007~\AA\ doublet profile from an MMT spectrum taken 8417 days past explosion (Table~\ref{table:archiv-obs-sum}). 
This spectrum had a lower S/N than the day~6101 spectrum of \citet{Milisavljevic2012} which was modelled by B17, enabling only the [O~{\sc iii}] doublet to be modelled
by us. We corrected the spectrum for a recessional velocity for M~81 of 140~km~s$^{-1}$ and used the same dust compositions as B17 in our models. Our best-fit clumped silicate and AmC dust models are shown in Figure~\ref{fig:1993j-fits}, with parameters listed in Table~\ref{table:4sn-params}.

For a 100~per~cent clumped silicate dust distribution with a grain radius of 0.04~$\mu$m, as in B17, 
we found a best-fit day~8417 dust mass of 0.50~M$_\odot$, from manually fitting the [O~{\sc iii}] doublet, versus B17's day~6101 dust mass of 0.15~M$_\odot$. 
Like B17, we were not quite able to fit the red wing of the [O~{\sc iii}] doublet, which could be due to the over-simplification of assuming that all of the emitting oxygen is distributed with a uniform density power-law, as discussed by B17 in their Section 4.3.

A Bayesian model of the [O~{\sc iii}] doublet using a 100~per~cent silicate dust composition returned a median dust mass of 0.21$^{+0.67}_{-0.20}$~M$_\odot$. However, as the grain species or radius could not be determined for SN~1993J, we adopt the dust mass of $4.00^{+13.0}_{-3.8}\times10^{-2}$~M$_\odot$ obtained for a 50:50 AmC to silicate dust mass ratio with a grain radius of 0.1~$\mu$m, where the error limits are scaled from the percentage errors of the Bayesian best-fitting 100~per~cent silicate dust mass of 0.21~M$_\odot$ reported above.

\begin{table*}
\small
\centering
\caption{
Parameters used in the {\sc damocles} models for SN~1986E's and SN~2004et's broad and intermediate width components of H$\alpha$ and SN~1986E's [O~{\sc i}] 6300,6363-$\AA$ emission lines for spherically symmetric smooth and clumped dust models. "Sil" stands for 100~per~cent astronomical silicate dust, while "AmC" is 100~per~cent AmC dust, each of grain radius $a$. The optical depth is calculated from R$_{in}$ to R$_{out}$ at the central line wavelengths ([O~{\sc i}]=6300~$\AA$, H$\alpha$=6563~$\AA$).  All SN2004et models used 100~per~cent silicate dust, which was determined from fits to the mid-IR SEDs by \citet{Kotak2009} and \citet{Fabbri2011a}. The optical depth was calculated from R$_{in}$ to R$_{out}$ at the wavelength of H$\alpha$. The grain radius is fixed to 0.2~$\mu$m for the BC at both epochs from fits to the IWC at day 646.} 
\centering
\begin{tabular}{ccccccccccccccc}
\hline
SN & Epoch & Line & Clumped? & Comp. & Species & a  & V$_{max}$  & V$_{min}$  & $\beta_{gas}$ & R$_{out}$          & R$_{in}$  & M$_{dust}$  & $\tau$  & $\chi^2$   \\
& days & & & & & $\mu$m & km~s$^{-1}$ & km~s$^{-1}$ & & 10$^{15}$~cm & 10$^{15}$~cm & $10^{-2}$~M$_\odot$ & & \\
\hline
1986E & 3712 & H$\alpha$ & yes & BC  & AmC & 0.20 & 7100 & 3337 & 2.3  & 225.8  & 106.1 & 1.7 & 6.55 & 1.14 \\
1986E & 3712 & H$\alpha$ & no  & BC  & AmC & 0.20 & 7100 & 3337 & 3.0  & 225.8  & 106.1 & 0.7 & 2.34 & 1.04 \\
1986E & 3712 & H$\alpha$ & yes & BC  & sil & 0.55 & 7100 & 3337 & 2.3  & 225.8  & 106.1 & 25  & 12.2 & 1.11 \\
1986E & 3712 & H$\alpha$ & no  & BC  & sil & 0.55 & 7100 & 3337 & 3.0  & 225.8  & 106.1 & 7.1 & 3    & 1.06 \\
1986E & 3712 & H$\alpha$ & yes & IWC & sil & 0.20 & 2300 & 575  & -0.5 & 948.0  & 237.0 & 5   & 1.75 & \\
1986E & 10619 & [O~{\sc i}] & yes & BC  & AmC & 0.20 & 6600 & 3102 & 3.0  & 603.8  & 241.5 & 3.5 & 1.39 & 1.02 \\
1986E & 10619 & [O~{\sc i}] & no  & BC  & AmC & 0.20 & 6600 & 3102 & 3.0  & 603.8  & 241.5 & 2.5 & 0.97 & 0.99 \\
1986E & 10619 & [O~{\sc i}] & yes & BC  & sil & 0.55 & 6600 & 3102 & 3.0  & 603.8  & 241.5 & 25  & 1.33 & 1.05 \\
1986E & 10619 & [O~{\sc i}] & no  & BC  & sil & 0.55 & 6600 & 3102 & 3.0  & 603.8  & 241.5 & 15  & 0.78 & 1.04 \\
1986E & 10619 & [O~{\sc i}] & yes & IWC & sil & 0.20 & 2300 & 1725 & 2.0  & 1000 & 650.0 & 5   & 0.34 &      \\
1986E & 10619 & H$\alpha$ & yes & BC  & AmC & 0.20 & 6600 & 3102 & 3.0  & 603.8  & 283.8 & 2.5 & 1.13 & 1.07 \\
1986E & 10619 & H$\alpha$ & no  & BC  & AmC & 0.20 & 6600 & 3102 & 3.0  & 603.8  & 283.8 & 2   & 0.86 & 1.03 \\
1986E & 10619 & H$\alpha$ & yes & BC  & sil & 0.55 & 6600 & 3102 & 3.3  & 603.8  & 283.8 & 25  & 1.51 & 1.16 \\
1986E & 10619 & H$\alpha$ & no  & BC  & sil & 0.55 & 6600 & 3102 & 3.3  & 603.8  & 283.8 & 15  & 0.86 & 1.09 \\
1986E & 10619 & H$\alpha$ & yes & IWC & sil & 0.20 & 2100 & 1365 & 1.0  & 1000.0 & 650.0 & 5   & 0.35 & \\  
1986E & 11723 & H$\alpha$ & yes & BC  & AmC & 0.20 & 6600 & 2640 & 3.0  & 668.3  & 314.1 & 3.5 & 1.39 & 0.92 \\
1986E & 11723 & H$\alpha$ & no  & BC  & AmC & 0.20 & 6600 & 2640 & 3.0  & 668.3  & 314.1 & 2.5 & 1.04 & 0.93 \\
1986E & 11723 & H$\alpha$ & yes & BC  & sil & 0.55 & 6600 & 2640 & 3.0  & 668.3  & 314.1 & 35  & 1.77 & 0.91 \\
1986E & 11723 & H$\alpha$ & no  & BC  & sil & 0.55 & 6600 & 2640 & 3.0  & 668.3  & 314.1 & 25  & 1.34 & 0.93 \\
1986E & 11723 & H$\alpha$ & yes & IWC & sil & 0.20 & 1900 & 1330 & 2.0  & 1100.0 & 770.0 & 5   & 0.23 &   \\
 & & & & & & & & & & & \\
2004et & 646   & H$\alpha$   & no  & BC & sil & 0.2 & 9500 & 7410 & 1.7 & 53.02  & 39.24  & 0.03 & 0.5 &\\
2004et & 646 & H$\alpha$   & no  & IWC & sil & 0.2 & 3000 & 450  & 1.6 & 170.0 & 51.00  & 0.12 & 0.39 & \\
2004et & 646 & H$\alpha$   &  yes & BC & sil  & 0.2 & 9500 & 7410 & 1.7 & 53.02  & 39.24  & 0.05 & 0.57  & \\
2004et & 646 & H$\alpha$   &  yes & IWC & sil  & 0.2 & 3000 & 450  & 1.6 & 170.0 & 51.00  & 0.2  & 0.73 &  \\
2004et & 3849 & H$\alpha$    & no  & BC & sil   & 0.2 & 6000 & 5400 & 2   & 200.5 & 180.5 & 0.6  & 0.81 & 1.22  \\
2004et & 3849  & H$\alpha$   & yes & BC & sil  & 0.2 & 6000 & 5400 & 2   & 200.5 & 180.5 & 1.0 & 1.31 & 1.25 \\
\hline
\label{table:1986e-params}
\end{tabular}
\end{table*}

\begin{figure*}
\centering

    \includegraphics[width=1.5\columnwidth]{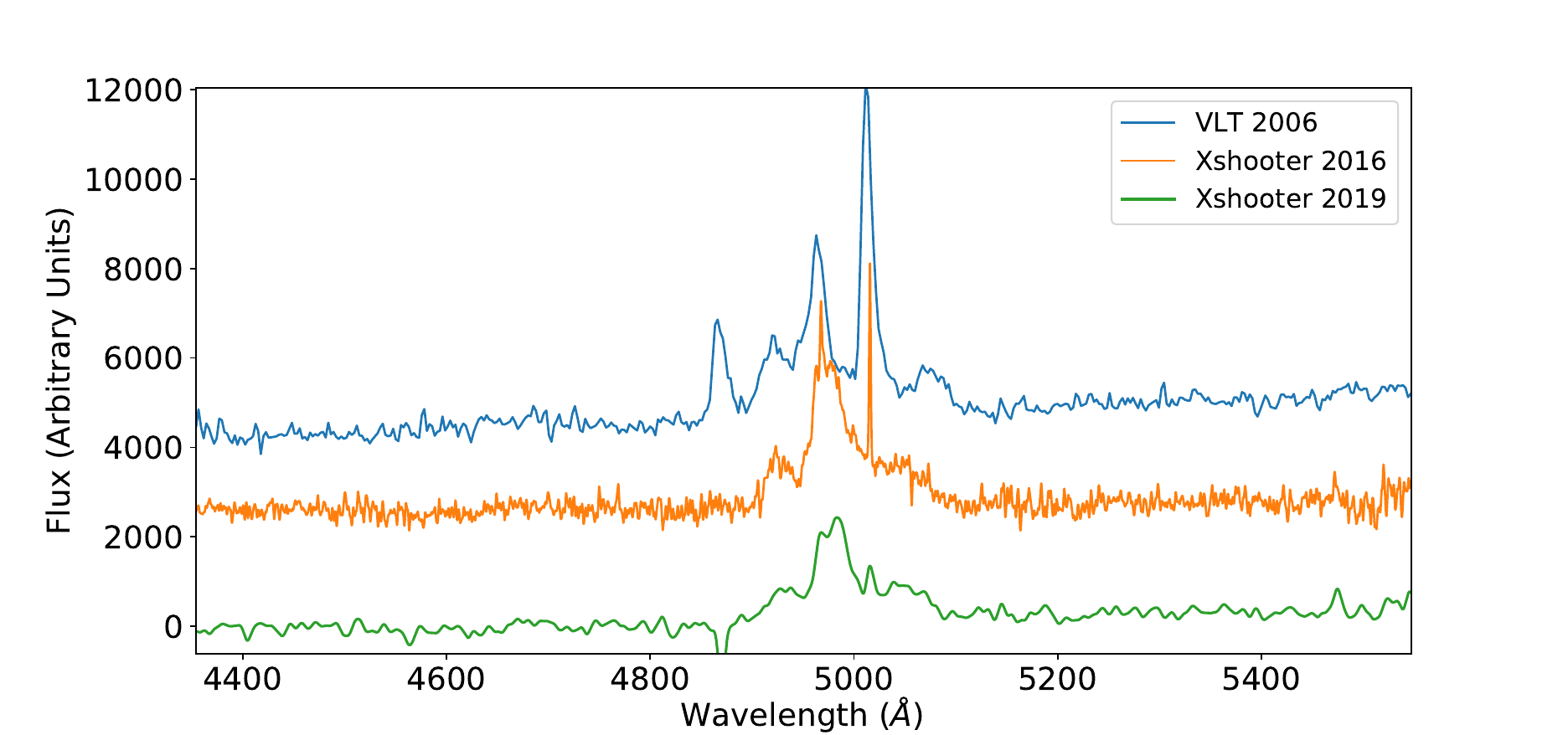}
%
    \includegraphics[width=1.5\columnwidth]{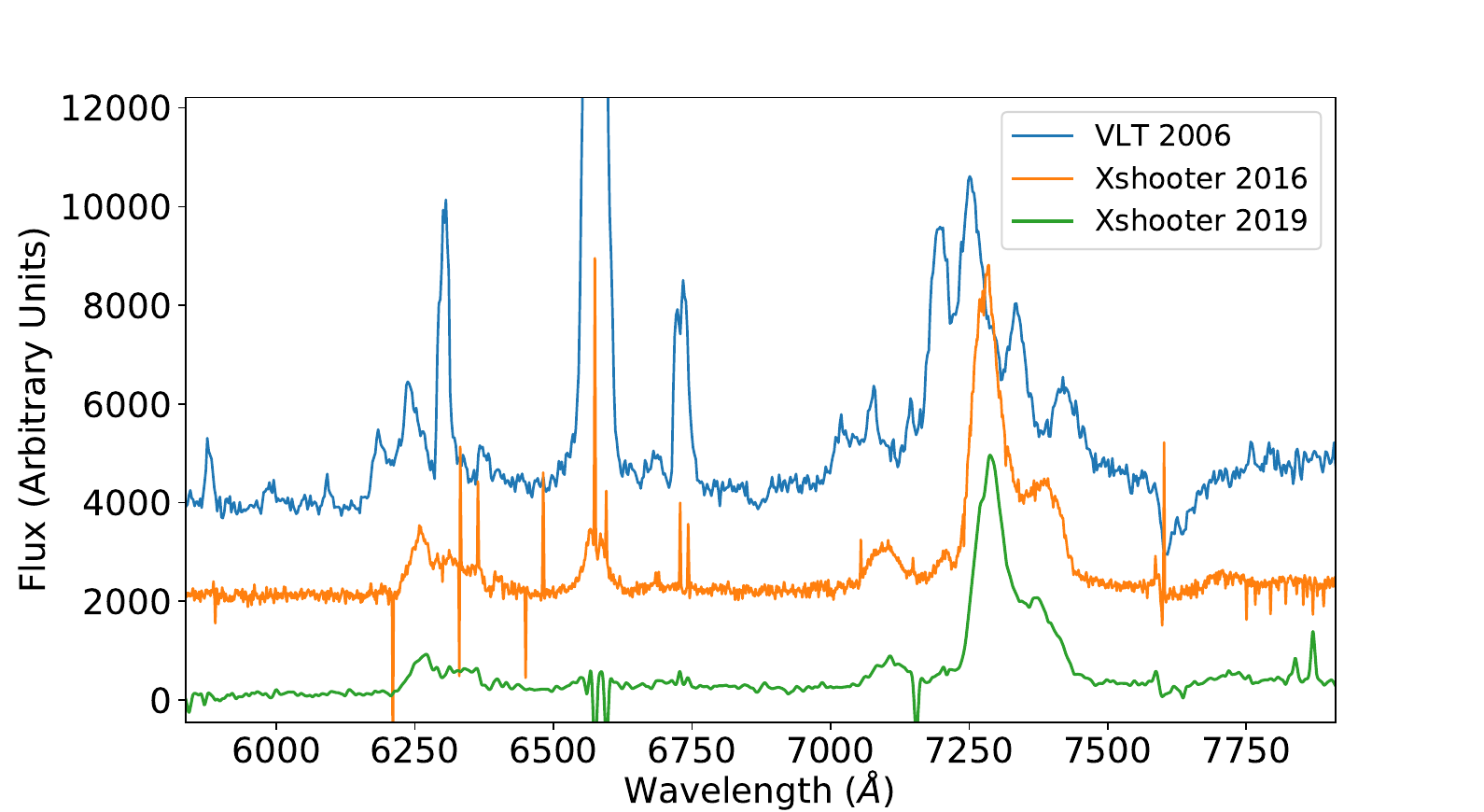}
\caption{The evolution of the optical spectrum of SN~1996cr between 2006 and 2019. Information on the spectra is summarised in Tables~\ref{table:xshoot-dat} and \ref{table:archiv-obs-sum}.  }
\label{fig:1996cr-evoln}
\end{figure*}

\subsubsection{SN 1986E}
SN~1986E was discovered on April 13th 1986 in NGC~4302, by G. Candeo at the Asiago Observatory \citep{Rosino1986}. 
An optical spectrum taken a month later of SN~1986E led  \citet{C.Pennypacker1986} to infer that it was around 2 months past maximum brightness, and they classified SN 1986E as a Type~II supernova from the presence of strong, broad H$\alpha$ and [O~{\sc i}] 6300,6363 $\AA$ lines with P-Cygni profiles in the spectra. We adopt an explosion date of 13/03/1986 for our models. \citet{Cappellaro1990} fitted photometric measurements of SN 1986E, taken over the course of 2 years, with the light curve of the typical Type~II-L SN~1979C.


Our optical spectra were corrected for a recessional velocity of NGC~4302 of 1148~km~s$^{-1}$. We have modelled the evolving optical emission H$\alpha$ line profile of SN 1986E at ages of 3712, 10619 and 11723 days, and also modelled the [O~{\sc i}] 6300,6363-$\AA$ doublet at day 10619, as the shape of the doublet at the other epochs is too unclear. Our best-fitting models can be found in Figure \ref{fig:86e-allfits}. The day 3712 spectrum was taken from the WISeREP archive\footnote{\url{https://wiserep.weizmann.ac.il/}} and appears to be unpublished: it has a spectral resolution of 12~$\AA$.

There is a strong contribution to both the H$\alpha$ and [O~{\sc i}] 6300,6363-\AA\ line profiles from an IWC and a BC, where the brightness of the IWC relative to that of the BC increases between days 3712 and 10619 past-explosion, and decreases between days 10619 to 11723. The IWC of SN~1986E is a much more prominent component to the profiles than for the emission lines of SN~2004et and SN~1979C at similar epochs, and it also varies much less in relative brightness over a longer period of time than for either SN~2004et or SN~1979C. The evolving spectra suggest there could be an interaction between the SN ejecta and an extended, smooth CSM. We assume the broad component (BC) in both the H$\alpha$ and [O~{\sc i}] profiles represents the freely expanding ejecta, whereas the IWC represents a collision-region between CSM and the ejecta, where the outer radius of the IWC is a free parameter.

The spectrum at day 10619 has the highest S/N and we consider the model parameters derived for this epoch the most reliable. As much of the BC shape is not visible due to the overlapping IWC and the presence of a scattering wing cannot be determined due to low S/N, we could not accurately constrain the grain species or grain radius of the dust inducing the red-blue asymmetry in the BC. Model parameters using conservative grain sizes for either a 100~per~cent AmC or a 100~per~cent silicate dust composition in the BC are listed in Table \ref{table:1986e-params}. We derived clumped dust masses which minimized the $\chi^2$ values between the model and observed H$\alpha$ BC profiles of SN~1986E for the three epochs for a 50:50 carbon to silicate composition with a single grain radius of 0.1~$\mu$m. These dust masses were 0.03$_{-0.024}$~M$_\odot$ at day 3712 and $\sim$0.07$_{-0.065}$~M$_\odot$ at 10619 and 11723 days, where the lower limits were determined by a 35~per~cent variation to the best-fitting $\chi^2$ value. We could not determine accurate upper limits for these epochs due to optically thick conditions for dust masses higher than the best fitting value but the above values indicate some dust mass growth in the ejecta of SN~1986E between days 3712-11723.

It is difficult to independently determine a dust mass from the BC of the [O~{\sc i}] doublet profile at day 10619 due to low S/N. However, we find we can fit the [O~{\sc i}] BC and IWC components with the same dust mass as that used for the H$\alpha$ model, implying that the oxygen- and hydrogen-emitting gas could be co-located in SN 1986E, differing from the cases of SN 1980K, 1979C and 1970G.

We tried a model for the H$\alpha$ line profile at all 3 epochs where the dust was coupled to the freely expanding ejecta represented by the BC, and the IWC outer radius was set so that R$_{in,IWC}$ = R$_{out,BC}$, as in the case for our models for SN~2004et at day 646. However, this setup led to over-absorption of the red wing of the IWC by the dust in the BC. Therefore, the inner radius of the IWC needs to be slightly larger than the outer radius of the BC. A diagram of the distributions adopted for the SN~1986E H$\alpha$ and oxygen-line emitting regions can be found in Figure A7. The IWC showed a red scattering wing, most noticeably in the H$\alpha$ profile at day 10619
which could be best fit using a mass of 0.05$^{+0.13}_{-0.047}$~M$_\odot$ for 0.2-$\mu$m silicate dust grains coupled to the IWC gas. As this dust would be located outside the BC-emitting shell, it would not induce a red-blue asymmetry in the broad component.

\subsubsection{SN 1996cr}
SN~1996cr in the Circinus galaxy was first identified as an ultra-luminous X-ray source by \citet{Bauer2001}. VLT optical spectra of the source taken by \citet{Bauer2008} revealed broad oxygen emission lines, and strong narrow H$\alpha$ emission which allowed them to classify the source as a supernova remnant created by a Type~IIn supernova explosion. Multi-wavelength archival observations helped them constrain the explosion date to between 28/02/1995 and 16/03/1996. We have adopted the latter date in order to calculate the post-explosion ages corresponding to our X-Shooter spectra of SN~1996cr.

The evolution of the optical spectrum of SN~1996cr from 2006-2019 is shown in Figure~\ref{fig:1996cr-evoln}. The oxygen doublets display red-blue asymmetries at every epoch. The 2006 VLT FORS~I spectrum was first presented by \citet{Bauer2008}. The oxygen doublets at this epoch have multiple peaks, indicative of ejecta interactions with a complex, likely asymmetric CSM. Ten years later, some of the peaks had disappeared from both the blue and red-shifted parts of the oxygen doublets, and the strength of the H$\alpha$ line has decreased substantially between 2006 and 2016, disappearing almost entirely by 2019. We assume that the spectrum of SN~1996cr in 2006 is dominated by the complicated ejecta-CSM interactions, whereas by 2016 the spectrum is dominated by ejecta emission, hence making it an ideal epoch to model with {\sc damocles}.  

\begin{figure*}
\centering

\includegraphics[width=0.9\linewidth]{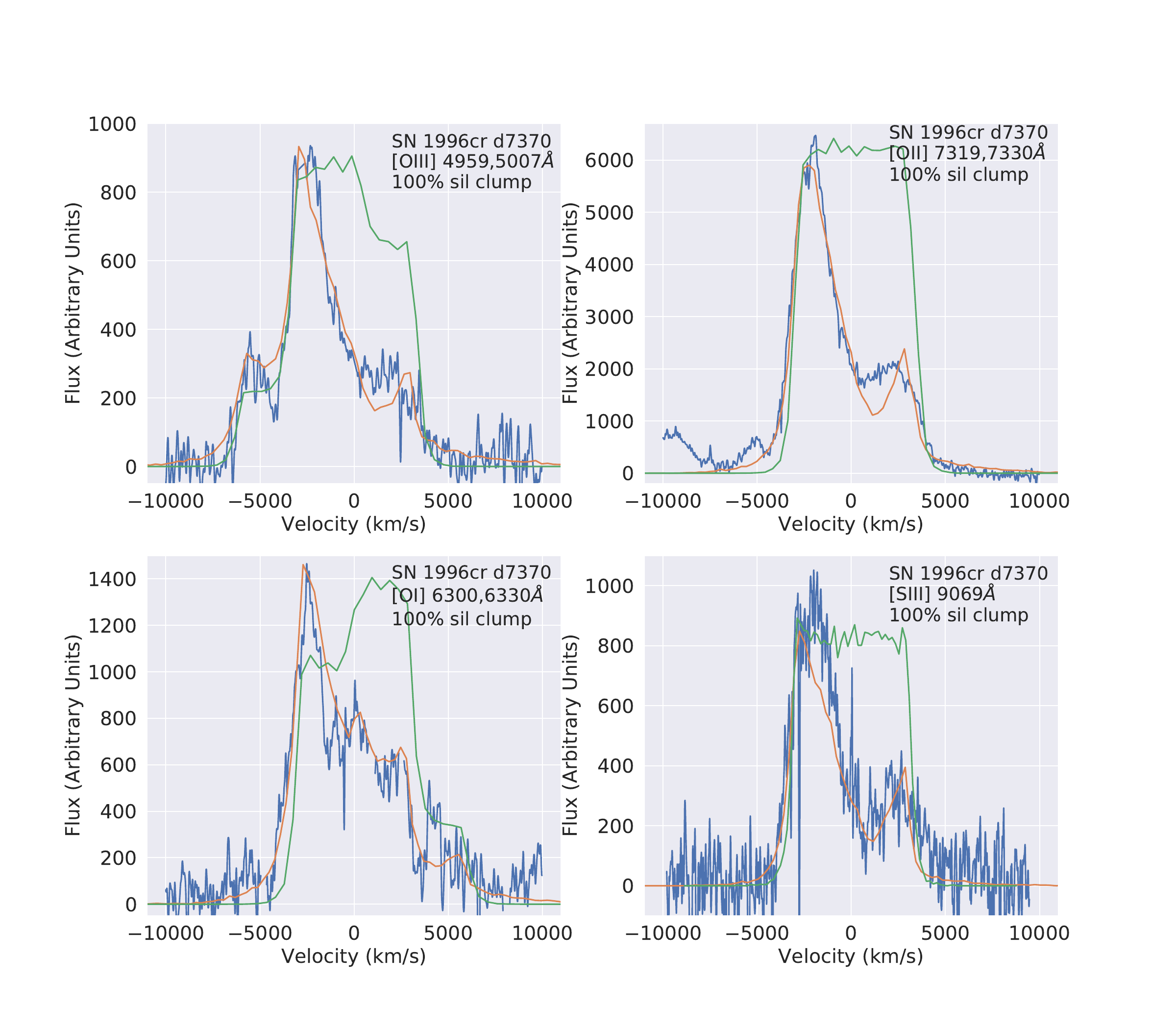}

\caption{{\sc damocles} models for the [O~{\sc i}], [O~{\sc ii}], [O~{\sc iii}] and [S~{\sc iii}] X-shooter line profiles of SN~1996cr, 7370 days post-explosion. The green lines are the dust-free {\sc damocles} models and the orange lines are the dust-affected model. Clumped dust models of 100~per~cent astronomical silicates are shown.}
\label{fig:96cr-7400-fits}
\end{figure*}
\begin{figure*}
\centering

\includegraphics[scale=0.3]{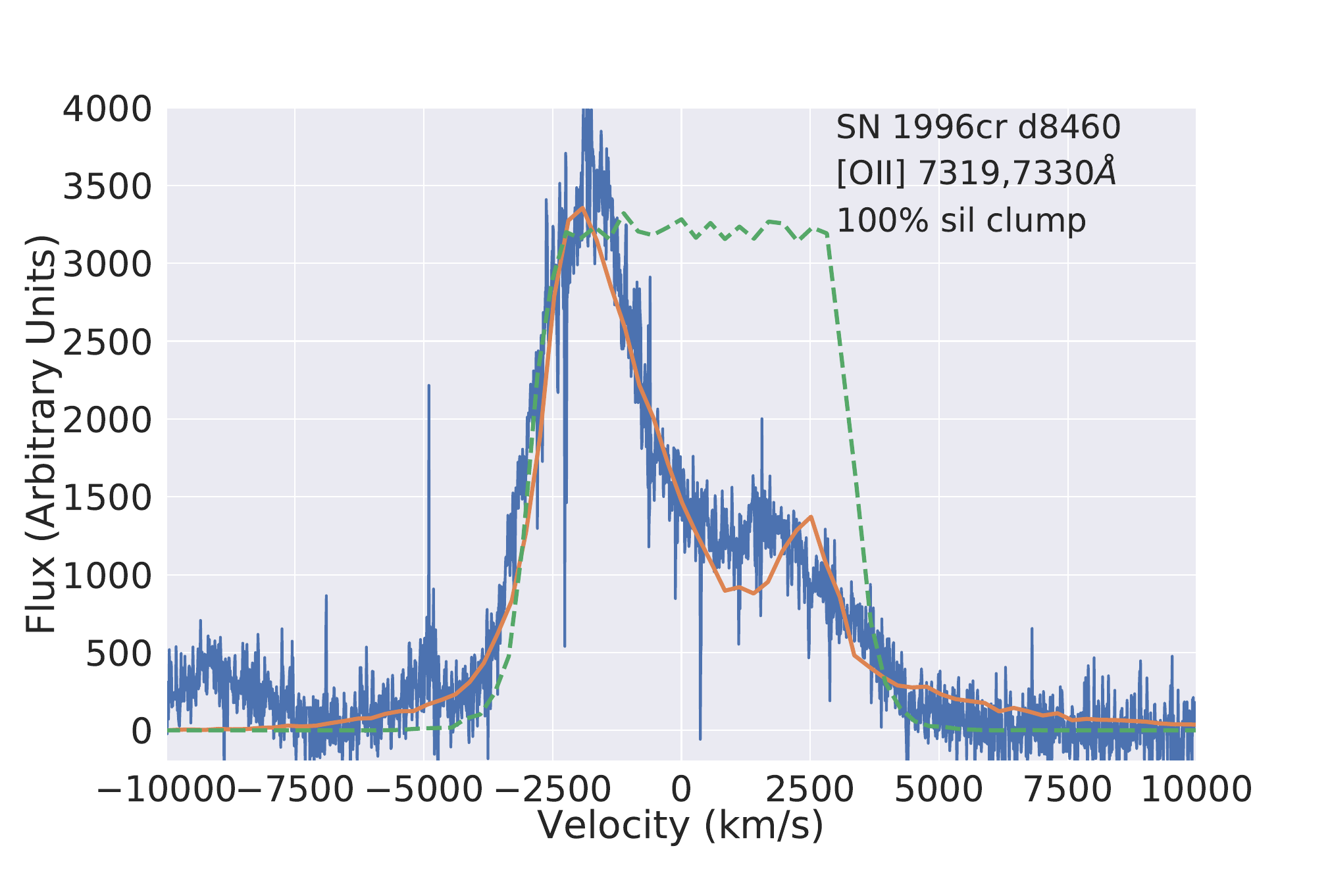}

\caption{{\sc damocles} models of the [O~{\sc ii}] X-Shooter line profile of SN~1996cr 8460 days post-explosion. The green line is the dust-free {\sc damocles} model and the orange line is the dust-affected model.}
\label{fig:96cr-8400-fits}
\end{figure*}

\begin{figure*}
\centering

\includegraphics[scale=0.8]{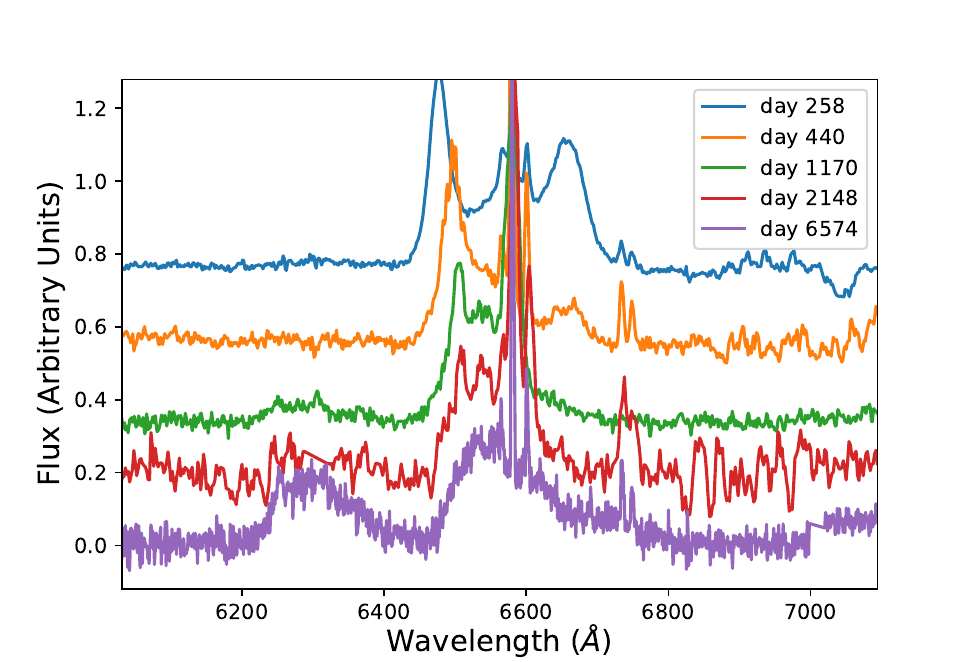}
\caption{The evolution of the red region of the optical spectrum of SN~1998S between days 258 and 6574. Information on the sources of the spectra is summarised in Table~\ref{table:archiv-obs-sum}.  }
\label{fig:1998s-evoln}
\end{figure*}

We merged our day 7360, 7362 and 7389 X-Shooter spectra to form a `day 7370' spectrum. Similarly, we merged our day 8446 and 8474 X-Shooter spectra to form a `day 8460' spectrum.
Our best fitting {\sc damocles} models to the [O~{\sc i}] 6300,6363-\AA , [O~{\sc ii}] 7319,7330-\AA , [O~{\sc iii}] 4959,5007-\AA\ and [S~{\sc iii}] 9069-\AA\ lines at 7370 days are shown in Figure~\ref{fig:96cr-7400-fits}, 
and for the [O~{\sc ii}] doublet 8460 days after explosion in Figure~\ref{fig:96cr-8400-fits}. A recessional velocity of 510 km/s was adopted for the Circinus galaxy. Model parameters for both epochs are listed in Table~\ref{table:2012-3sn-params}. We were able to fit all the lines at day 7370 with similar values for V$_{max}$, $\beta$ and R$_{in}$/R$_{out}$, implying a mixing of ionisation states in the remnant. Since dust absorption has a grain-radius to wavelength dependence, we were able to find a single dust grain radius which required a matching dust mass that fit all the doublets simultaneously. However, due to the lack of a red scattering wing, which if present could have ruled out high AmC dust proportions, we were not able to constrain the dust species. Therefore, we present model fits for smooth and clumped 100~per~cent AmC and 100~per~cent silicate dust, where in both cases we were able to constrain the dust grain radius for a clumped distribution to 0.12$\mu$m and 0.05$\mu$m. Due to the strength of the oxygen lines from the ejecta of SN~1996cr, a high proportion of silicates, from 50-100~per~cent, seems likely.

To try and rigorously constrain the error limits on the grain radius and dust mass, we ran a Bayesian simulation fitting the day 7370 [O~{\sc iii}] and [O~{\sc ii}] lines simultaneously using 100~per~cent silicate dust. The distributions of the [O~{\sc iii}]- and [O~{\sc ii}]-emitting ions were assumed to be co-located, and coupled to the dust. The resulting corner plot can be seen in Figure~A4. Small silicate grains ($<0.1~\mu$m) are seen to be marginally preferable over larger ones. The median grain radius is 0.07~$\mu$m, very close to our initial estimate of 0.05~$\mu$m. The silicate dust mass probability distribution is strongly peaked at 0.47$^{+0.63}_{-0.37}$~M~$_{\odot}$, and the error limits are well defined. For a 50:50 ratio of carbon to silicate dust with a single grain radius of 0.11~$\mu$m, we obtain a dust mass of 0.10$^{+0.13}_{-0.088}$ M$_{\odot}$. The quoted uncertainties on this value are derived from the Bayesian errors, which are taken as percentage errors on the median value and then used for the dust mass derived for the 50:50 carbon to silicate dust mix in order to obtain absolute uncertainties.

We only modelled the [O~{\sc ii}] 7319,7330-\AA\ doublet in the day 8460 spectrum, as all the other lines had a low signal to noise. For a 50:50 ratio of AmC to silicate dust, using a single grain radius of 0.11~$\mu$m, 0.15$_{-0.14}^{+0.30}$ M$_{\odot}$ of dust was required to fit the profile, where error limits were determined by a Bayesian analysis. The formal uncertainties on the dust masses are too high to infer anything about the evolution of the dust mass of SN~1996cr between 2016 and 2019.

\subsubsection{SN 1998S}
SN~1998S was a Type IIn object discovered in NGC~3877 on March 2nd 1998 \citep{li1998}. 
Its optical spectrum has changed extensively over the first 6000 days post-explosion. The spectrum was dominated by a CSM interaction at very early times. 
After 100 days the fairly symmetric line profiles changed drastically, and the H$\alpha$ and He~{\sc i} 10830-\AA\ lines showed a highly asymmetrical triple-peaked structure \citep[][]{Gerardy2000}. 
\citet{Fransson2005} 
attributed the multi-peaked H$\alpha$ line profile to a geometrically thin cool dense shell (CDS) behind the reverse shock, at large optical depths. Between days 258-440 (see Figure~\ref{fig:1998s-evoln}) this multi-peaked structure can be seen to exhibit a pronounced red-blue asymmetry which could be attributable to newly formed dust.


\citet{Mauerhan2012} noted that after ten years the optical spectrum had become more dominated by ejecta emission, with the oxygen forbidden lines brightening relative to H$\alpha$ (Fig.~\ref{fig:1998s-evoln}), although the high H$\alpha$ luminosity indicated that SN~1998S was still interacting with dense circumstellar material.

\begin{figure*}
\centering
\includegraphics[width=0.9\linewidth]{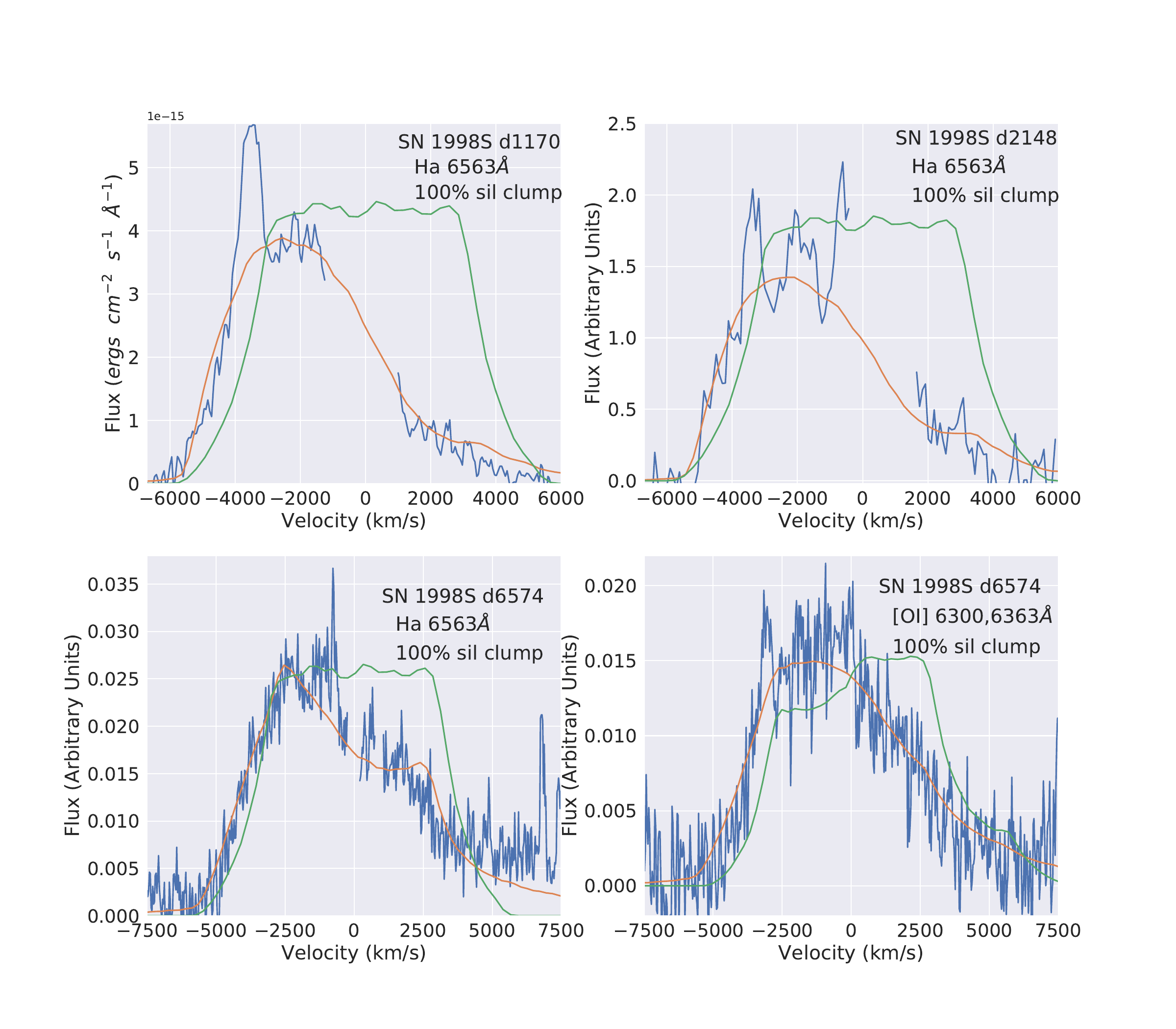}
\caption{The orange lines represent the best fitting clumped silicate dust-affected models for the H$\alpha$ profiles of SN~1998S at three epochs and the [O~{\sc ii}]~6300,6363-$\AA$ lines at day 6574. Parameters of the best fitting models can be found in Table \ref{table:2012-3sn-params}. The dust-free model profiles are shown by the green lines.}

\label{fig:1998s-fits}
\end{figure*}

There is potential evidence for dust formation in SN~1998S. NIR photometry and spectra between days 305–1242 (\citet{Gerardy2000} and \citet{Pozzo2004}) showed CO emission and a significant IR excess in the spectral energy distribution, although it was unclear how much this could be due to IR echoes reflecting off pre-existing dust clouds in the outer CSM \citep{Sugerman2012}. However, \citet{Mauerhan2012} also noted blue-shifted peaks to the H$\alpha$ and oxygen line profiles and the disappearance of a red-shifted peak in the H$\alpha$ spectral line between days 1093 and 2900, which could indicate dust formation. It is unclear as to what quantities of dust were being formed in the ejecta or in a CDS around SN~1998S.

We model the H$\alpha$ line from archival spectra at days 1170, 2148 and 6574 days past explosion; [O~{\sc i}] 6300,6363~$\AA$ is bright enough at day 6574 to model as well. The sources of these published archival data are listed in Table~\ref{table:archiv-obs-sum}. The spectra were corrected for a recessional velocity of NGC~3877 of 847~km~s$^{-1}$ \citep{Fassia2000}.
The day 1170 and 2148 H$\alpha$ lines show a sharp blue-shifted emission spike around -4000~km~s$^{-1}$, in the same location as that seen on day 440. We do not try to fit this feature in our models, as it is likely a feature leftover from the CSM-ejecta collision. The spectra at these epochs have fairly low resolution, so much of the profile is also taken up by narrow H$\alpha$ emission features. We try and fit our ejecta models to the underlying shape similar to that seen at day 6574 in the more ejecta-dominated era, where the narrower peaks have disappeared. 

At day 6574 the H$\alpha$, and to a lesser extent the [O~{\sc i}] profile, exhibited a red scattering wing, and we found that a 100~per~cent silicate dust model improved the $\chi^2$ value by a factor of 1.5 compared to the best fitting 100~per~cent AmC dust model. Hence we only show models consisting of a 100~per~cent silicate dust distribution for all epochs in Fig~\ref{fig:1998s-fits}, with parameters listed in Table~\ref{table:2012-3sn-params}. We note, however, that it is possible for the earlier epochs to be tracing dust in a CDS region. We cannot determine the degree of similarity of the underlying H$\alpha$ profile between earlier and later epochs due to the earlier line profiles being affected by narrow components between -2000 and 1000~km~s$^{-1}$, but it is clear that a red scattering wing is much more evident at day 6574 than at days 1170 and 2148. Due to S/N limitations, for the day 1170 and 2148 line profiles the $\chi^2$ value is the same for a 100~per~cent carbon or 100~per~cent silicate dust species. 

\begin{figure*}
\centering

\includegraphics[width=\linewidth]{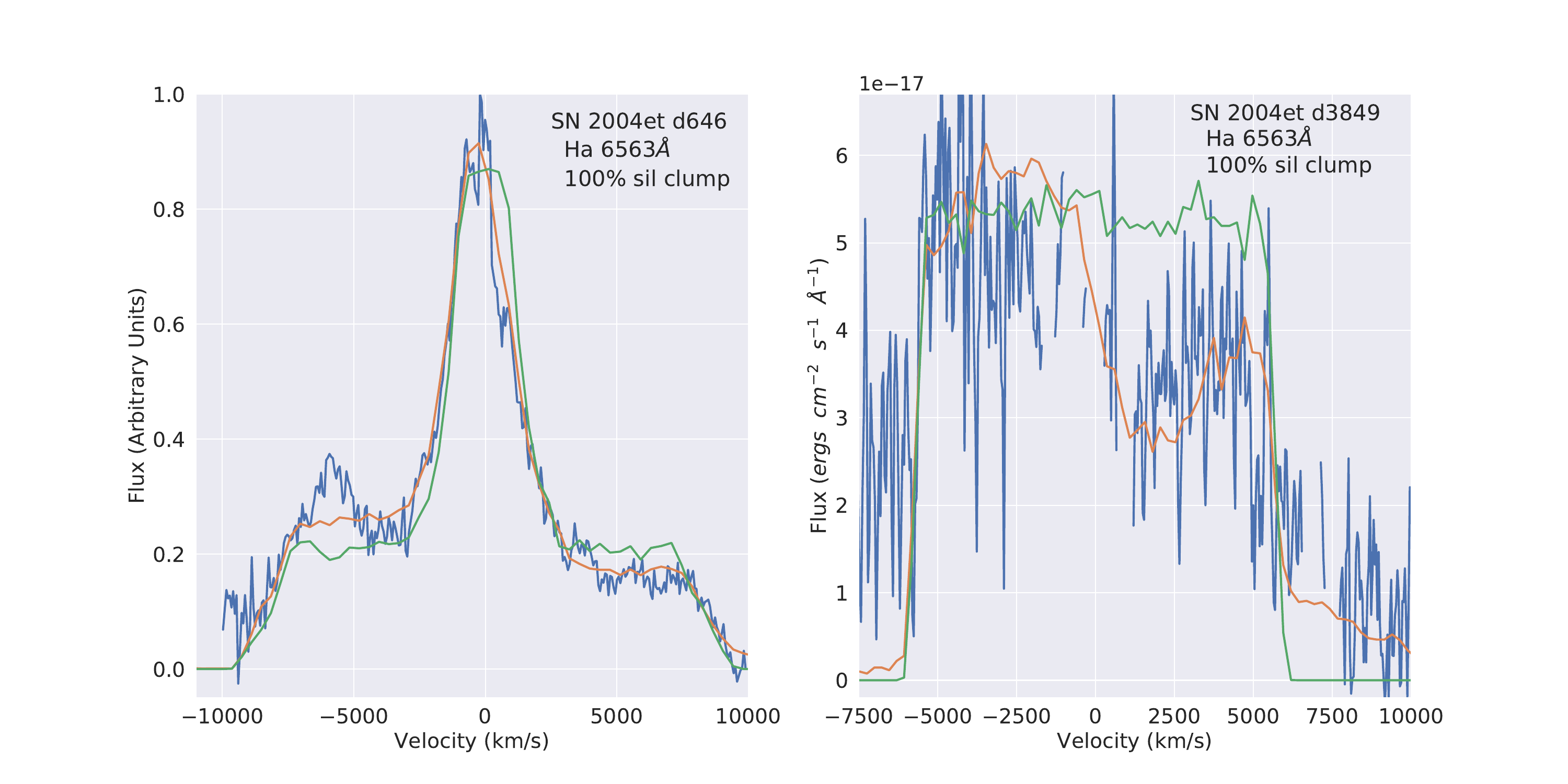}

\caption{{\sc damocles} models of the H$\alpha$ line profiles of SN~2004et at 646 and 3849 days post-explosion. The green line is the dust-free {\sc damocles} model, and the orange line is the dust-affected model. For day 646, the orange line consists of a broad and intermediate width component summed together. Clumped dust models for 100~per~cent astronomical silicates and 100~per~cent AmC are shown.}
\label{fig:2004et-ha-fits}
\end{figure*}

We cannot determine whether SN~1998S's dust mass grew between days 1170 and 2148, as the dust required to attenuate the line profiles is optically thick at both epochs (see $\tau$ values in Table \ref{table:2012-3sn-params}), and the line profiles have significant "gaps" as discussed above. Due to this, our clumped models use a clump mass fraction of 0.85, with the rest of the dust being smoothly distributed. The optical depth of a single dust clump at day 1170 was calculated to be 28. We find a dust mass of roughly 2.5$^{+4.10}_{-1.60}\times10^{-2}$~M$_{\odot}$ at day 1170 and 4.0$^{+6.00}_{-3.30}\times10^{-2}$~M$_{\odot}$ at day 2148, with uncertainties at day 2148 given by a 35~per~cent variation about the best fitting $\chi^2$ value, and those at day 1170 taken from the 1D posterior distribution of the dust mass found from a Bayesian model of the H$\alpha$ line for a 100~per~cent silicate dust species.

At day 6574 we found that the H$\alpha$ and [O~{\sc i}] lines could be well fit with matching parameters, implying a co-location of the emitting species. We found that a grain radius of 0.08~$\mu$m for clumped silicate dust fitted both profiles best, and this grain radius and dust species was fixed for the other two epochs. It was close to the median grain radius of 0.20~$\mu$m found from a Bayesian analysis run with 100~per~cent silicate dust for H$\alpha$ and [O~{\sc i}] simultaneously.
We can fit both lines with a clumped silicate dust mass of around 0.15$^{+0.54}_{-0.14}$~M$_{\odot}$, with the uncertainties also taken from the Bayesian model referenced previously. This represents a significant increase in dust mass from the earlier epochs (a factor of ten).

\subsubsection{SN 2004et}
SN 2004et was a Type II-P supernova discovered on 2004 September 22 \citep{zwitter2004} in NGC 6946, the same galaxy that hosts SN~1980K.
We adopt the discovery date as the explosion date. 
The optical spectrum was monitored from day 8-451 by \citet{Sahu2006}, who noted that the emission peaks became blue-shifted at around day 300, suggesting early dust-formation.

\citet{Kotak2009} and \citet{Fabbri2011a} continued to monitor the late-time optical spectra, where the H$\alpha$ line became box-like around 2 years past explosion, perhaps indicative of an ejecta collision with a CSM, forming a cool dense shell (CDS). \citet{Kotak2009} presented {\em Spitzer} observations between days 64-1406, where they inferred from modelling the SED that $\sim10^{-4}$~M$_\odot$ of newly formed radioactively heated silicate dust was present in the ejecta. \citet{Fabbri2011a} also modelled the optical and mid-IR SED of SN 2004et, and found dust masses, made from 80~per~cent silicate and 20~per~cent AmC, that increased from $\sim10^{-4}$~M$_{\odot}$ at day 300 to $\sim10^{-3}$~M$_{\odot}$ at day 690.

We have modelled the H$\alpha$ line profile of SN~2004et at days 646 and 3686 post-explosion (information on the spectra can be found in Tables~\ref{table:gmos-dat} and \ref{table:archiv-obs-sum}.) The spectra were corrected for a recessional velocity of NGC~6946 of 40~km~s$^{-1}$. As both the work of \citet{Kotak2009} and \citet{Fabbri2011a} found that a high proportion of silicates was required to fit the IR SED of SN 2004et, we only use 100~per~cent silicate dust grains for our {\sc damocles} models. From a visual inspection, the H$\alpha$ line appears very different between the two epochs. The spectrum at day 646 also has a clear IWC not present at day 3686, possibly due to a cool dense shell (CDS) formed by ejecta-CSM interaction, or else due to flash-ionized CSM gas, as well as a broad component (BC) representing the fast-expanding ejecta.

We assumed a simple model setup where the ejecta emits the BC, and a separate component emits the IWC, where R$_{in,IWC}$ = R$_{out,BC}$.
The dust required for a good fit to the day 646 H$\alpha$ BC, around $5\times10^{-4}$ M$_{\odot}$, barely introduced an asymmetry in the IWC profile and we found that a separate dust mass component of around $\sim2\times10^{-3}$~M$_{\odot}$ was required to fit the IWC. 
Parameters for our models can be found in Table~\ref{table:1986e-params}, with the line profile fits shown in Figure~\ref{fig:2004et-ha-fits}.

\begin{figure*}
\centering

\includegraphics[width=\linewidth]{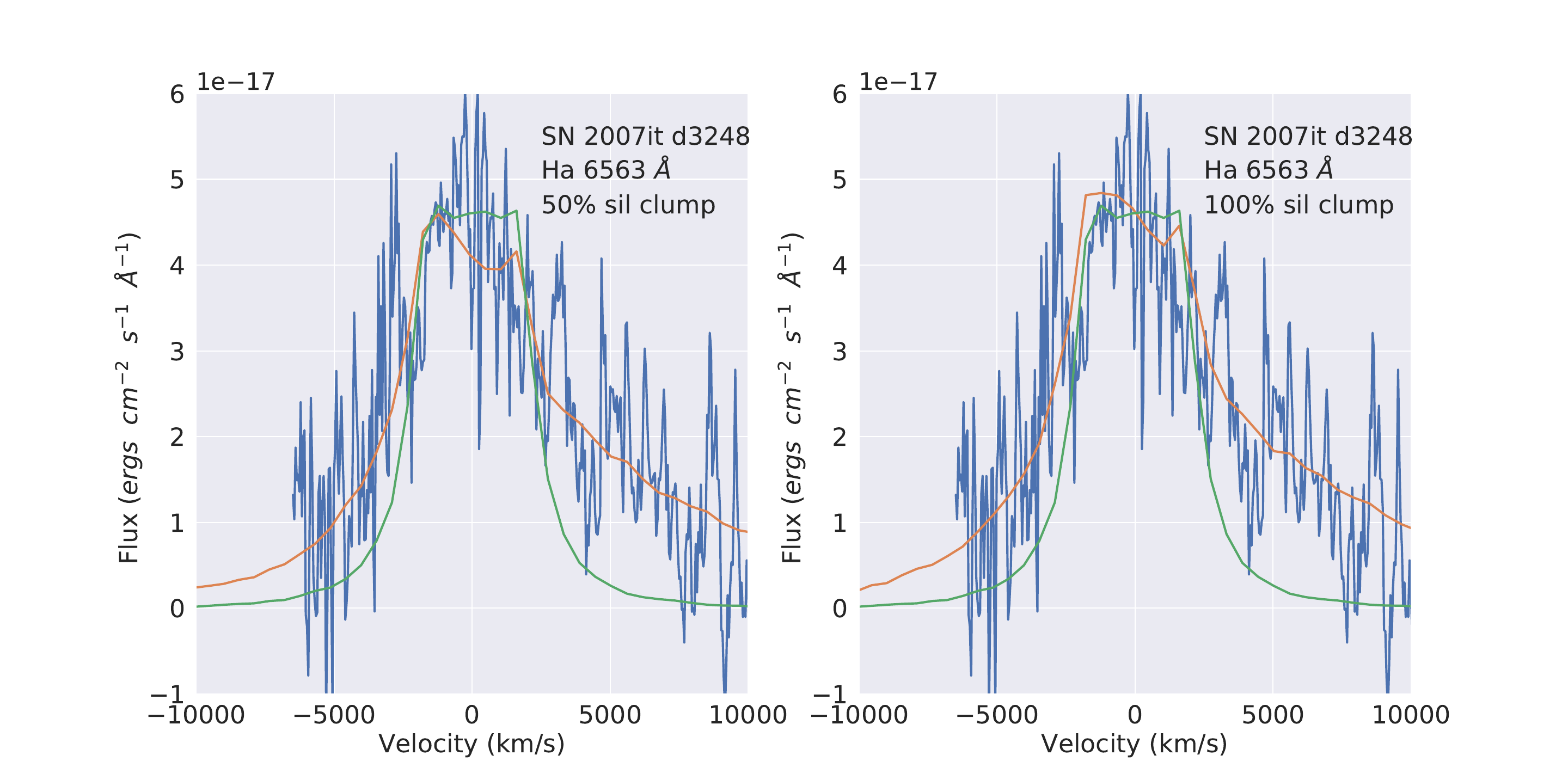}

\caption{{\sc damocles} models SN~2007it's H$\alpha$ line profiles 3248 days post-explosion. The green line is the dust-free {\sc damocles} model, and the orange line is the dust-affected model. Clumped dust models of 100~per~cent astronomical silicates (left) and for a 50:50 astronomical silicate to AmC ratio (right) are shown. The gap blueward of -6000~km~s$^{-1}$ is where contaminating [O~{\sc i}] emission has been removed.}
\label{fig:2007it-ha-fits}
\end{figure*} 

Due to the presence of a scattering wing in the day 646 intermediate width component, we were able to constrain the silicate grain radius to around 0.2~$\mu$m. 
From a Bayesian analysis of the BC component (Figure~A5), we were able to quantify the uncertainties on the dust mass that had formed in the ejecta, the lower and upper limits being 4$\times10^{-5}$~M$_\odot$ and 2.9$\times10^{-3}$~M$_\odot$,
but we were not able to constrain the grain radius due to uncertainties in the continuum level. We have therefore adopted the same grain radius for the BC dust component as was required to fit the IWC. The dust mass derived for this grain radius functions roughly as a lower limit, as smaller or larger silicate grains require more dust to produce the same optical depth.

By day 3849, the H$\alpha$ line-emission from the IWC had faded, and the H$\alpha$ emission is dominated by the ejecta. The signal to noise in this spectrum is low
(Fig~\ref{fig:2004et-ha-fits}), but there does appear to be a red scattering wing in the profile, which limits the silicate dust grains to $>$0.1~$\mu$m in radius, and rules out larger grain radii $>$0.5$\mu$m. A grain radius of 0.2~$\mu$m and a dust mass of around 0.01~M$_{\odot}$ gave a $\chi^2$ value of 1.2 when comparing the observed spectra and the model, with a lower limit of $2\times10^{-4}$~M$_{\odot}$ and an upper limit of 0.9~M$_{\odot}$ taken from a varying the best-fitting $\chi^2$ of 35~per~cent when all other parameters were fixed. This indicates a likely growth by a factor of 20 of the dust mass in the ejecta of SN 2004et between days 646 and 3849.

\subsubsection{SN 2007it}
SN~2007it was discovered in NGC~5530 by R. Evans on 13/09/2007 at V=13.5 mag \citep{evans2007, Itagaki2007}. Pre-discovery images taken with the All Sky Automated Survey (ASAS-3) constrained the explosion
date to be between September 4 and 6 2007 \citep{Pojmanski2007}. We adopt an explosion date of 5/09/2007. 
\citet{Andrews2011} studied SN~2007it in the optical and IR regime between 10-944 days after explosion. They classified it as a Type~II-P supernova based on the plateau in its optical lightcurve between days 20-107. 

At day~540 the optical lines of SN 2007it showed an intermediate width component in the [O~{\sc i}] and H$\alpha$ profiles. \citet{Andrews2011} also analysed the IR SEDs at four epochs between days 351-944. 
Day 351 showed the presence of an IR excess, well before the estimated time of new dust formation from the optical lightcurves, which they attributed to scattering dust in a CSM producing a light echo.
They also found that the IR flux doubled in intensity at 5$\mu$m between days 351-561, and at the same time saw a dimming in the visible light curve, which they attributed to dust formation in the ejecta. They modelled the 0.1-10-$\mu$m SED with {\sc Mocassin} between days 351 and 944 and found that up to $1\times10^{-4}$~M$_\odot$ of dust had formed in the ejecta by day~944.

We combined our day 3136 X-Shooter spectrum with our day 2780 and 3830 GMOS spectra, as all 3 spectra initially had fairly low S/N ratios and overplotting the 3 spectra did not show significant differences in line profile shape over time. The spectra were corrected for a recessional velocity of NGC~5530 of 1170~km~s$^{-1}$. {\sc damocles} models were made for `day 3248' post-explosion, the average of the three epochs. The presence of a red scattering wing implies a silicate dust proportion larger than 50~per~cent. The best fitting models for a 50:50 AmC to silicate dust mixture and for 100~per~cent silicate dust are plotted in Figure~\ref{fig:2007it-ha-fits}, with model parameters listed in Table~\ref{table:2012-3sn-params}. A model using 50:50 AmC to silicate dust species with a grain radius of 0.3~$\mu$m and a dust mass of $\sim0.008_{-0.007}^{+0.05}$M$_{\odot}$, minimized the $\chi^2$ value. The uncertainties on this value are taken from a Bayesian model of the H$\alpha$ line of SN 2007it. 
It should be noted that it is difficult to accurately delineate the shape of the H$\alpha$ profile, due to the low S/N of the combined spectra and the fact that the blue edge of the H$\alpha$ profile is blended with the [O~{\sc i}]~6300,6363-\AA\ profile. Therefore, there is a large degree of uncertainty in the values of the model parameters.

\subsubsection{SN 2010jl}
SN 2010jl was discovered in UGC 5189A on November 3rd 2010. \citet{Stoll2011} established an explosion date for SN~2010jl of 09/10/10, from its detection in pre-discovery images, and noted that SN~2010jl was very luminous with a peak absolute magnitude of $\sim-20$. It was classified as a Type~IIn near maximum light, from the presence of narrow lines emitted by flash-ionized CSM 
\citep{Benetti2010, Yamanaka2010}.
It has been well studied at optical and infrared wavelengths, with many authors estimating dust masses present in SN~2010jl. Several authors have modelled the optical and NIR line profiles first noted by \citet{Smith2011} after 30 days, and others have modelled the IR excess, also  commencing 30 days after maximum light. \citet{Andrews2011a, Fransson2014, Sarangi2018} and \citet{bevan2020} all concluded that a light echo from pre-existing CSM dust was responsible for the IR excess before ~day 400, while \citet{bevan2020} found that ejecta/CSM dust dominated the IR SED after this time. Estimates of the newly formed dust mass, likely present in the CDS/ejecta, range from $\sim(2.5-9.0) \times 10^{-4}$~M$_{\odot}$ at day~500, to ($1.5-3.0) \times 10^{-3}$~M$_{\odot}$ at day~800 \citep{Maeda2013, Gall2014, Chugai2018, Sarangi2018, bevan2020}.

\begin{figure*}
\centering

\includegraphics[scale=0.5]{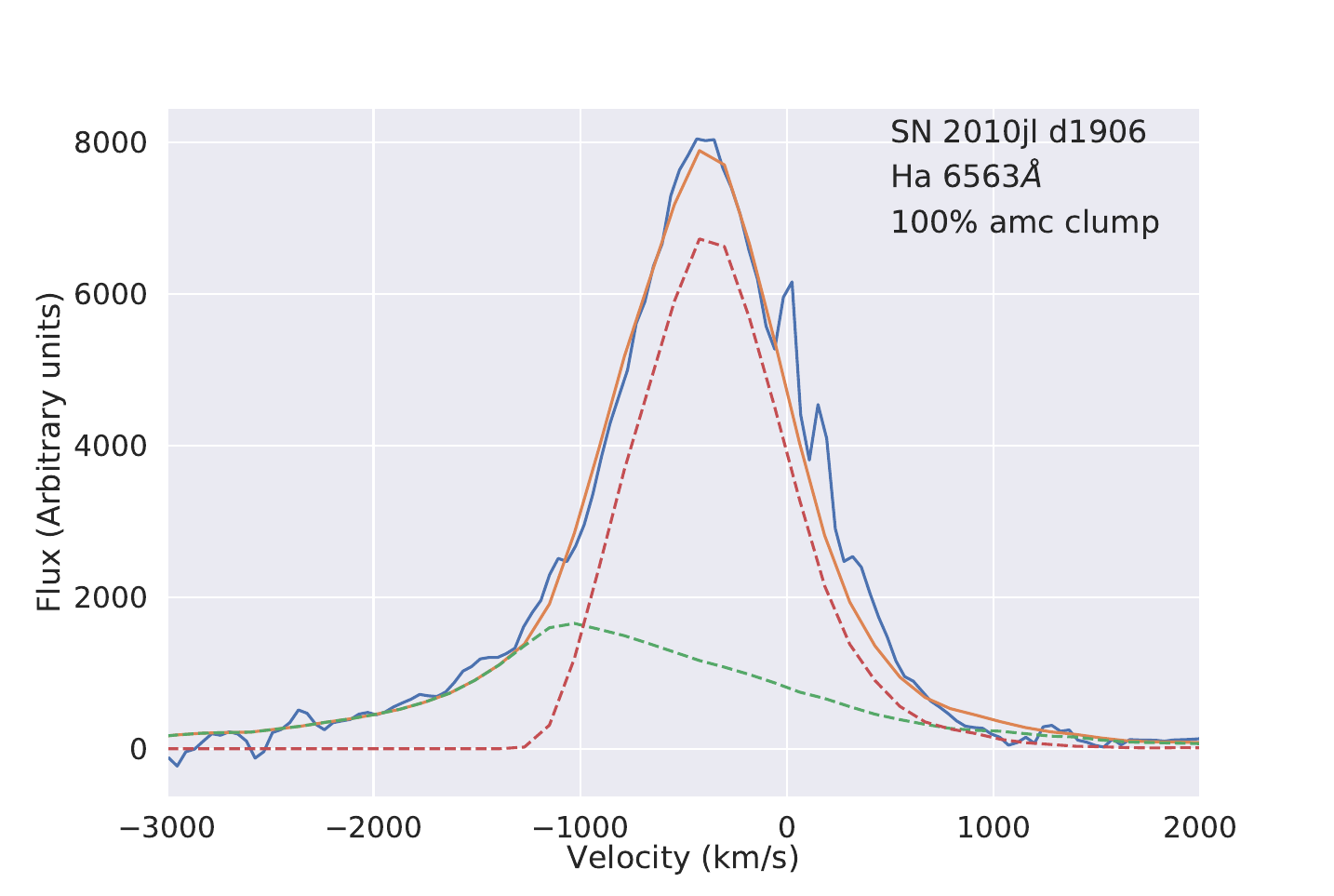}

\caption{{\sc damocles} models of SN~2010jl's  H$\alpha$ profile in the Gemini GMOS spectrum taken 1906 days post-explosion. The red and green dashed lines are the dust-affected {\sc damocles} models of the Intermediate Width and Broad Width components, and the orange line is the sum of these two components. The clumped dust models are for 100~per~cent AmC dust.}
\label{fig:2010jl-fits}
\end{figure*}

\begin{figure*}
\centering

\includegraphics[width=0.9\linewidth]{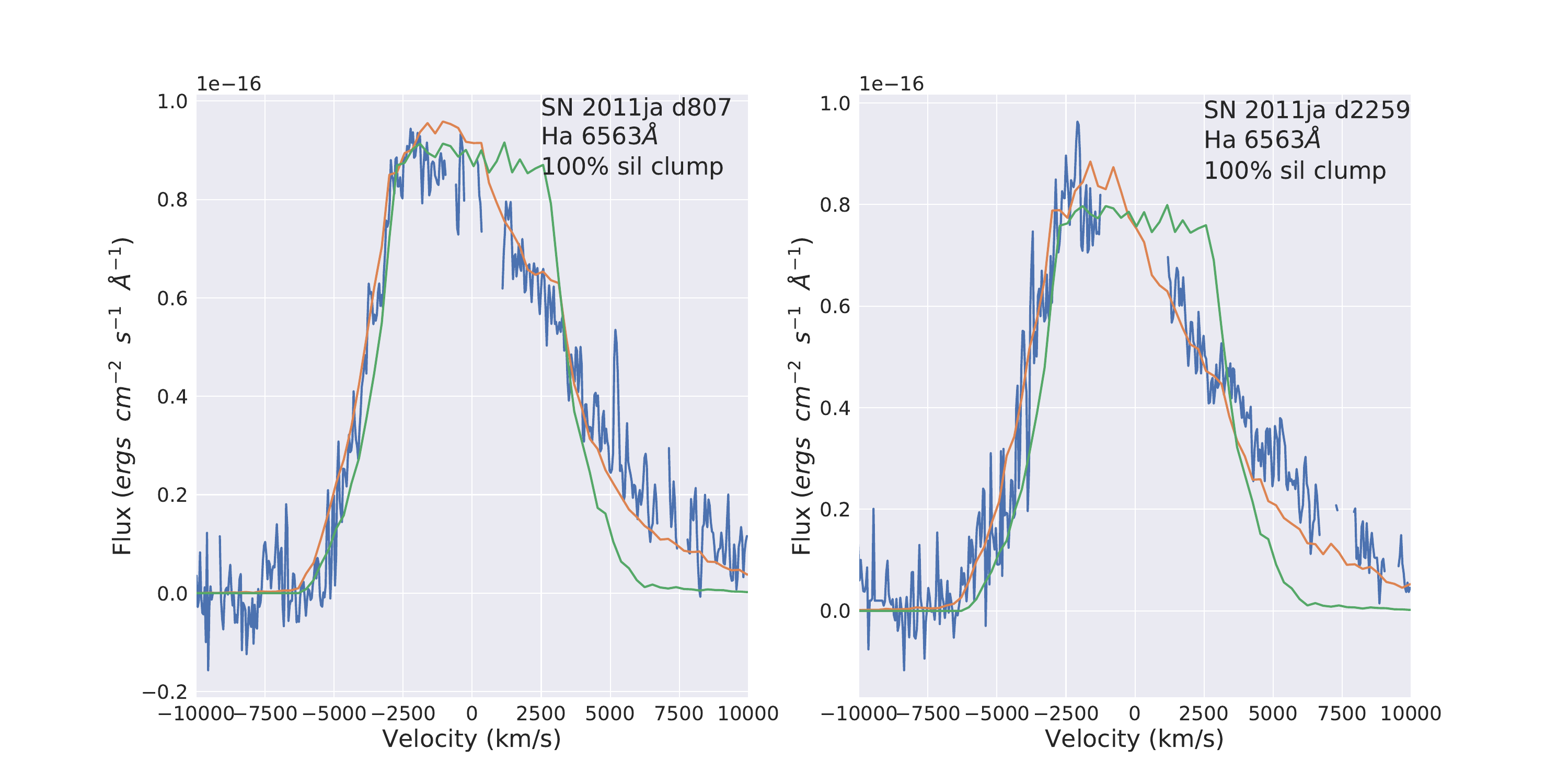}

\caption{{\sc damocles} models of SN~2011ja's H$\alpha$ profile (orange line) in Gemini GMOS spectra (blue line) taken 807 days (left) and 2259 days (right) post-explosion. The green line is the dust free model. Clumped dust models of 100~per~cent silicate dust are shown.}
\label{fig:2011ja-fits}
\end{figure*}

\begin{table}
\centering
\caption{
Parameters used in the {\sc damocles} models of the  H$\alpha$ line of SN~2010jl 1906 days past explosion for the best-fitting clumped AmC dust model. R$_{out}$ and R$_{in}$ are the outer and inner radii of the post-shock region.
The optical depth is calculated from R$_{in}$ to R$_{out}$ at a wavelength of 6563~$\AA$. $\alpha$ is the radius-independent
gradient of the velocity distribution for the IWC, and $\beta$ is the gradient of the velocity distribution for the BC.  } 
\centering
\begin{tabular}{cc}
\hline
V$_{max,IWC}$ (km~s$^{-1}$) & 1150  \\
V$_{min,IWC}$ (km~s$^{-1}$)& 370 \\
V$_{max,BC}$ (km~s$^{-1}$) & 15000\\
V$_{min,BC}$ (km~s$^{-1}$) & 1150\\
$\alpha$ & 0.8\\
$\beta$ & -2.5\\
a ($\mu$m) & 0.15 \\
Mdust ($10^{-2}$~M$_{\odot}$) & 1.8 \\
$\tau$ & 12.6 \\
R$_{out}$ (10$^{15}$ cm)& 230 \\
R$_{in}$ (10$^{15}$ cm)& 26.4 \\

\hline
\end{tabular}
\label{table:2010jl-params}
\end{table}

We obtained a good S/N detection of the H$\alpha$ emission profile in our day 1906 Gemini GMOS spectrum. Narrow nebular lines of H$\alpha$ and [N~{\sc ii}] 6584,6548~$\AA$ were removed using the {\sc dipso} ELF routine.
The spectrum was corrected for a recessional velocity of UGC~5189A of 3167~km~s$^{-1}$. 
We extrapolated the \citet{bevan2020} day 1286 H$\alpha$ model parameters for SN~2010jl to model our day 1906 H$\alpha$ profile. We summarise their model below, which is described in more detail in Section 4.2 of their paper. Their model consists of 3 components: a smooth slowly expanding CSM gas region emitting an IWC, a rapidly expanding smooth ejecta gas component producing a BC, and a clumped dust component. The bounds of the model were set by R$_{out}$, representing the location of the forward shock, and R$_{in}$, the location of the reverse shock. Instead of assuming free expansion, the velocity of emitting packets in the CSM were sampled from a power law velocity distribution p(v) $\propto$ v$^{\alpha}$, between the limits V$_{max,IWC}$ and V$_{min,IWC}$.  The BC used a radius-independent power-law velocity distribution, p(v) $\propto$ v$^{\beta}$, between the limits V$_{max,BC}$ and V$_{min,BC}$ where for continuity V$_{min,BC}$=V$_{max,IWC}$.
A gas emissivity law $\rho$ $\propto$ r$^{-4}$ was used for both the ejecta and CSM components. The AmC dust was clumped with a filling factor of 0.1.

\begin{table*}
\centering
\caption{
Parameters used in the {\sc damocles} models of the broad emission lines of SN~1996cr, SN~1998S, SN~2007it, SN~2011ja and SN~2012au for spherically symmetric smooth and clumped dust models. The "\% Sil" column stands for percentage of the dust species that is astronomical silicate, while the remainder of the dust is AmC dust. The optical depth is calculated from R$_{in}$ to R$_{out}$ at the central line wavelengths ([O~{\sc iii}]=5007~$\AA$, [O~{\sc ii}]=7319~$\AA$, [O~{\sc i}]=6300~$\AA$, Ha=6563~$\AA$, [S~{\sc iii}]=9069-$\AA$)).} 
\centering
\begin{tabular}{cccccccccccccc}
\hline
SN & Epoch & Line & Clumped? & \% Sil & a & V$_{max}$ & V$_{min}$ & $\beta_{gas}$ & R$_{out}$ & R$_{in}$ & M$_{dust}$ & $\tau$  & $\chi^{2}$ \\
 & days & & & & $\mu$m & km~s$^{-1}$ & km~s$^{-1}$ & & 10$^{15}$~cm & 10$^{15}$~cm & $10^{-2}$~M$_\odot$ & & \\
\hline
1996cr & 7370 & [O~{\sc iii}] & no  & 0   & 0.15 & 7000 & 2940 & 6.0   & 442.8 & 186.0 & 0.80   & 1.46  \\
1996cr & 7370 & [O~{\sc iii}] & yes & 0   & 0.12 & 7000 & 2940 & 6.0   & 442.8 & 186.0 & 3.5   & 5.18   \\
1996cr & 7370 & [O~{\sc iii}] & no  & 100 & 0.04 & 7000 & 2940 & 6.0   & 442.8 & 186.0 & 20  & 0.89   \\
1996cr & 7370 & [O~{\sc iii}] & yes & 100 & 0.05 & 7000 & 2940 & 6.0   & 442.8 & 186.0 & 120 & 7.97  \\
1996cr & 7370 & [O~{\sc ii}]  & no  & 0   & 0.15 & 6900 & 2898 & 6.0   & 436.5 & 183.3 & 0.80   & 0.97  \\
1996cr & 7370 & [O~{\sc ii}]  & yes & 0   & 0.12 & 6900 & 2898 & 6.0   & 436.5 & 183.3 & 3.5   & 5.33   \\
1996cr & 7370 & [O~{\sc ii}]  & no  & 100 & 0.04 & 6900 & 2898 & 6.0   & 436.5 & 183.3 & 40  & 0.66   \\
1996cr & 7370 & [O~{\sc ii}]  & yes & 100 & 0.05 & 6900 & 2898 & 6.0   & 436.5 & 183.3 & 120 & 2.89  \\
1996cr & 7370 & [O~{\sc i}] & no  & 0   & 0.15 & 6900 & 2760 & 5.6 & 436.5 & 174.6 & 0.80   & 1.13  \\
1996cr & 7370 & [O~{\sc i}]  & yes & 0   & 0.12 & 6900 & 2760 & 5.6 & 436.5 & 174.6 & 3.5   & 5.52  \\
1996cr & 7370 & [O~{\sc i}] & no  & 100 & 0.04 & 6900 & 2760 & 5.6 & 436.5 & 174.6 & 20    & 0.99 \\
1996cr & 7370 & [O~{\sc i}] & yes & 100 & 0.05 & 6900 & 2760 & 5.6 & 436.5 & 174.6 & 120   & 4.58 \\
1996cr & 7370 & [S~{\sc iii}] & no  & 0   & 0.15 & 7000 & 2940 & 6.0   & 442.8 & 186.0 & 0.80   & 1.03 \\
1996cr & 7370 & [S~{\sc iii}] & yes & 0   & 0.12 & 7000 & 2940 & 6.0   & 442.8 & 186.0 & 3.5   & 4.21  \\
1996cr & 7370 & [S~{\sc iii}] & no  & 100 & 0.04 & 7000 & 2940 & 6.0   & 442.8 & 186.0 & 40    & 0.50  \\
1996cr & 7370 & [S~{\sc iii}] & yes & 100 & 0.05 & 7000 & 2940 & 6.0   & 442.8 & 186.0 & 120   & 3.10  \\
1996cr & 8460 & [O~{\sc ii}] & no  & 0   & 0.15 & 6600 & 2640 & 5.5 & 482.48   & 192.97   & 1.0     & 1.10    \\
1996cr & 8460 & [O~{\sc ii}] & yes & 0   & 0.15 & 6600 & 2640 & 5.5 & 482.48   & 192.97   & 3.5   & 2.80   \\
1996cr & 8460 & [O~{\sc ii}] & no  & 100 & 0.06 & 6800 & 2380 & 5.0   & 497.04   & 173.96   & 50    & 1.30   \\
1996cr & 8460 & [O~{\sc ii}] & yes & 100 & 0.06 & 6800 & 2380 & 5.0   & 497.04   & 173.96   & 80    & 2.40    \\
&&&&&&&&&&&& \\
1998S & 1170 & H$\alpha$       & no  & 100 & 0.08 & 5500 & 3410 & 0.5 & 55.60  & 34.47  & 1.00  & 2.45  & 1.26 \\
1998S & 1170 & H$\alpha$       & yes & 100 & 0.08 & 5500 & 3410 & 0.5 & 55.60  & 34.47  & 2.50  & 10.6 & 1.28 \\
1998S & 2148 & H$\alpha$       & no & 100 & 0.08 & 5300 & 3339 & 0.5 & 97.63  & 61.51  & 1.00  & 1.51  & 1.91 \\
1998S & 2148 & H$\alpha$       & yes & 100 & 0.08 & 5300 & 3339 & 0.5 & 97.63  & 61.51  & 4.00  & 5.44  & 1.88 \\
1998S & 6574 & H$\alpha$       & no  & 100 & 0.08 & 5600 & 2688 & 2.3 & 260.0 & 124.0 & 14.0 & 2.32  & 1.37 \\
1998S & 6574 & H$\alpha$       & yes & 100 & 0.10  & 5600 & 2688 & 2.3 & 260.0 & 124.0 & 15.0 & 4.99  & 1.29 \\
1998S & 6574 & [O~{\sc i}] & no  & 100 & 0.08 & 5600 & 2688 & 2.3 & 260.0 & 124.0 & 14.0 & 2.47  & 1.24 \\
1998S & 6574 & [O~{\sc i}] & yes & 100 & 0.10  & 5600 & 2688 & 2.3 & 260.0 & 124.0 & 15.0 & 4.46  & 1.24\\
&&&&&&&&&&&& \\
2007it & 3248 & H$\alpha$  & no  & 100 & 0.40 & 14000 & 6860 & 2.4  & 39.29  & 5.11  & 0.700   & 0.36 \\
2007it  & 3248 & H$\alpha$  & yes & 100 & 0.40 & 14000 & 6860 & 2.4  & 39.29  & 5.11  & 0.900   & 0.46 \\
2007it & 3248 & H$\alpha$   & no  & 50  & 0.25 & 14000 & 6860 & 2.4  & 39.29  & 5.11  & 0.600   & 0.56 \\
2007it & 3248 & H$\alpha$   & yes & 50  & 0.30 & 14000 & 6860 & 2.7  & 39.29  & 5.11  & 0.800   & 0.61 \\
&&&&&&&&&&&& \\
2011ja  & 807 & H$\alpha$  & no  & 100 & 0.15 & 6200  & 3038 & 2.00 & 42.91  & 21.02 & 0.021 & 0.72 \\
2011ja &  807 & H$\alpha$  & yes & 100 & 0.15 & 6200  & 3038 & 2.00 & 42.91  & 21.02 & 0.023 & 0.96 \\
2011ja  & 1275  & H$\alpha$ & no  & 100 & 0.15 & 6100  & 2928 & 2.00 & 66.88  & 32.10 & 0.060 & 0.81 \\
2011ja & 1275  & H$\alpha$  & yes & 100 & 0.15 & 6100  & 2928 & 2.00 & 66.88  & 32.10 & 0.060 & 0.81 \\
2011ja & 1611  & H$\alpha$  & no  & 100 & 0.15 & 6100  & 2745 & 2.00 & 84.91  & 38.21 & 0.070 & 0.52 \\
2011ja & 1611  & H$\alpha$  & yes & 100 & 0.15 & 6300  & 2835 & 2.00 & 87.69  & 39.46 & 0.080 & 0.70 \\
2011ja & 2259  & H$\alpha$  & no  & 100 & 0.15 & 6100  & 2928 & 2.00 & 113.5 & 54.47 & 0.300 & 1.51 \\
2011ja & 2259  & H$\alpha$  & yes & 100 & 0.20  & 6100  & 2928 & 2.00 & 113.5 & 54.47 & 0.250 & 1.40 \\
&&&&&&&&&&&& \\
2012au & 2277 & [O~{\sc iii}]  & no  & 0   & 0.20 & 4900  & 1274 & 3.10 & 96.40  & 25.06 & 0.025 & 0.57 \\
2012au & 2277 & [O~{\sc iii}] & yes & 0   & 0.35 & 4900  & 1274 & 3.10 & 96.40  & 25.06 & 0.040 & 0.49 \\
2012au  & 2277 & [O~{\sc iii}] & no  & 100 & 1.50 & 4700  & 1222 & 3.10 & 92.46  & 24.04 & 0.500 & 0.66 \\
2012au & 2277 & [O~{\sc iii}] & yes & 100 & 1.00 & 4900  & 1274 & 3.30 & 96.40  & 25.06 & 1.200 & 1.03 \\
2012au & 2277 & [O~{\sc ii}]  & no  & 0   & 0.20 & 5000  & 1150 & 3.10 & 98.37  & 22.62 & 0.025 & 0.48 \\
2012au  & 2277 & [O~{\sc ii}] & yes & 0   & 0.35 & 5000  & 1150 & 3.10 & 98.37  & 22.62 & 0.040 & 0.41 \\
2012au  & 2277 & [O~{\sc ii}] & no  & 100 & 1.50 & 5000  & 1150 & 3.30 & 98.37  & 22.62 & 0.400 & 0.48 \\
2012au & 2277 & [O~{\sc ii}] & yes & 100 & 1.10 & 5000  & 1150 & 3.30 & 98.37  & 22.62 & 0.500 & 0.75 \\
2012au & 2277 & [O~{\sc i}]   & no  & 0   & 0.20 & 5100  & 1224 & 3.50 & 100.3 & 24.08 & 0.023 & 0.51 \\
2012au & 2277 & [O~{\sc i}] & yes & 0   & 0.35 & 5100  & 1224 & 3.50 & 100.3 & 24.08 & 0.046 & 0.53 \\
2012au & 2277 & [O~{\sc i}] & no  & 100 & 1.50 & 5100  & 1224 & 3.50 & 100.3 & 24.08 & 0.450 & 0.60 \\
2012au & 2277 & [O~{\sc i}] & yes & 100 & 1.10 & 5100  & 1224 & 3.50 & 100.3 & 24.08 & 0.500 & 0.75 \\
\hline
\end{tabular}
\label{table:2012-3sn-params}
\end{table*}

The best fitting H$\alpha$ model is shown in Figure \ref{fig:2010jl-fits}, with the  model parameters listed in Table \ref{table:2010jl-params}. We derive a day~1906 dust mass of 0.018~M$_{\odot}$, corresponding to a large optical depth of $\tau$=12.6. This dust mass is larger than the dust mass of 0.005~M$_{\odot}$ derived by \citet{bevan2020} for day 1286. 
As we could not quantify the uncertainty on the parameters using Bayesian inference due to the complex nature of the line profile, the uncertainties on the dust mass were found by fixing the other parameters and varying the dust mass until the best-fitting $\chi^2$ value changed by 35~per~cent. We also determined an uncertainty on the grain radius using the same method. The grain radius and its error limits were found to be 0.15$^{+0.21}_{-0.14}~\mu$m, which is in agreement with the grain radius of 0.2$\mu$m used by \citet{bevan2020} in their {\sc damocles} models.

\subsubsection{SN~2011ja}

\begin{figure*}
\centering

\includegraphics[width=\linewidth]{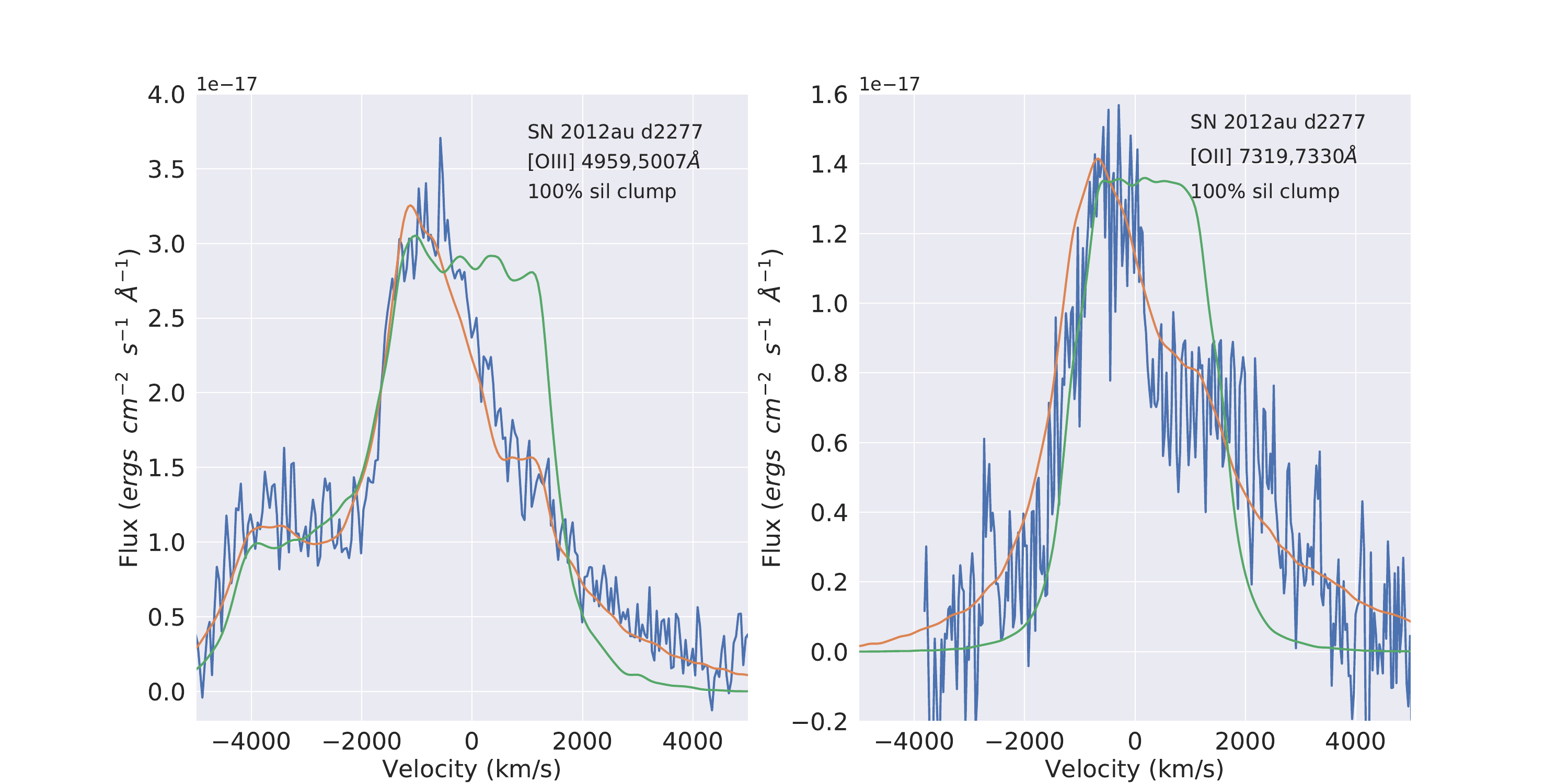}

\caption{{\sc damocles} models of SN~2012au's H$\alpha$ profile (orange line) from the Gemini GMOS spectra (blue line) takem 2277 days post-explosion. The green line is the dust free model. Clumped dust models of 100~per~cent silicate dust are shown.}
\label{fig:2012au-fits}
\end{figure*}

SN~2011ja was discovered on 18/12/2011 in the galaxy NGC~4945, and was determined to be a Type~II-P supernova with an explosion date of 12/12/2011 after a spectrum taken on 19/12/2011 bore a strong resemblance to the spectrum of SN~2004et at an age of 7 days (\citet{Monard2011}. 

\citet{Andrews2016} studied the optical emission from SN~2011ja between days 84-807, from Gemini GMOS photometry and spectroscopy, and the near- and mid-IR emission using {\em Spitzer} IRAC and NTT/SOFI data for days 8-857. They also noted signs of an early interaction with CS material, from the emergence of a double peak in the broad H$\alpha$ line in the optical spectrum, around the same time as noted by \citet{Chakraborti2013}. They compared it to several Type~IIn supernovae, stating it could belong in between Type~II-P and IIn. \citet{Andrews2016} also noticed several indicators of dust formation starting around the same time as the CSM interaction. The optical line profiles started becoming visibly more blueshifted between days 84-112, and a near-IR excess was observed. To quantify the newly-formed dust mass they modelled the dust emission SEDs from days 100-857 using the radiative transfer code {\sc mocassin}, finding $\sim1\times10^{-4}$~M$_{\odot}$ of clumpy AmC dust at the latest epoch. 
They surmised this dust to have been formed in a CDS, 
similar to SN 1998S \citep{Pozzo2005} and SN 2010jl \citep{Gall2014}. 
\citet{Tinyanont2016a} also modelled the {\em Spitzer} SED of SN~2011ja over several epochs, including at day 1382, as part of their large mid-IR survey SPIRITS. They estimated the presence of  $4\times10^{-4}$~M$_{\odot}$ of graphite dust at day 807 and $10^{-3}$~M$_{\odot}$ of graphite dust at day 1382 post explosion.

From a manual parameter investigation, we found best fitting models to the H$\alpha$ line in the GMOS spectra of SN~2011ja over four epochs: 807, 1275, 1611 and 2259 days past explosion. All spectra were corrected for a recessional velocity of NGC~4945 of 560~km~s$^{-1}$ \citep{Crook2006}. We found the presence of a persistent red scattering wing in the H$\alpha$ line at all 4 epochs, which became more pronounced with time, and the best fitting models required a silicate dust composition with a grain radius of 0.15~$\mu$m. This disagrees with \citet{Tinyanont2016} and \citet{Andrews2016}, who did not detect a 9-10-$\mu$m silicate emission feature in their IR SEDs of SN 2011ja. Our best fitting models for the day 807 and 2259 H$\alpha$ spectra using a 100~per~cent silicate dust composition can be found in Figure \ref{fig:2011ja-fits} while parameters for the models for all four epochs are listed in Table~\ref{table:2012-3sn-params}).

We ran a Bayesian simulation of the H$\alpha$ line at 2259 days using 100~per~cent silicate dust, the corner plot for which can be seen in Figure~A6. The median values for the dust mass and grain radius in the 1-D posterior probability distributions closely matched our manually found best fitting model values. Due to this, and the fact all line profiles had a similar S/N, we extrapolated the errors calculated from the Bayesian MCMC model on the dust mass to all other epochs. There is a steady increase of dust mass with time in SN~2011ja, with the mass increasing by a factor of $\sim$10 between day 807 (2.3$^{+0.67}_{-1.9}$ x10$^{-4}$ M$_{\odot}$) and day 2259 (2.5$^{+0.75}_{-2.1}$ x10$^{-3}$ M$_{\odot}$). Despite the difference in dust composition used, our dust masses at days 807 and 1275 agree well with those derived by \citet{Andrews2016} and by \citet{Tinyanont2016a} at similar epochs, when taking into account our upper and lower limits.

\subsubsection{SN 2012au}
SN 2012au in NGC~4790 was discovered in on 14/03/2012 by the SNHunt project \citep{Howerton2012} and shortly after was classified as a Type~Ib supernova \citep{Soderberg2012, Silverman2012}. \citet{Milisavljevic2013b} and \citet{Takaki2013} noted its similarity to a Type Ic.
\citet{Milisavljevic2018} presented optical spectra of SN~2012au taken 2277 days after explosion, and noted the appearance of strong [O~{\sc iii}]~4959,5007-\AA\ and [O~{\sc ii}]~7319,7330-\AA\ emission, and a stark difference in the [O~{\sc i}]~6300,6363-\AA\ profile when compared to day 321 spectra. They attributed this to the turn-on of a pulsar wind nebula exciting oxygen-rich ejecta.

Our best fitting {\sc damocles} models for the day 2277 [O~{\sc ii}]~7319,7330-\AA\ and [O~{\sc iii}] 4959,5007~\AA\ spectra published by  \citet{Milisavljevic2018} for SN~2012au are shown in Figure~\ref{fig:2012au-fits} and the parameters can be found in Table~\ref{table:2012-3sn-params}. The [O~{\sc iii}] and [O~{\sc ii}] features required similar model parameters for the best fitting models, implying a co-location of ionisation species similar to that found for SN~1996cr. We found that large silicate grains of 1.0-$\mu$m radius; carbon grains of 0.4-$\mu$m radius; or a 50:50 carbon to silicate mixture of 0.5-$\mu$m radius grains, fitted the oxygen line profiles best. The observed [O~{\sc i}]~6300,6363-\AA\ profile on day~2277 contained what we assumed to be a contaminating emission feature located between -300 and -800 km~s$^{-1}$, of uncertain origin. When it was removed the resulting [O~{\sc i}] profile was well fit with similar parameters to those required to fit the [O~{\sc ii}] and [O~{\sc iii}] profiles
(Table~\ref{table:2012-3sn-params}).

We ran a Bayesian model assuming 100~per~cent AmC dust, which fit the [O~{\sc iii}] and [O~{\sc ii}] lines in SN~2012au simultaneously. The median grain radius was 0.3-$\mu$m, whose large uncertainty limits comfortably included our initially derived grain radius. Scaling the uncertainties found from the Bayesian analysis to our best fitting 50:50 AmC to silicate dust mass, gives a dust mass at day 2277 in SN 2012au of ($1.0^{+0.43}_{-0.90})\times10^{-3}$ M$_{\odot}$, which is similar to the dust mass derived for SN 2011ja at a similar epoch.

\section{Dust masses and dust properties}

\begin{table*}
\caption{A summary of the best-fitting ejecta clumped dust masses for all the observation epochs of the CCSNe modelled in this sample. These are the values plotted in Figure \ref{fig:total-dm-sig}. The "\% Sil" column stands for the mass percentage of the dust species that consists of astronomical silicate, where the remainder of the dust is AmC dust. The listed optical depths $\tau_\lambda$ are calculated from R$_{in}$ to R$_{out}$ at the central line wavelengths given in column 4.}
\begin{tabular}{lllllllllllll}

\hline
SN & Type & Epoch & $\lambda$ & $a$ & \% Sil & V$_{max}$ & V$_{min}$ & $\beta$ & R$_{out}$ & R$_{in}$ & $\tau_{\lambda}$ & M$_{dust}$ \\
 & & (days) & (\AA) & ($\mu$m) & & km~s$^{-1}$& km~s$^{-1}$ & & 10$^{15}$cm & 10$^{15}$cm & & (M$_{\odot}$) \\
\hline

1957D  & II   & 11371 & 5007 & 0.40 & 50  & 5800  & 870  & 2.5  & 1079.00 & 140.30 & 0.42  & $1.20^{+4.8}_{-1.18}\times10^{-2}$    \\
-      & -    & 19459 & 5007 & 0.40 & 50  & 6800  & 1020 & 2.4  & 1265.00 & 140.30 & 0.13  & $3.50^{+7.50}_{-3.00}\times10^{-2}$    \\
-      & -    & 21535 & 5007 & 0.40 & 50  & 7500  & 962  & 2.5  & 1377.00 & 140.30 & 0.29  & $5.00^{+49.80}_{-4.78}\times10^{-2}$     \\
1970G  & II-L & 16733 & 6563 & 0.10 & 100 & 6600  & 3960 & -0.1 & 954.20  & 572.50 & 1.12  & 0.10$^{+1.41}_{-0.097}$     \\
1979C  & II-L & 5146  & 7319 & 0.10 & 50  & 6800  & 5440 & 4.5  & 297.00  & 237.00 & 15.00 & 0.10$^{+0.40}_{-0.07}$      \\
-      & -    & 10575 &  7319 & 0.15 & 50  & 6600  & 4950 & 5.5  & 603.00  & 452.30 & 11.00 & 0.65$^{+0.87}_{-0.43}$     \\
-      & -    & 13903 &  7319 & 0.15 & 50  & 6600  & 3696 & 2.2  & 793.00  & 444.10 & 4.58  & 0.30$^{+0.13}_{-0.15}$       \\
1980K  & II-L & 13706 &  6563 & 0.10 & 100 & 5300  & 4505 & 1.5  & 593.10  & 504.00 & 2.91  & 0.60$^{+3.29}_{-0.57}$      \\
1986E  & II-L & 3712 &  6563 & 0.10 & 50  & 7100  & 3337 & 3.0  & 225.80  & 106.10 & 2.15  & $3.00^{+0.00}_{-2.40}\times10^{-2}$     \\
-      & -    & 10619 &  6563 & 0.10 & 50  & 6600  & 3102 & 3.0  & 603.80  & 283.80 & 1.07  & $7.00^{+0.00}_{-6.50}\times10^{-2}$ \\
-      & -    & 11723 &  6563 & 0.10 & 50  & 6600  & 2640 & 3.0  & 668.30  & 314.10 & 1.02  & $7.00^{+0.00}_{-6.50}\times10^{-2}$     \\
1993J  & IIn  & 8417  & 5007 & 0.10 & 50  & 5800  & 4408 & 5.0  & 421.80  & 320.60 & 2.17  & $4.00^{+13.0}_{-3.8}\times10^{-2}$     \\
1996cr & IIn  & 7370 & 7319 & 0.11 & 50  & 6900  & 2898 & 6.0  & 436.50  & 183.30 & 5.47  & 0.10$^{+0.13}_{-0.078}$     \\
-      & -    & 8460  & 7319 & 0.11 & 50  & 6900  & 2380 & 5.0  & 497.04  & 173.96 & 4.11  & 0.15$^{+0.30}_{-0.14}$     \\
1998S  & IIn  & 1170  & 6563 & 0.08 & 100 & 5500  & 3410 & 0.5  & 55.60   & 34.47  & 10.60 & $2.50^{+4.10}_{-1.60}\times10^{-2}$     \\
-      & -    & 2132 & 6563  & 0.08 & 100 & 5300  & 3339 & 0.5  & 97.63   & 61.51  & 5.44  & $4.00^{+6.00}_{-3.30}\times10^{-2}$    \\
-      & -    & 6574 & 6563  & 0.10 & 100 & 5600  & 2688 & 2.3  & 260.00  & 124.00 & 4.99  & 0.15$^{+0.54}_{-0.14}$     \\
2004et & II-P & 646  & 6563 & 0.20 & 100 & 9500  & 7410 & 1.7  & 53.02   & 39.24  & 0.57  & $5.00^{+24.00}_{-0.60}\times10^{-4}$ \\
-      & -    & 3868 & 6563 & 0.20 & 100 & 6000  & 5400 & 2.0  & 200.50  & 180.50 & 1.31  & $1.00^{+89}_{-0.98}\times10^{-2}$ \\
2007it & II   & 3248 & 6563 & 0.30 & 50  & 14000 & 6860 & 2.7  & 39.29   & 5.11   & 0.61  & $8.00^{+3.00}_{-5.00}\times10^{-3}$    \\
2010jl & IIn  & 1906 & 6563 & 0.15 & 0   & 1150  & 370  & 2.0  & 230.00  & 26.40  & 12.60 & $1.80^{+0.75}_{-0.75}\times10^{-2}$     \\
2011ja & II   & 801 & 6563  & 0.15 & 100 & 6200  & 3038 & 2.0  & 42.91   & 21.02  & 0.72  & $2.30^{+7.11}_{-1.93}\times10^{-4}$  \\
-      & -    & 1269 & 6563 & 0.15 & 100 & 6100  & 2928 & 2.0  & 66.88   & 32.10  & 0.81  & $6.00^{+18.50}_{-5.04}\times10^{-4}$   \\
-      & -    & 1611 & 6563 & 0.15 & 100 & 6300  & 2835 & 2.0  & 87.69   & 39.46  & 0.70  & $8.00^{+24.72}_{-6.72}\times10^{-4}$   \\
-      & -    & 2153 & 6563 & 0.20 & 100 & 6100  & 2928 & 2.0  & 113.50  & 54.47  & 1.40  & $2.50^{+7.73}_{-2.10}\times10^{-3}$   \\
2012au & Ib/c & 2277 & 5007 & 0.50 & 50  & 5000  & 1150 & 3.3  & 98.37   & 22.62  & 0.75  & $1.20^{+4.29}_{-0.95}\times10^{-3}$ \\ 
2012aw & II-P & 1158 & 6563 & 0.30 & 100  & 5500  & 1100 & 1.3  & 55.03  & 11.01  & 0.70  & $2.00^{+8.86}_{-1.55}\times10^{-4}$ \\ 
- & - & 1863 & 6563 & 0.30 & 100  & 5100  & 663 & 1.5  & 82.09  & 10.67  & 1.02  & $6.00^{+21.9}_{-3.60}\times10^{-4}$ \\ 
\hline
\label{table:final-dm}
\end{tabular}
\end{table*}

\subsection{CCSN dust masses as a function of ejecta age}

The ejecta dust masses derived for the CCSNe in our sample from their best fitting models are summarised in Table~\ref{table:final-dm}. The values plotted for SN 2012aw, which were derived as part of the ORBYTS outreach programme, are presented in more detail by Niculescu-Duvaz et al. 2022, in prep. The dust mass derived for the IWC component of SN~2004et is not included, as there is the possibility that this dust could have formed before the supernova explosion.
We plot these ejecta dust masses against age since explosion in Figure~\ref{fig:total-dm-sig}. As discussed in the individual sections, for epochs where we could not determine the dust species, we use a dust mass for a 50:50 carbon to silicate mass ratio, and where we cannot determine the grain radius, we have plotted the dust mass for a 50:50 carbon to silicate mass ratio using a grain radius of 0.10~$\mu$m in order to constrain a lower limit to the dust mass. 
Some error bars are taken as percentage errors of the 16th and 84th quartiles on the median dust mass value in the marginalised 1-D dust mass probability distributions derived from the Bayesian modelling process. As our Bayesian models were generally run for 100~per~cent AmC or 100~per~cent silicate dust, for supernovae where we could not determine the dust species we then scaled these percentage errors to the best-fitting dust mass derived from a manual fit using a 50:50 AmC to silicate dust ratio, in order to derive upper and lower limits. We justify this approach by the fact that when we ran Bayesian models for 100~per~cent AmC or 100~per~cent silicate for the [O~{\sc ii}] feature of SN~1979C at day 13903, the error bars from the 1D posterior distribution of the dust mass in both cases only varied by a factor of $\sim$1.3. Other error limits are taken from a 35~per~cent variation in $\chi^2$ when changing the dust mass whilst keeping all other parameters fixed. These values in general tend to have tighter error limits, but this does not mean that the dust mass determined is more accurate. We consider the size of the uncertainties determined from the Bayesian analyses to be a more accurate representation of the dust mass uncertainties.

\begin{figure*}
\centering

\includegraphics[width=0.85\linewidth]{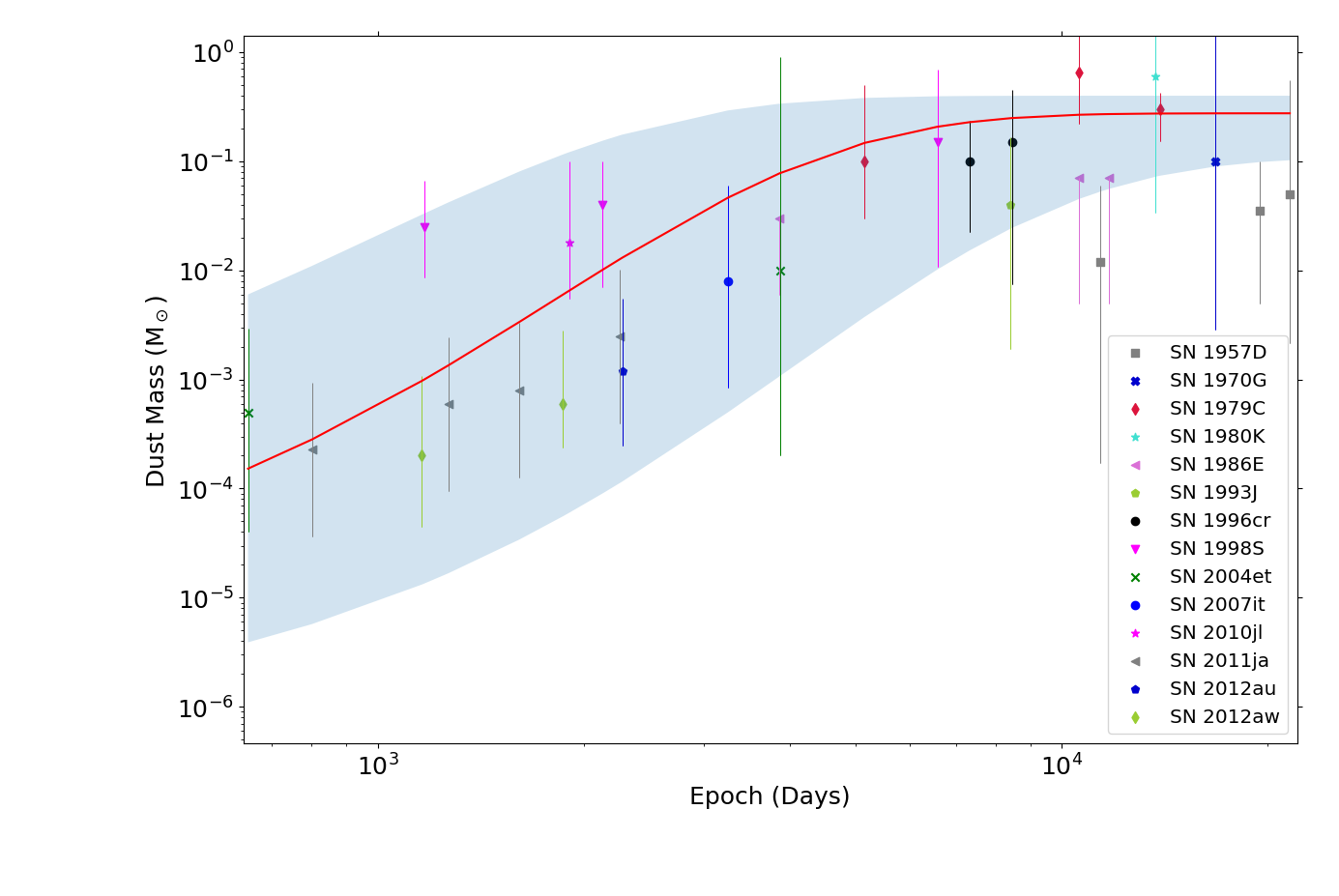}

\caption{Dust mass against time past explosion, for the sample of CCSNe modelled in this work (see Table~\ref{table:final-dm}), with the least-squares minimized sigmoid curve defined by equation (3) overplotted. The grey band encloses the error region on the best-fitting sigmoid curve parameters, where the errors are derived from a Monte Carlo bootstrap simulation and are provided in the text.}
\label{fig:total-dm-sig}
\end{figure*}

\begin{figure*}
\centering

\includegraphics[width=0.85\linewidth]{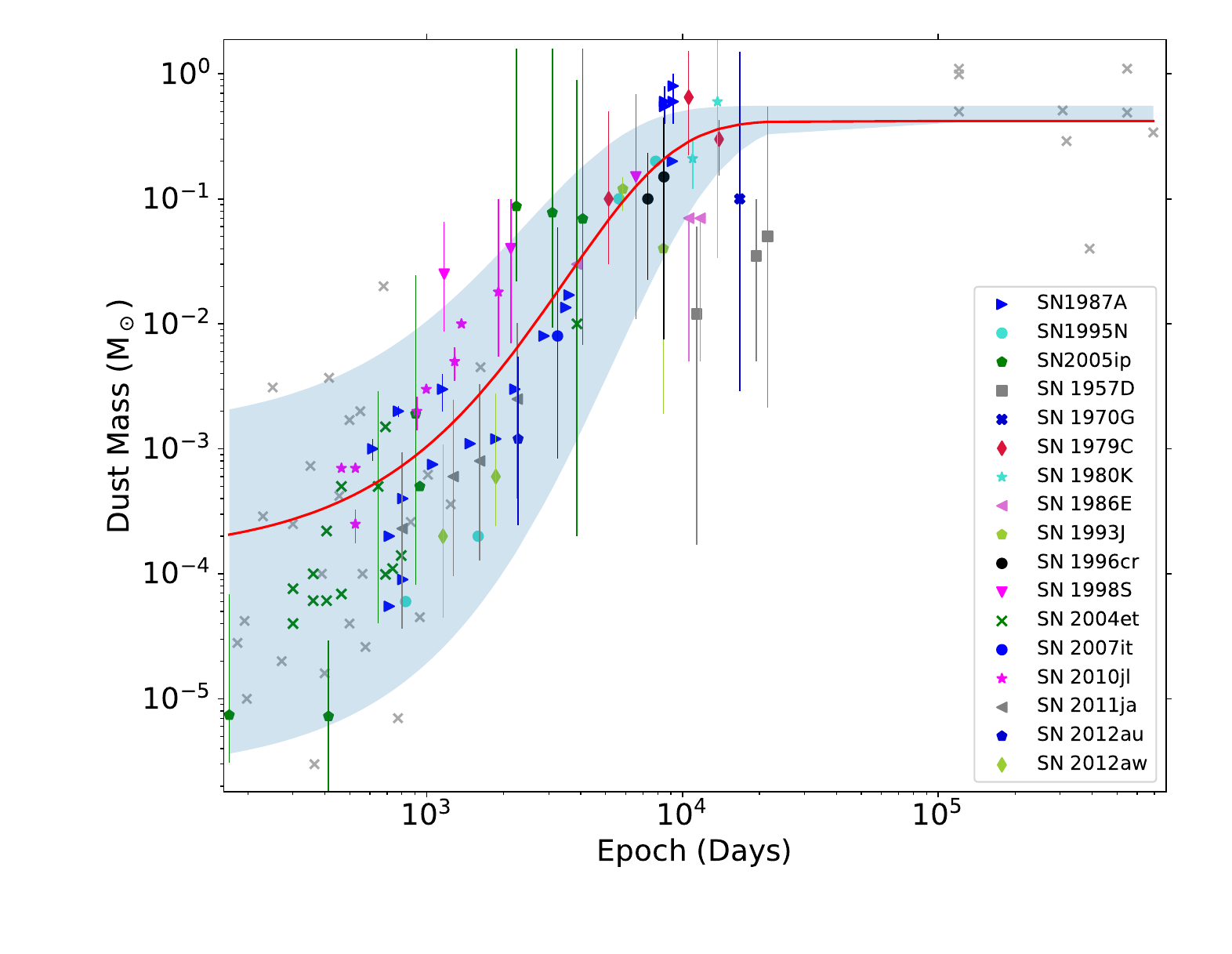}

\caption{Dust mass against time (age past explosion) for the sample of CCSNe modelled with {\sc damocles} (coloured points), combined with literature dust mass determinations (grey crosses) with details summarised in Table~A1. A least-squares minimized sigmoid curve defined by equation~(\ref{eq:sigeq}) is overplotted in red. The grey band encloses the error region on the best-fitting sigmoid curve parameters, where the errors are derived from a Monte Carlo bootstrap simulation and are provided in the text. Individual dust mass error limits were not taken into account in the fitting, since many of the literature values lacked such error estimates. The CCSNe for which {\sc damocles} dust mass determinations exist are identified by the key in the inset box. As a function of increasing age, the six supernova remnants for which dust masses are plotted are Cas~A, G29.7-0.3, G21.5-0.9, the Crab Nebula, G54.1+0.3 and G11.2-0.3, with ages $\sim$330, 850, 880,1100,1500,1900 years respectively. Note that many of the supplementary data-points plotted below 1500 days are based on {\em Spitzer} mid-infrared flux measurements;
we only used such dust masses when the spectral coverage extended to at least 8~$\mu$m.}
\label{fig:total-dm-sig-all}
\end{figure*}

\begin{figure*}
\centering

\includegraphics[width=0.75\linewidth]{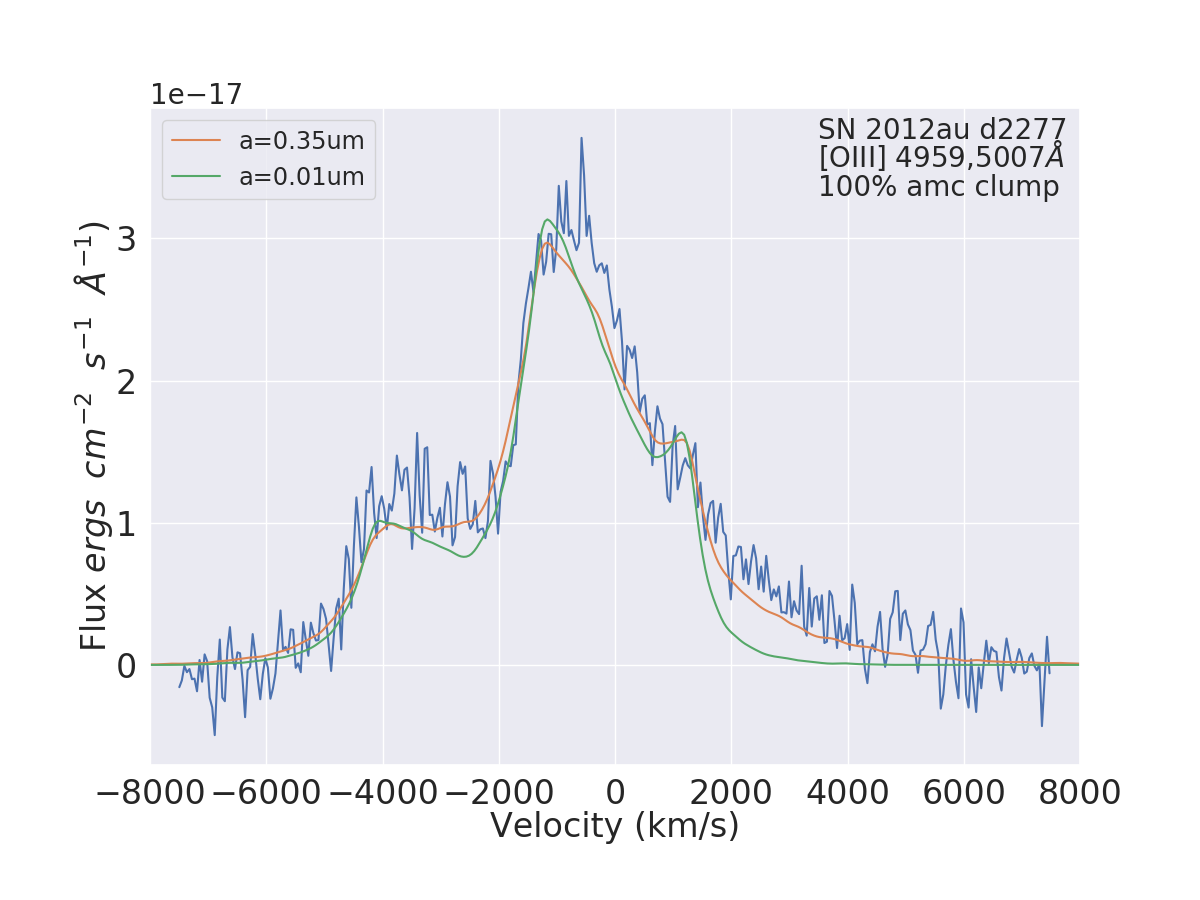}

\caption{The best fitting model for SN~2012au's day~2277 [O~{\sc iii}] 4959,5007~\AA\ doublet feature, using a clumped amorphous carbon dust distribution with a grain radius of 0.35~$\mu$m and a dust mass of $4\times10^{-4}$~M$_{\odot}$, with the other parameters described in Table \ref{table:2012-3sn-params}. This can be seen to provide a better fit to the observed profile than the best fitting model using clumped amorphous carbon dust with a smaller grain radius of 0.01~$\mu$m and a dust mass of $1\times10^{-4}$~M$_{\odot}$, with otherwise matching model parameters.}
\label{fig:2012au-gs-diff}
\end{figure*}

We used the least$\_$squares function found in the python scipy.optimize package to fit our data points in Figure~\ref{fig:total-dm-sig} with a sigmoid curve similar to that used by \citet{Wesson2015}, 
to fit the time evolution of SN~1987A's dust mass, defined by the equation
\begin{equation} \label{eq:sigeq}
M_d(t) = ae^{be^{ct}}
\end{equation}
Such a curve provides a natural limit to the growth of the parameter being fitted, which might correspond in this case to the exhaustion of condensable heavy elements.
$a$ can be interpreted as the dust mass at saturation, in M$_\odot$, that is formed by CCSNe,
$b$ describes when the peak dust formation epoch occurs such that a more negative $b$ leads to a later onset of dust formation, and $c$ describes the rate of peak dust formation, such that a more negative value of $c$ corresponds to faster dust formation. We used a Monte Carlo simulation to bootstrap our data plotted in Figure~\ref{fig:total-dm-sig}. We generated 1000 datasets randomly drawn from a normal distribution to quantify the best fitting curve parameters and their associated uncertainties. We derived curve fit parameters of $a$=0.23$^{+0.17}_{-0.12}$~M$_\odot$, $b$= -10.51$^{+4.51}_{-1.49}$ and $c$=(-4.73$^{+2.25}_{-5.37})\times10^{-4}$ days$^{-1}$, where the best-fitting parameter values for a and c were taken as the median of the resulting parameter distributions of the bootstrapped samples, and all uncertainties are taken as the 16$^{th}$ and 84$^{th}$ quartiles of this distribution. The mean for the b value was taken, as the least squares fitting algorithm bimodally preferred values between the upper and lower quartiles. The median and quartiles were preferred to the mean and standard deviation as  the parameter distributions were non-gaussian.

Figure~\ref{fig:total-dm-sig-all} supplements our {\sc damocles}-derived dust masses with dust masses from various sources in the literature, as listed with references in Table~A1. 
As discussed in Section~1,
for supernovae observed during their photospheric phase (up to $\sim$150 days post-explosion)
we discount dust mass measurements made via line profile fitting, as blue-shifted emission line peaks before these times can be created by occultation of red-shifted emission by the optically thick, steeply distributed ejecta gas \citep[see][]{Anderson2014}.

Although some of the dust masses taken from the literature were determined using {\sc damocles},  
many were determined from optical-infrared SED fitting. For the latter cases we excluded dust masses based on `warm' {\em Spitzer} photometry that only extended to 4.5-$\mu$m or less, potentially missing cooler dust
emitting at longer wavelengths.
Overall, the data points that we took from the literature
consisted of dust masses at additional epochs for six of our Figure~\ref{fig:total-dm-sig} sample, together with multi-epoch dust mass determinations for eleven further CCSNe and dust mass estimates for six CC-SNRs older than 300 years. Since a large number of the literature dust masses plotted in Figure~\ref{fig:total-dm-sig-all} did not have any uncertainty limits associated with them, the fit to Figure~\ref{fig:total-dm-sig-all} did not take individual error limits into account.

A sigmoid curve using equation (\ref{eq:sigeq}) was fit to the data-points in Figure~\ref{fig:total-dm-sig-all}, using the same method as described for the curve fit procedure on the data in Figure~\ref{fig:total-dm-sig} , yielding $a$=0.42$^{+0.09}_{-0.05}$~M~$_\odot$, $b$= -8.0$^{+4.0}_{-2.0}$ and $c$=(-2.88$^{+1.03}_{-1.27})\times10^{-4}$ days$^{-1}$. 
The sigmoid curve fits plotted in
Figures~\ref{fig:total-dm-sig} and \ref{fig:total-dm-sig-all} require very similar values for $b$ and $c$. At 0.42$^{+0.09}_{-0.05}$~M~$_\odot$, the value of $a$, the dust mass at saturation, is derived to be larger for the Figure~\ref{fig:total-dm-sig-all} sample
than the value of $a$ = 0.23$^{+0.17}_{-0.12}$~M$_\odot$ found for our {\sc damocles} dataset plotted in Figure~\ref{fig:total-dm-sig}.
For the sigmoid curve plotted in
Figure~\ref{fig:total-dm-sig-all}, the dust mass reaches 75~per~cent of its saturation value within 30 years.

The uncertainties on the value of $b$ are large for the sigmoid curve plotted in Figure~\ref{fig:total-dm-sig} due to the relative lack of dust mass samples at early times ($<$1000 days past explosion), such that it is difficult to determine the floor of the sigmoid curve: this explains why the error band on the best fitting sigmoid curve at early times is so broad.
However, the value of the dust mass parameter $a$ is well constrained in both cases (as is the value of $c$ in the case of the larger sample).

There appears to be a real dispersion in some of the SNR dust masses plotted in Figure~\ref{fig:total-dm-sig-all}. Cas~A has three dust mass determinations from independent methods that all lie between 0.5 and 1.1~M$_\odot$ \citep{DeLooze2017, Bevan2017, Niculescu-Duvaz2021}, while the dust mass determined for the Crab Nebula by \citet{DeLooze2019} was $0.040\pm0.008$~M$\odot$.
The sigmoid curve fits plotted in Figures~\ref{fig:total-dm-sig} and \ref{fig:total-dm-sig-all} have similar slopes to that found by \citet{Wesson2015} for SN~1987A alone, but have a lower value for the saturation dust mass parameter compared to the \citet{Wesson2015} value of $a=1.0$~M$_\odot$ derived from a range of observations of SN~1987A at multiple epochs. It is therefore possible that SN~1987A has formed a larger dust mass than other SNe of similar ages. The curve plotted by \citet{Bevan2019} in their Figure 10, which was a fit to the dust mass evolution for just SN~1987A and SN~2005ip, does not fit most of the datapoints in our Figure \ref{fig:total-dm-sig}. Their curve exhibits larger dust masses at early times, a less steep dust mass growth with time, a later dust mass growth plateau as well as a larger dust mass saturation than that found by our best-fitting sigmoid curve. However, their curve well fit our Type IIn dust mass datapoints for SN~2010jl and SN~1998S. Their curve would also fit the dust masses found here for SN~1996cr assuming a 100~per~cent silicate dust composition.

The dust mass growth plot in Figure~1 of \citet{Gall2018} has a plateau value of $0.40\pm0.07$~M$_\odot$, similar to the saturation value of 0.42$^{+0.09}_{-0.05}$~M$_\odot$ for the curve in Figure~\ref{fig:total-dm-sig-all}. The latter curve predicts that SNe produce less dust on average at ages $<40$ years than that plotted by \citet{Gall2018}, which has a steeper dust growth rate before this time, and plateaus sooner at an age of 20 years rather than at 40 years.

Of the fourteen CCSNe in our sample, four were Type~II-L supernovae, namely SN~1970G, SN~1979C, SN~1980K and SN~1986E. All had spectra taken beyond 30 years after outburst, so from the sigmoid curve fit to the dust mass versus ejecta age distribution in Fig~\ref{fig:total-dm-sig-all}, we should expect dust masses of 0.42$^{+0.09}_{-0.05}$~M$_\odot$, while
the average dust mass from the latest observed epochs for these four Type~II-L SNe was 0.35$\pm0.33$~M$_\odot$.
We found that there was a large spread in the radial gas/dust distributions of these Type II-L supernovae, perhaps indicating a diversity in the evolutionary history of their progenitor stars. SN~1970G and SN~1980K both exhibit similar optical spectra and required similar [O~{\sc i}] and H$\alpha$ emitting gas distributions, however SN~1970G was found to have a lower dust mass than SN 1980K by a factor of $\sim$6. 
The dust mass found for SN~1979C is the highest of the four Type~II-L objects and its optical spectrum exhibits the highest degree of ionization of the four. It has been suggested to have a black hole at its centre \citep[][]{Patnaude2011}, which would imply a large progenitor mass and this could possibly account for its larger dust mass when compared to other CCSNe of a similar age.



Four of the CCSNe in our sample were of Type~IIn (SN~1993J, SN~1996cr, SN~1998S and SN~2010jl), a classification indicating the presence of circumstellar material that produces a narrow emission line, usually superposed on
a broader profile from the supernova ejecta.
Of the four, SN~1998S and SN~2010jl are outliers to the sigmoid dust growth curve, presenting high dust masses at early epochs of up to 2000 days in comparison to other Type IIL/IIP supernovae. 
Their data points are well fit by the SN 2005ip dust growth curve seen in Figure~11 of \citet{Bevan2019}. The increased dust mass could be due to the fact that there is an additional Cool Dense Shell component at early times which produces extra dust. For SN~1998S, at the late epoch of 6500 days the dust mass derived is comparable to those of SN~1979C, SN~1987A and SN~1996cr, the latter another Type~IIn.

The use of the nucleosynthetic yields of \citet{Woosley2007}, as extracted by \citet{Temim2017}, predicts that a star with an initial mass of 15~M$_\odot$ can produce 0.13~M$_\odot$ of carbon and 0.23~M$_\odot$ of silicate constituents, while an 18~M$_\odot$ star can produce 0.20~M$_\odot$ of carbon and 0.35~M$_\odot$ of silicates.
Comparing these limits with the carbon and silicate dust masses that can be found in Table~\ref{table:final-dm} indicates that the only values that may exceed these constraints are the day~10575 dust mass for SN~1979C ($0.65^{+0.87}_{-0.43}$~M$_\odot$ split equally between amorphous carbon and silicates) and the day 13706 silicate dust mass for SN~1980K ($0.60^{+3.3}_{-0.57}$~M$_\odot$).

\subsection{CCSN dust grain properties}

We were able to put some constraints on the dust grain albedo for most of the CCSNe in our sample, either from the presence of an extended red scattering wing in the line profiles, or from exploiting the wavelength dependence of dust absorption. 
For some objects, such as SN 1979C, we also obtained 1-D probability density distribution curves for the grain radii from a Bayesian analysis. 
Six of the CCSNe in our sample (SN~1970G, SN~1980K, SN~1998S, SN~2004et, SN~2007it and SN~2011ja) showed large red scattering wings in their line profiles, which required close to 100~per~cent silicate dust to fit. For all these SNe, silicate grain radii of the order of 0.1-0.3~$\mu$m were required, which yield an optical albedo of $\sim0.9$. For the SNe where we were able to constrain the dust grain albedo but not the species, we could constrain the grain size for the cases of either 100~per~cent AmC, 100~per~cent silicate or 50:50 carbon to silicate mixes. From their 1-D probability distributions values derived from a Bayesian analysis, SN~2012au and SN~1957D require median grain radii of 0.2~$\mu$m and 0.3~$\mu$m respectively for 100~per~cent AmC grains, where the probability distributions plateau at large grain radii of $\sim1~\mu$m. 

Figure \ref{fig:2012au-gs-diff} demonstrates how the grain radius affects the presence of a scattering wing in a model of the [O~{\sc iii}] 4959,5007~\AA\ profile of SN~2012au at 2277 days past explosion. The orange line shows the best fitting model of SN~2012au for a 100~per~cent amorphous carbon dust species using a grain radius of 0.35~$\mu$m and a dust mass of $4\times10^{-4}$~M$_\odot$, whereas the green line is the best fitting model using a smaller grain radius of 0.01~$\mu$m, a dust mass of 1$\times$10$^{-4}$ M$_{\odot}$ and otherwise matching parameters as those generating the orange line. It is clear that in this case, as was the case for other SNe modelled in this work, a visual inspection can rule out small dust grain radii based on goodness-of-fit to the red scattering wing in emission line profiles.

SN~1979C and SN~1996cr are the  two objects in our sample that are most likely to  have grain radii <0.2~$\mu$m. In the case of SN~1996cr, for dust grain radii $\sim0.1~\mu$m there should be a wavelength dependence of dust extinction for wavelengths $>2 \pi a$ = 0.63~$\mu$m, while its spectra had broad emission lines ranging between 6300 and 9069~\AA. This, and the fact that in our models multiple emitting ionic species were found to be co-located, allowed us to constrain the 100~per~cent clumped AmC grain radius to be 0.12~$\mu$m, and 0.05~$\mu$m for SN~1996cr for clumped 100~per~cent silicate dust. The high signal to noise of the line profiles in our spectra of SN~1979C allowed us to use a Bayesian analysis to constrain the median 100~per~cent AmC grain radius to be 0.1~$\mu$m, with an upper limit of 0.3~$\mu$m. This translates for a 50:50 AmC to silicate mix to a single grain radius of $\sim$0.2~$\mu$m. For SN~1970G, SN~1993J and SN~1986E we were unable to constrain the dust grain species or radius, while for our model of SN~2010jl at day 1906, we utilised previous grain radius information from \citet{bevan2020} for our models.
Therefore, we find that the majority of the objects in our sample require "medium-sized" dust grains with radii from 0.1-0.5~$\mu$m, regardless of composition. 
The large grain radii that we find are in agreement with other works \citep[e.g.][]{Gall2014, owen2015, Wesson2015, Bevan2017, priestley2020}.

From hydrodynamic grain destruction modelling, \citet{Slavin2020} found that, considering grain sputtering only, 30~per~cent of carbonaceous grains with radius of $>0.1~\mu$m and 10~per~cent of silicates with a radius of 0.2~$\mu$m, increasing up to 20~per~cent for larger silicate grains, can survive processing through a typical supernova reverse shock. From their hydrodynamic grain destruction models \citet{Kirchschlager2019} and \citet{Kirchschlager2020} found that for initial log-normal grain size distributions peaking at 0.1~$\mu$m, between 15 and 50 per~cent of silicate grains would survive an encounter with the reverse shock, for a range of clump density contrasts, where their simulations treated sputtering, grain-grain collisions and ion impact trapping. Currently it is difficult to predict theoretically what fraction of CCSNe will have surviving dust masses $>0.1$~M$_\odot$, the minimum dust mass needed per CCSN, according to
\citet{Morgan2003a} and \citet{Dwek}, in order for them to be candidates as the main stellar dust producers in the Universe. 
 

\section{Conclusions}

We have modelled the red-blue asymmetries in the optical line profiles of thirteen supernovae,  using {\sc damocles} to determine dust masses and other ejecta parameters.
Prior to this work, many dust mass measurements existed for CCSNe aged $<4$ years, corresponding to when infrared dust emission may have been detectable by {\em Spitzer}, but relatively few dust mass measurements existed for CCSNe older than this. We have greatly increased the sample of dust mass measurements available for CCSNe with ages between five and sixty years, Our measurements, combined with literature values, are plotted in Figure \ref{fig:total-dm-sig-all} to generate the most well-sampled and informative supernova dust mass growth curve to date.

A striking aspect of our results is that apart from SN~1998S, which had a dust mass of 2.5$^{+4.1}_{-1.6} \times10^{-2}$~M$_\odot$ at day 1170, at epochs earlier than 2200 days none of the derived dust masses listed in Table~\ref{table:final-dm}
exceeds 
2.5$\times10^{-3}$~M$_\odot$. This stands in contrast with predictions from a range of recent theoretical studies of dust formation in CCSN ejecta, which have typically predicted that $\geq0.1$~M$_\odot$ of dust should have formed within 1000 days after outburst \citep[e.g.][]{Sarangi2013, Biscaro2014, Sarangi2015, Sarangi2018, Mauney2018, Sluder2018, Brooker2021}. Our results indicate that CCSNe can indeed form dust masses
$>0.1$~M$_\odot$, but over much longer timescales.

The dust mass growth with time for all the CCSN epochs modelled in this work is well described by a sigmoid curve similar to that fitted to SN~1987A's dust mass evolution by \citet{Wesson2015}. The sigmoid curve plotted in Figure~\ref{fig:total-dm-sig} implies that 90~per~cent of the saturation dust mass of 0.23$^{+0.17}_{-0.12}$~M$_\odot$ is reached after about 10$^4$~days ($\sim$30~years) post-outburst.
For the expanded sample that includes supplementary dust mass estimates from the literature, we find a similar period to reach 75~per~cent of the saturation dust mass of 0.42$^{+0.09}_{-0.05}$~M$_\odot$
(Figure~\ref{fig:total-dm-sig-all}). 

For most of the CCSNe modelled we have put constraints on the dust grain radii for dust species consisting of either 100~per~cent AmC; 100~percent
astronomical silicates; or a 50:50 mixture of the two species. Inferred grain radii $>0.1~\mu$m for either carbon or silicate dust species are common in our models.  For SN~1970G, SN~1986E and SN~1993J, where we could not constrain the grain radii or dust species, we have provided a lower limit for the dust mass, assuming the dust is made from a 50:50 AmC to silicate dust grain mixture with a grain radius of 0.1~$\mu$m. For SNe~1980K, 1998S, 2007it and 2011ja we constrained the grain species to have a high silicate proportion.  \citet{Slavin2020} and \citet{Kirchschlager2020} predicted a dust destruction rate for grain radii $>0.1~\mu$m of $\sim$20-50~per~cent for a silicate composition, and \citet{Slavin2020} predicted a dust destruction rate of $\sim$30~per~cent for an amorphous carbon composition. Given our value of 0.42~M$_{\odot}$ produced by an average CCSN and given these dust destruction rates, 0.1-0.2~M$_{\odot}$ would survive the passage of the reverse shock. This suggests that CCSNe can contribute a large proportion of the dust formed in high-redshift galaxies. The model parameters found for the CCSNe studied here, particularly their dust masses, compositions and grain radii, will hopefully help contextualise further calculations of dust destruction rates, which are crucial to determining the overall contribution of CCSNe to the dust budget of the Universe.

Out of the eight SNe in our {\sc damocles} sample that are over 20 years old, six have formed dust masses $>0.1$~M$_\odot$.
Amongst the four Type~II-L SNe studied here (SNe 1970G, 1979C, 1980K and 1986E), there is a large range in the dust masses that have formed 30 years or more after outburst (0.07-0.60~M$\odot$; see Table~\ref{table:final-dm}) and in their deduced geometries (see Figure~\ref{fig:3sn}).
For SN~1979C we find that the oxygen- and hydrogen-emitting regions are not co-located. We find a best-fitting model in each case where the dust must be coupled to the freely-expanding oxygen ejecta, and where the H$\alpha$ emitting shell is outside the dusty oxygen-emitting regions.
For the other Type II-L's, the hydrogen- and oxygen-emitting regions were found to be co-located, although SN~1986E exhibited an intermediate width component in both its hydrogen and oxygen emission line profiles, which was not the case for SN~1980K or SN~1970G.




Our work indicates that at similar epochs the four Type~IIn objects may have formed more dust than the five Type~II and II-P objects (see Table~\ref{table:final-dm}). More dust masses are needed for each CCSN sub-type at a range of ages, while tighter progenitor mass limits are required if we are to establish clear links between CCSN sub-types, progenitor masses and ejecta dust properties. 

Many of the objects modelled in this work warrant follow-up observations over an extended period of time, since for many their dust masses are inferred to have not finished increasing. 


\section*{Acknowledgements}

We thank Nathan Smith and Franz Bauer for providing us with spectra of SN~1998S and SN~1996cr, respectively. We also thank Knox Long for permission to use spectra of SN~1957D for this work. MND thanks Haley Gomez and Amelie Saintonge for suggestions that have improved this work.
MND, MJB, AB and RW acknowledge support from European Research Council (ERC) Advanced Grant 694520 SNDUST. DM acknowledges National Science Foundation support from grants PHY-1914448 and AST-2037297. IDL acknowledges support from European Research Council (ERC) Starting Grant 851622 DustOrigin. We thank the anonymous referee for constructive comments on the paper.

\section{Data Availability}
The {\sc damocles} code, with Bayesian implementation, is available at https://github.com/damocles-code/damocles. Wavelength-calibrated copies of our Gemini GMOS and VLT X-Shooter spectra for the SNe listed in Tables 1 and 2 are available in the WISeREP archive (https://wiserep.weizmann.ac.il/). 


\bibliographystyle{mnras}
\bibliography{damoc-paper} %


\appendix*
\section{}

In addition to the discussion below of some of the limitations of our {\sc damocles} modelling, this Appendix contains Table~A1, which summarises those supplementary dust mass data-points taken from the literature and plotted in Figure \ref{fig:total-dm-sig-all}. Bayesian corner plots for six of the simulations run for this work are also included. The CCSNe modelled with a Bayesian inference and plotted in Figures~A1-A6 are SN~1957D, SN~1979C at two epochs, SN~1996cr, SN~2004et and SN~2011ja. We also include a schematic, in Figure~A7, of the oxygen and hydrogen spatial distributions deduced for three of the Type II-L CCSNe that were modelled, SN~1979C, SN~1986E and SN~1980K. Finally, we include plots in Figures A8 and A9 that compare the best-fitting {\sc damocles} model parameters derived by the manual-fitting process and by the Bayesian process.

\subsection{Limitations to our {\sc damocles} modelling}

Firstly, our method is only sensitive to dust distributed co-spatially or internally to the emitting gas. For those CCSNe where we only modelled oxygen lines (SN~1957D, SN~1993J, SN~1996cr and SN~2012au) it is possible our modelling has missed some dust present in any cool dense shell (CDS) outside of the oxygen shell, a region which would typically be traced by hydrogen gas. However, in our spectra of SN~1979C at 13150 and 13907 days, presented in Section \ref{subsubsec:1979c}, where we have high signal-to-noise line profiles of H$\alpha$ and the forbidden oxygen line doublets, we find that the hydrogen shell is external to the oxygen ejecta and that it is not coupled to any dust, and that red-wing attenuation of the H$\alpha$ line is caused by dust co-located with the ejecta. It is difficult to say how applicable this scenario is to the other CCSNe for which we have only modelled the oxygen lines.

Secondly, our models are simplified by the fact that they only use amorphous carbon and astronomical silicate dust. \citet{Niculescu-Duvaz2021} showed that varying the type of silicate species does not change optically derived dust masses significantly. Grains such as Al$_2$O$_3$, which has been detected in Cas~A \citep{Rho2008}, are unlikely to make up a large part of the dust species, due to aluminium being a trace element. Therefore, we do not include Al$_2$O$_3$ dust in our models. The observed line profiles could also be fit with iron grains, which behave similarly to amorphous carbon. Due to its higher mass density, 100~per~cent iron grain dust would require a much larger dust mass to fit the line profiles of our sample CCSNe than 100~per~cent amorphous carbon.

Degeneracies exist between some of the {\sc damocles} model parameters. The clearest way to visualise these is in the corner plots in the Appendix. One of the main degeneracies is between the dust mass, M$_d$, and grain radius, $a$. This is because the radius of a dust grain affects its absorption and scattering efficiencies, which changes the cross-sectional area available for interaction and hence its opacity. The degeneracy between M$_d$ and $a$ is particularly clear in the bottom left panel in the corner plot for the {[}O~{\sc ii}{]} 7319,7330~\AA\ doublet in SN 1979C at day 13903 (Figure A3), where it can be seen that larger amorphous carbon dust grains require a higher dust mass to fit the line profile. In this case, as there is no scattering wing in the observed profile, it is difficult to constrain the grain size for a single line. However, from manually fitting the {[}O~{\sc i}{]}, {[}O~{\sc ii}{]} and {[}O~{\sc iii}{]} lines simultaneously we were able to constrain the grain radius, which was well within the uncertainties of the 1D posterior for $a$. This then allowed us to constrain the dust mass. This was also the case for SN 1996cr. In some cases where the grain radius could be well constrained from the scattering wing, this degeneracy is far less prominent (e.g. bottom left panel in Figure A6, the Bayesian model for SN 2011ja at day 2259). 

If the posterior curve is above zero for all possible values of a given parameter, this means that the model is unable to fully exclude any of these values, and that parameter cannot be considered to be well constrained. In some cases, the grain radius cannot be constrained well at all, such as for the case of SN 2004et at day 646, for which the corner plot can be found in Figure A5. In cases like this, the minimum dust mass can still be determined for the grain radius at which a dust grain is at its most absorbing, but the maximum value of the dust mass cannot be quantified.

For more complex multi-component profiles we were not able to run a Bayesian model for every CCSN in our sample. We also did not run Bayesian models for every SN epoch due to time restrictions. In these cases, we derived uncertainties on our dust masses from manual fits by varying the dust mass until the $\chi^2$ value increased by 35~per~cent, whilst fixing all other parameters. We did not quantify uncertainties in these cases on the other parameters, and the uncertainty on the dust mass is likely to be smaller using this method than for a Bayesian method, where the dust mass posterior distribution is marginalised over other parameters with which it can be degenerate. Figure~A8 compares the V$_{\rm max}$, R$_{in}$/R$_{out}$, $\beta$ and grain radius parameters derived from a manual fit (black arrowheads), with the median values (circles) of the parameters taken from the 1D posterior distribution from Bayesian models, for a range of CCSNe studied in this work. The 16th and 84th quartiles of the 1D distribution for each parameter are plotted as the error bars. Figure~A9 is the same but for the dust mass. For all observations, the manually selected parameters were generally well within the uncertainties determined from the Bayesian models.

\begin{table*}

\centering

\caption{Sources for all supplementary data-points plotted in Figure~\ref{fig:total-dm-sig-all}.} 
\begin{tabular}{p{2.5cm}p{2.5cm}p{2.5cm}p{3.8cm}}
Object          & Epoch (Days)    & M$_d$ (M$_{\odot}$)   & Reference                                  \\
\hline
G11.2-0.3     & 693500 &    0.34$\pm0.14$ & \citet{Chawner2019} \\
G54.1+0.3       & 547900   & 1.1$\pm0.8$         &  \citet{Temim2017}    \\
$"$       & 547900   & 0.49$\pm0.41$        &  \citet{Rho2018}    \\
Crab            & 348955 & (4.0$\pm^{0.9}_{0.8}$)$\times10^{-2}$        & \citet{DeLooze2019}               \\
G21.5-0.9     & 317550 &     0.29$\pm0.08$ & \citet{Chawner2019} \\ 
G29.7-0.3     & 306600 &     0.51$\pm0.13$ & \citet{Chawner2019} \\
Cas A           & 120300 & 0.5$\pm0.1$         &  \citet{DeLooze2017}    \\
$"$           & 119700 & 1.1         &  \citet{Bevan2017}    \\
$"$           & 117600 & 0.99$\pm0.10$        &  \citet{Niculescu-Duvaz2021}   \\
SN~1980K         & 10958  & 0.21        &  \citet{Bevan2017}    \\
SN~1987A         & 615      & (1.0$\pm0.2)\times10^{-3}$       &  \citet{Wesson2015}\\
$"$         & 714      & 2.0$\times10^{-4}$      &  \citet{bevan2016}    \\
$"$         & 714      & 5.5$\times10^{-5}$     &  \citet{bevan2016}    \\
$"$         & 775      & (2.0$\pm0.20)\times10^{-3}$       &  \citet{Wesson2015}    \\
$"$         & 806      & 4.0$\times10^{-4}$      &  \citet{bevan2016}    \\
$"$         & 806      & 9.0$\times10^{-5}$       &  \citet{bevan2016}    \\
$"$         & 1054     & 7.5$\times10^{-4}$     &  \citet{bevan2016}    \\
$"$         & 1153     & (3.0$\pm1.0)\times10^{-3}$       &  \citet{Wesson2015}    \\
$"$         & 1478     & 1.1$\times10^{-3}$      &  \citet{bevan2016}    \\
$"$         & 1862     & 1.2$\times10^{-3}$      &  \citet{bevan2016}    \\
$"$         & 2211     & 3.0$\times10^{-3}$       &  \citet{bevan2016}    \\
$"$         & 2875     & 8.0$\times10^{-3}$       &  \citet{bevan2016}    \\
$"$         & 3500     & 1.4$\times10^{-2}$      &  \citet{bevan2016}    \\
$"$         & 3604     & 1.7$\times10^{-2}$       &  \citet{bevan2016}    \\
$"$         & 8515     & 0.55$\pm0.15$           &  \citet{Matsuura}    \\
$"$         & 8515     & 0.6$\pm0.2$         &  \citet{Wesson2015}    \\
$"$         & 9131  & 0.20           &  \citet{Indebetouw2014}    \\
$"$         & 9200     & 0.8$\pm0.2$         &  \citet{Wesson2015}    \\
$"$         & 9200     & 0.6$\pm0.2$         &  \citet{Matsuura2015}    \\
SN~1993J         & 5844     & 0.12        &  \citet{Bevan2017}    \\
SN~1995N         & 827  & 6.0$\times10^{-5}$           &  Wesson et al. in prep    \\
$"$         & 1589  & 2.0$\times10^{-4}$           &  Wesson et al. in prep    \\
$"$         & 5648  & 0.10           &  Wesson et al. in prep    \\
$"$         & 7858  & 0.20           &  Wesson et al. in prep    \\
SN~1998S         & 359      & 1.9$\times10^{-3}$  &  \citet{Gall2014}    \\
$"$        & 499      & 1.7$\times10^{-3}$      &  \citet{Sugerman2006}    \\
$"$        & 499      & 4.0$\times10^{-5}$       &  \citet{Meikle2007ApJ}    \\
$"$        & 678      & 2.0$\times10^{-2}$        &  \citet{Sugerman2006}    \\
SN~2004dj        & 271      & 2.0$\times10^{-5}$       &  \citet{Szalai2011} \\
$"$        & 866      & (2.6$\pm0.6)\times10^{-4}$     &  \citet{Szalai2011} \\
$"$        & 1011     & (6.2$\pm1.4)\times10^{-4}$     &  \citet{Szalai2011} \\
$"$        & 1241     & (3.6$\pm^{1.0}_{0.6})\times10^{-4}$     &  \citet{Szalai2011} \\
SN~2004et        & 300      & 4.0$\times10^{-5}$       &  \citet{Kotak2009}    \\
$"$        & 300      & 7.6$\times10^{-5}$       &  \citet{Fabbri2011a}    \\
$"$        & 360      & 6.1$\times10^{-5}$     &  \citet{Kotak2009}    \\
$"$        & 360      & 1.0$\times10^{-4}$     &  \citet{Fabbri2011a}    \\
$"$        & 406      & 6.1$\times10^{-5}$     &  \citet{Kotak2009}    \\
$"$        & 406      & 2.2$\times10^{-4}$     &  \citet{Fabbri2011a}    \\
$"$        & 464      & 6.9$\times10^{-5}$     &  \citet{Kotak2009}    \\
$"$        & 464      & 5.0$\times10^{-4}$     &  \citet{Fabbri2011a}    \\
$"$        & 690      & 9.9$\times10^{-5}$     &  \citet{Kotak2009}    \\
$"$        & 690      & 1.5$\times10^{-3}$     &  \citet{Fabbri2011a}    \\
$"$        & 736      & 1.1$\times10^{-4}$     &  \citet{Kotak2009}    \\
$"$        & 795      & 1.4$\times10^{-4}$     &  \citet{Kotak2009} \\
SN~2005ad & 198 & 1.0$\times10^{-5}$ & \citet{Szalai2013}\\
\end{tabular}
    \end{table*}
    \begin{table*}
    \begin{tabular}{p{2.5cm}p{2.5cm}p{2.5cm}p{3.8cm}}
$"$ & 364 & 3.0$\times10^{-6}$& \citet{Szalai2013}\\
SN~2005af        & 194      & 4.2$\times10^{-5}$ &  \citet{Szalai2013}    \\
$"$        & 399       & 1.6$\times10^{-5}$ &  \citet{Szalai2013}    \\
$"$        & 576       & 2.6$\times10^{-5}$ &  \citet{Szalai2013}    \\
$"$        & 772       & 7.0$\times10^{-6}$ &  \citet{Szalai2013}    \\
$"$        & 940       & 4.5$\times10^{-5}$ &  \citet{Szalai2013}    \\
SN 2005ip        & 169      & (7.4$\pm^{61.8}_{4.3})\times10^{-6}$   &  \citet{Bevan2019}    \\
$"$        & 413      & (7.2$\pm^{22.3}_{6.2})\times10^{-6}$   &  \citet{Bevan2019}    \\
$"$        & 905      & (1.9$\pm^{22.6}_{1.8})\times10^{-3}$    &  \citet{Bevan2019}    \\
$"$        & 940      & 5.0$\times10^{-4}$      &  \citet{Fox2010}    \\
$"$        & 2242     & (8.7$\pm_{6.5}^{150.0})\times10^{-2}$      &  \citet{Bevan2019}    \\
$"$        & 3099     & (7.8$\pm_{6.8}^{151.0})\times10^{-2}$     &  \citet{Bevan2019}    \\
$"$        & 4075     & (6.9$\pm_{6.2}^{152})\times10^{-2}$    &  \citet{Bevan2019}    \\
SN~2006bc        & 550      & 2.0$\times10^{-3}$       &  \citet{Gallagher2012}    \\
SN~2006jc        & 229      & 2.9$\times10^{-4}$ &  \citet{Nozawa2008}    \\
SN~2007it        & 944      & 1.0$\times10^{-4}$      &  \citet{Andrews2011}    \\
$"$        & 944      & 4.6$\times10^{-5}$     &  \citet{Andrews2011}    \\
SN~2007oc        & 250      & 3.1$\times10^{-3}$     &  \citet{Szalai2013}    \\
$"$        & 415      & 3.7$\times10^{-3}$     &  \citet{Szalai2013}    \\
SN~2007od        & 300      & 2.5$\times10^{-4}$     &  \citet{Andrews2010}    \\
$"$        & 455      & 4.2$\times10^{-4}$     &  \citet{Andrews2010}    \\
SN~2010jl        & 464      & 7.0$\times10^{-4}$      &  \citet{bevan2020}    \\
$"$        & 526      & (2.50$\pm0.75)\times10^{-4}$     &  \citet{bevan2020}    \\
$"$        & 526      & 7.0$\times10^{-4}$      &  \citet{bevan2020}    \\
$"$        & 915      & (2.0$\pm0.6)\times10^{-3}$       &  \citet{bevan2020}          \\
$"$        & 996      & 3.0$\times10^{-3}$       &  \citet{bevan2020}    \\
$"$        & 1286     & (5.0$\pm1.5)\times10^{-3}$       &  \citet{bevan2020}    \\
$"$        & 1367     & 1.0$\times10^{-2}$        &  \citet{bevan2020}    \\
SN~2014C         & 1623     & (4.5$\pm0.2)\times10^{-3}$       &  \citet{Tinyanont2019b}    \\
SN~2018hfm    & 182      & (2.8$\pm^{99.5}_{2.75})\times10^{-5}$    & \citet{Zhang2022}                                        \\
$"$      & 389      & (1.0$\pm^{38.8}_{0.7})\times10^{-4}$    & \citet{Zhang2022}    \\
\hline
\label{table:dustmass-allworks-tab}
\end{tabular}
\end{table*}

\begin{figure*}
\centering
\includegraphics[width=\linewidth]{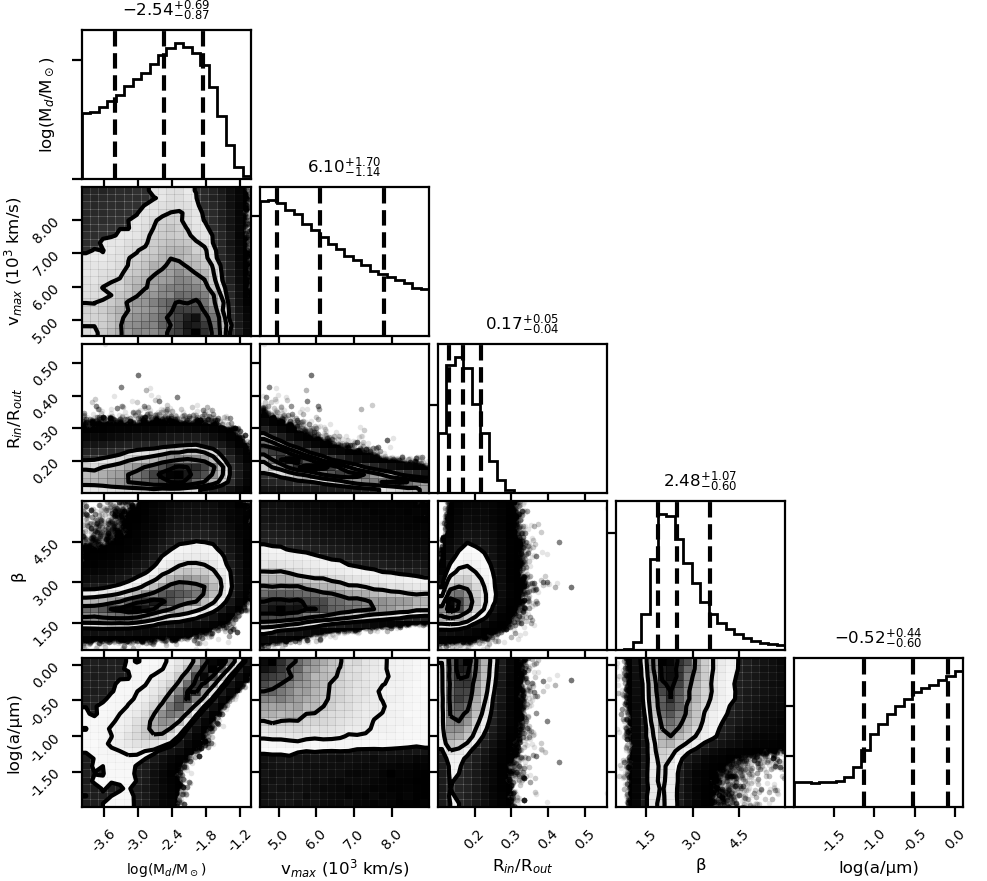}

\caption{The full posterior probability distribution for the 5D model of SN~1957D's {[}O {\sc iii}{]} profile at day 11371, using 100~per~cent carbon dust. The contours of the 2D distributions represent 0.5$\sigma$, 1.0$\sigma$, 1.5$\sigma$ and 2.0$\sigma$  and the dashed, black vertical lines represent the 16th, 50th, and 84th quantiles for the 1-D marginalized probability distributions. }
\label{fig:57d-bayesian-2011}
\end{figure*}

\begin{figure*}
\centering

\includegraphics[width=\linewidth]{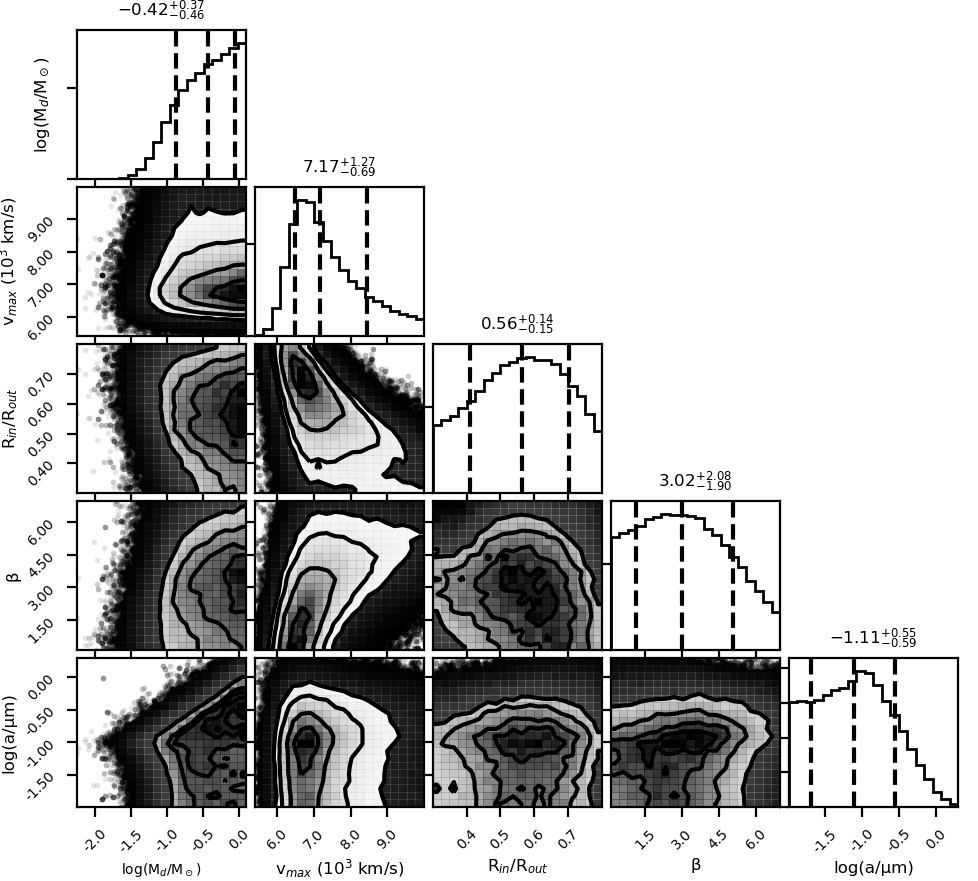}

\caption{The full posterior probability distribution for the 5D model of SN~1979C's [O {\sc ii}] profile at day 10575, using 100~per~cent carbon dust. The contours of the 2D distributions represent 0.5$\sigma$, 1.0$\sigma$, 1.5$\sigma$ and 2.0$\sigma$ and the dashed, black vertical lines represent the 16th, 50th, and 84th quantiles for the 1-D marginalized probability distributions.}
\label{fig:79c-bayesian-2008}
\end{figure*}

\begin{figure*}
\centering

\includegraphics[width=\linewidth]{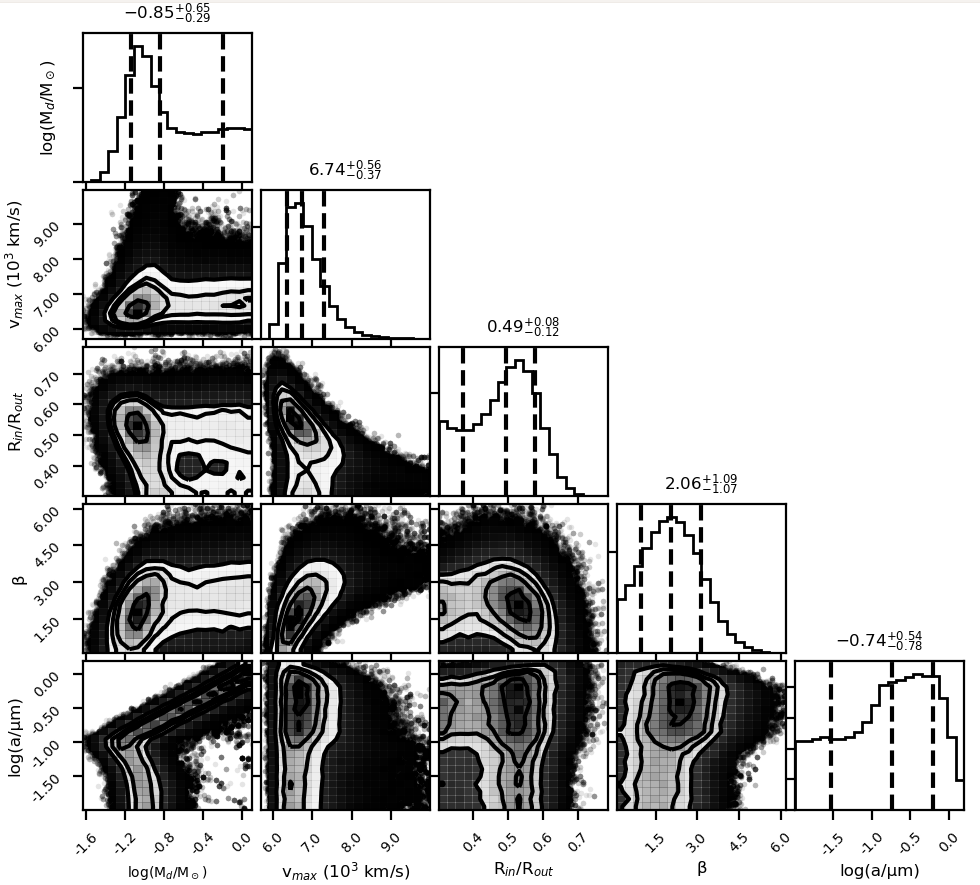}

\caption{The full posterior probability distribution for the 5D model of SN~1979C's [O~{\sc ii}] 7319,7330-$\AA$ profile at day 13903, using 100~per~cent carbon dust. The contours of the 2D distributions represent 0.5$\sigma$, 1.0$\sigma$, 1.5$\sigma$ and 2.0$\sigma$ and the dashed, black vertical lines represent the 16th, 50th, and 84th quantiles for
the 1-D marginalized probability distributions.}
\label{fig:79c-bayesian-2017}
\end{figure*}

\begin{figure*}
\centering
\includegraphics[width=\linewidth]{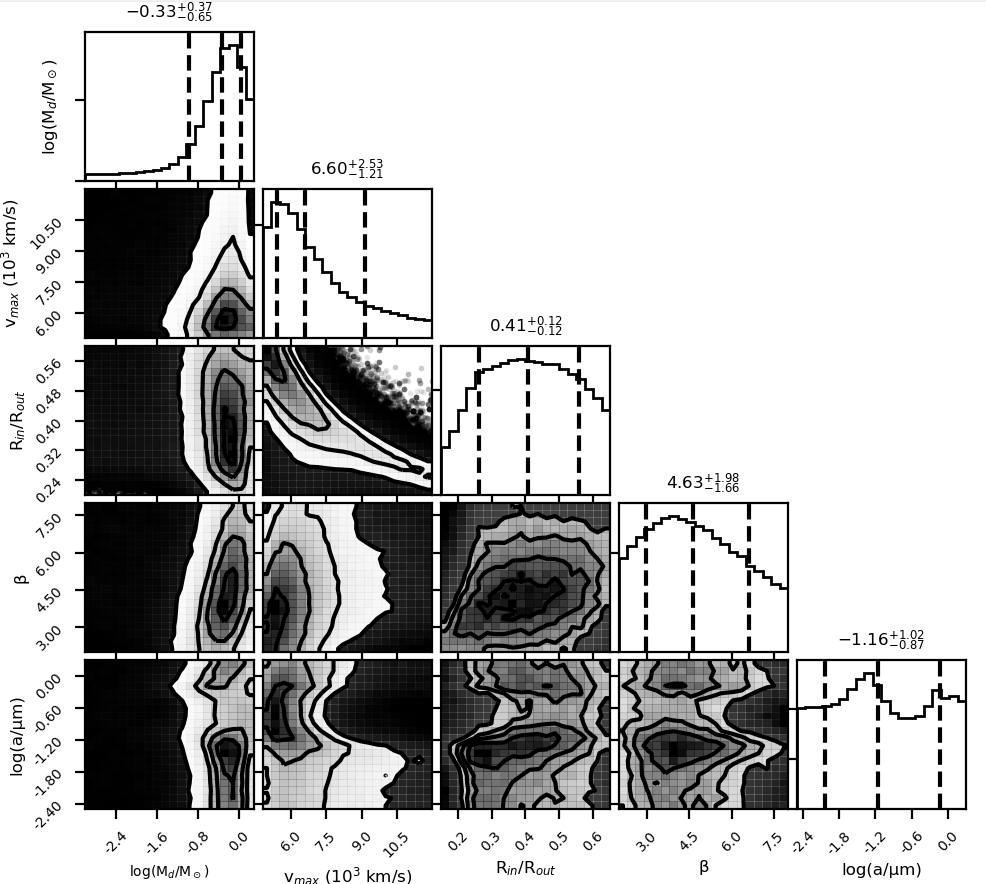}

\caption{The full posterior probability distribution for the 5D model of SN~1996cr's [O~{\sc ii}] and [O~{\sc iii}] profiles at day 7370, using 100~per~cent astronomical silicate dust. The contours of the 2D distributions represent 0.5$\sigma$, 1.0$\sigma$, 1.5$\sigma$ and 2.0$\sigma$ and the dashed, black vertical lines represent the 16th, 50th, and 84th quantiles for
the 1-D marginalized probability distributions. }
\label{fig:96cr-bayesian}
\end{figure*}

\begin{figure*}
\centering

\includegraphics[width=\linewidth]{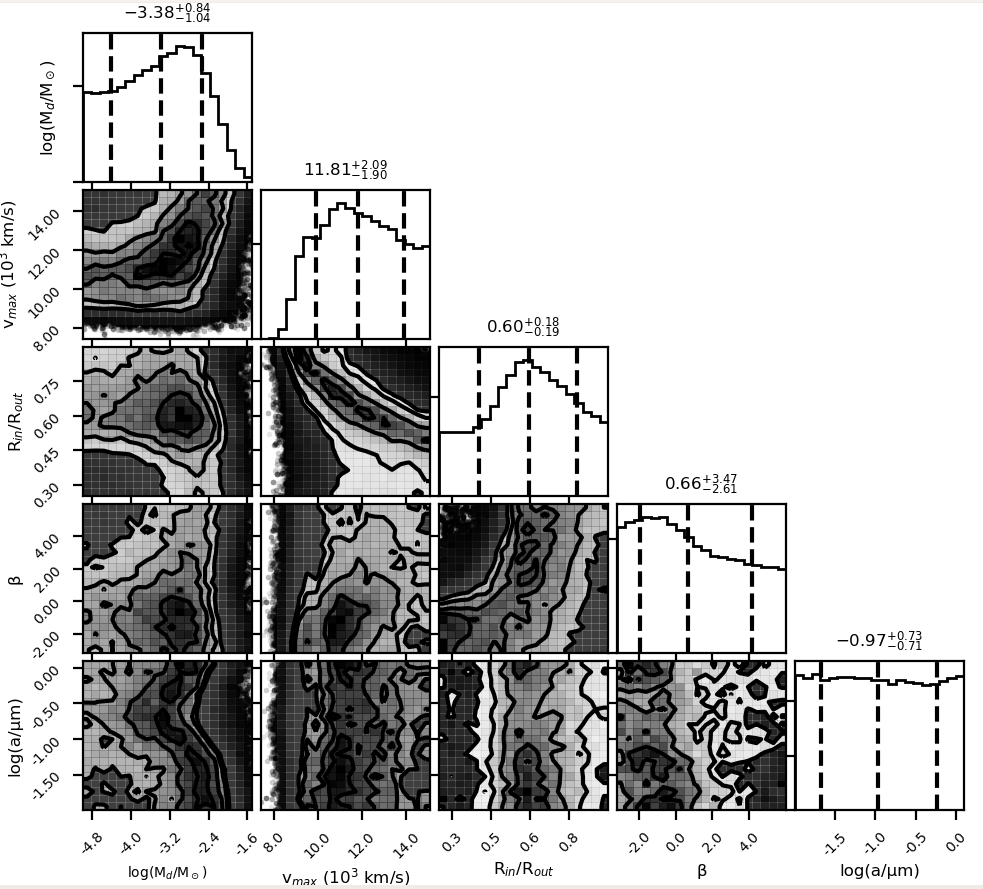}

\caption{The full posterior probability distribution for the 5D model of SN~2004et's H$\alpha$ profile at day 646, using 100~per~cent silicate dust. The contours of the 2D distributions represent 0.5$\sigma$, 1.0$\sigma$, 1.5$\sigma$ and 2.0$\sigma$ and the dashed, black vertical lines represent the 16th, 50th, and 84th quantiles for
the 1-D marginalized probability distributions.}
\label{fig:04et-bayesian}
\end{figure*}

\begin{figure*}
\centering
\includegraphics[width=\linewidth]{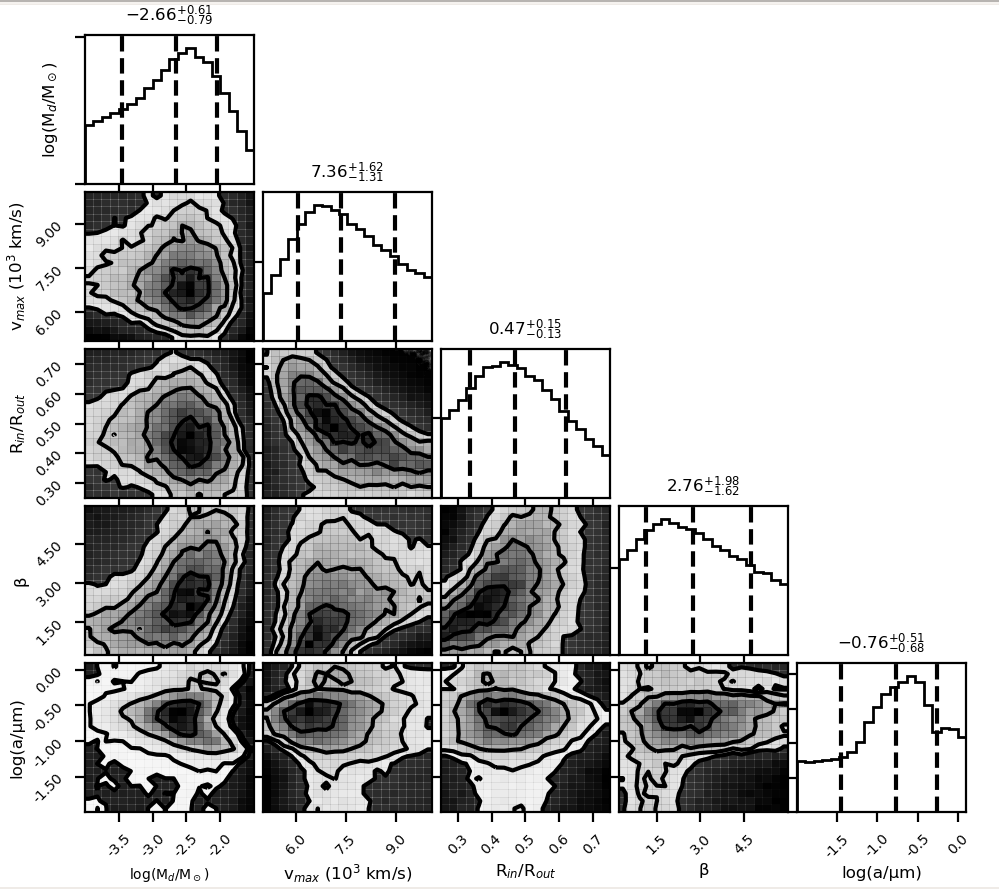}

\caption{The full posterior probability distribution for the 5D model of SN~2011ja's H$\alpha$ profile at day 2259, using 100~per~cent silicate dust. The contours of the 2D distributions represent 0.5$\sigma$, 1.0$\sigma$, 1.5$\sigma$ and 2.0$\sigma$ and the dashed, black vertical lines represent the 16th, 50th, and 84th quantiles for
the 1-D marginalized probability distributions. }
\label{fig:2011ja-bayesian}
\end{figure*}

\begin{figure*}
\centering

\includegraphics[width=\linewidth]{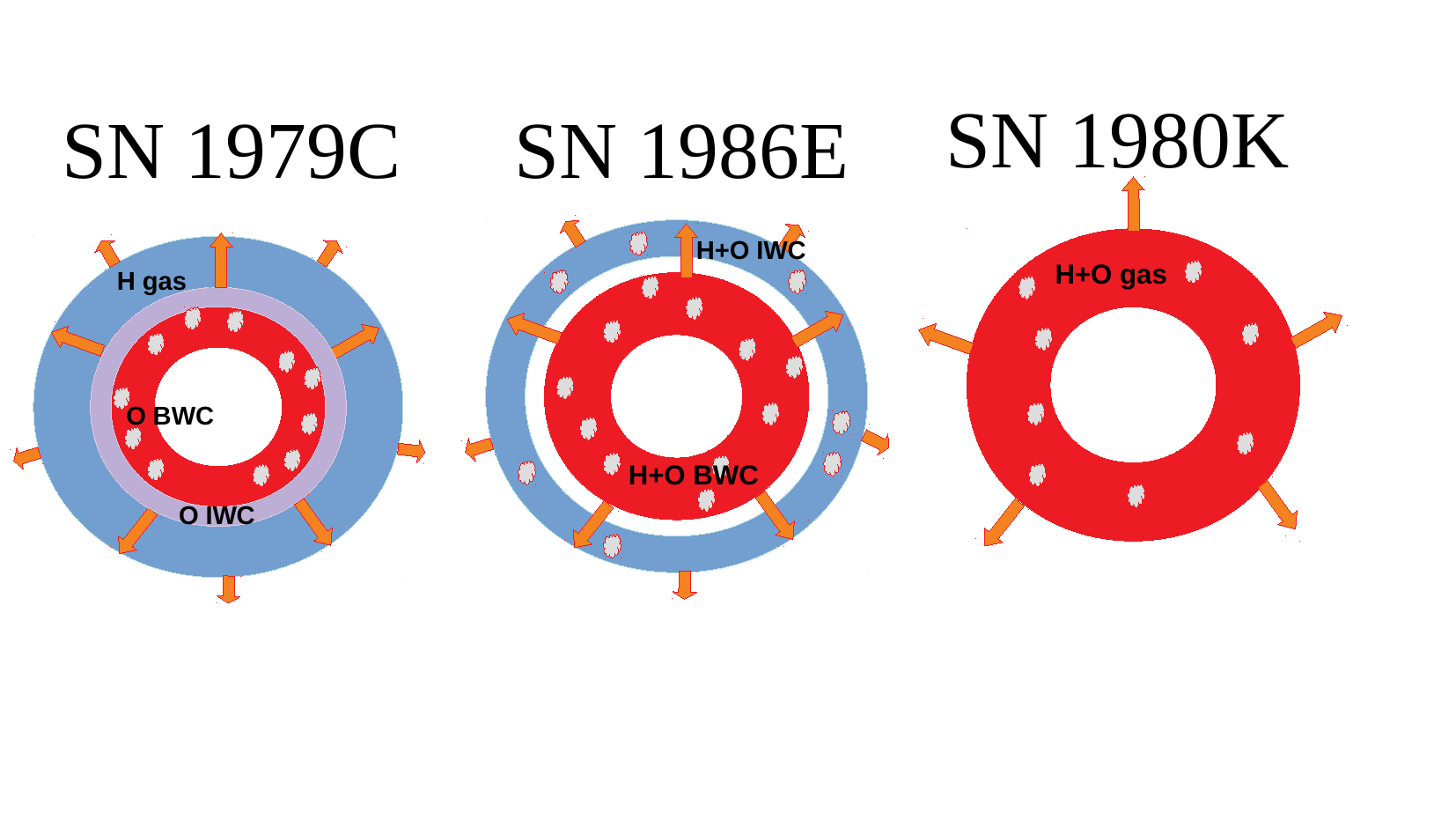}
\caption{Geometry of the H$\alpha$ and oxygen line emission shells in the Type~II-L SNe 1979C, 1986E SN 1980K, as deduced from our {\sc damocles} modelling. IWC stands for intermediate width component and BWC stands for broad width component. Small grey clouds represent the presence of dust. SN~1998S was deduced to have the same geometry as SN~1980K.}
\label{fig:3sn}
\end{figure*}

\begin{figure*}
\centering

\includegraphics[width=\linewidth]{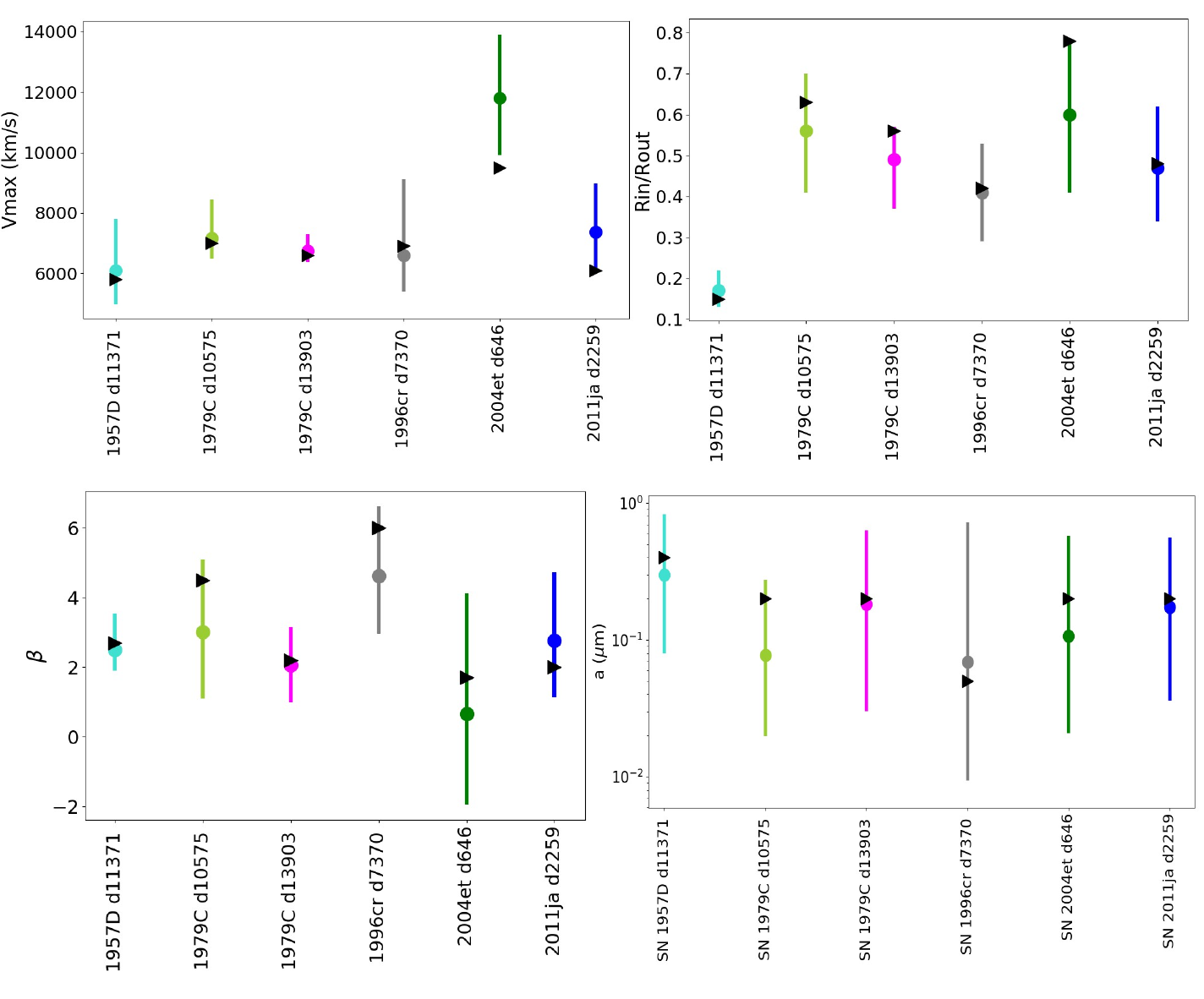}
\caption{The V$_{\rm max}$, R$_{in}$/R$_{out}$, $\beta$ and log($a$) parameters derived from manual fits (black arrow-heads), and the median values (circles) of the same parameters taken from the 1D posterior distribution from Bayesian models, with the 16th and 84th quartile of the distribution as the uncertainty limits, for a range of CCSNe studied in this work.}
\label{fig:bayes-param-com}
\end{figure*}

\begin{figure*}
\centering

\includegraphics[width=\linewidth]{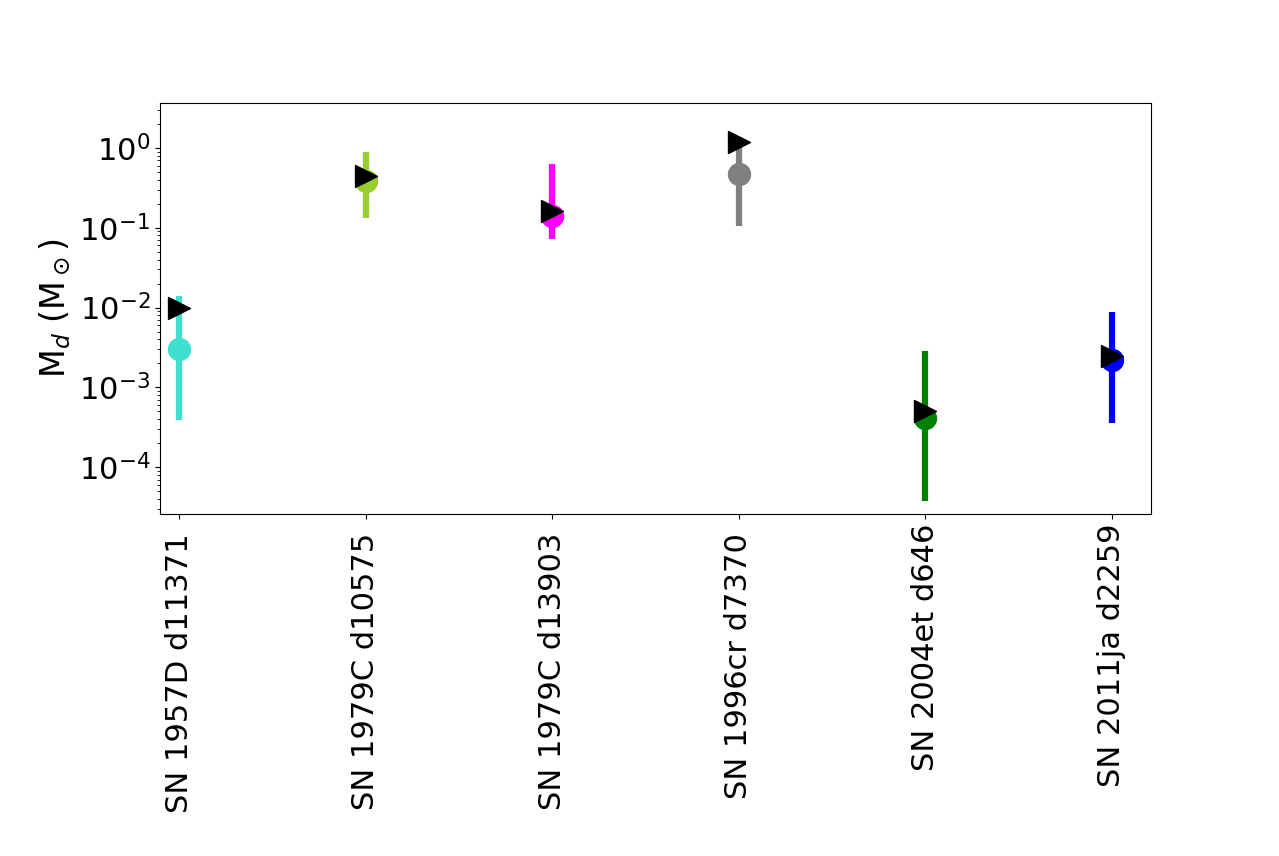}
\caption{The dust masses, M$_d$, derived from a manual fit (black arrow-heads), and the median values (circles) of the same parameter taken from the 1D posterior distribution from Bayesian models, with the 16th and 84th quartile of the distribution as the uncertainty limits, for a range of CCSNe studied in this work.}
\label{fig:bayes-param-com-dm}
\end{figure*}

\centering

\newpage
\newpage

\bibliography{damoc-paper} %


\label{lastpage}
\end{document}